%% file: arxiv_2018_tree_edit_distance.tex
\documentclass[a4paper,10pt]{article}

\usepackage{amsmath}
\usepackage{amsthm}
\usepackage{amssymb}
\usepackage{mathtools}
\usepackage{bm}

\usepackage{booktabs}

\usepackage{graphicx}
\usepackage{tango-colors}
\usepackage{tikz}
\usepackage{tikz-qtree}
\usetikzlibrary{backgrounds}
\usetikzlibrary{arrows}
\usetikzlibrary{shapes}
\usetikzlibrary{shapes.misc}
\usetikzlibrary{tikzmark}
\usetikzlibrary{matrix}

\usepackage{algorithm}
\usepackage{algpseudocode}

\usepackage{hyperref}

\usepackage[a4paper, left=3cm, right=2.5cm, top=3cm, bottom=3cm]{geometry}

\usepackage[backend=biber, style=authoryear]{biblatex}
\usepackage[english]{babel}
\usepackage[utf8]{inputenc}
\usepackage[T1]{fontenc}
\usepackage{csquotes}

\input{symbols.tex}

\input{tikz_styles.tex}

\pgfdeclarelayer{bg}    
\pgfsetlayers{bg,main}  

\title{Revisiting the tree edit distance and its backtracing: A tutorial}
\author{Benjamin Paaßen}
\date{Supplementary material for the ICML 2018 paper: Tree Edit Distance Learning via Adaptive Symbol Embeddings}

\begin{document}

\maketitle

\begin{abstract}
Almost 30 years ago, \textcite{Zhang1989} published a seminal paper describing an efficient dynamic programming algorithm
computing the tree edit distance, that is, the minimum number of node deletions, insertions, and replacements that are necessary
to transform one tree into another. Since then, the tree edit distance has been widely applied, for example in
biology and intelligent tutoring systems. However, the original paper of Zhang and Shasha can
be challenging to read for newcomers and it does not describe how to efficiently infer the optimal edit script.

In this contribution, we provide a comprehensive tutorial to the tree edit distance algorithm of Zhang and Shasha. We further
prove metric properties of the tree edit distance, and describe efficient algorithms to infer the cheapest edit script,
as well as a summary of all cheapest edit scripts between two trees.

A reference implementation of the algorithms presented in this work can be found at\\
\url{https://pypi.org/project/edist/}.
\end{abstract}

\textbf{Change Note:} On September 6, 2022, Daniel Germann and me
identified a problem in the prior version of Algorithm~\ref{alg:backtrace}.
In particular, we noticed that a purely iterative backtracing scheme
failed to appropriately take subtree boundaries into account.
Therefore, I now switched to a recursive writeup, which is consistent
with the reference implementation. Theorem~\ref{thm:backtrace}
and the following example have been updated, accordingly.
Thanks go to Daniel Germann for pointing me to this mistake.

\section{Introduction}

The tree edit distance \parencite[TED,][]{Zhang1989} between two trees $\ltree$ and $\rtree$ is defined as the minimum number of
nodes that need to be replaced, deleted, or inserted in $\ltree$ to obtain $\rtree$. This makes the TED an intuitive notion
of distance, which has been applied in a host of different application areas \parencite{Pawlik2011}, for example to
compare RNA secondary structures and phylogenetic trees in biology \parencite{Akutsu2010,Henikoff1992,McKenna2010,Smith1981},
or to recommend edits to students in intelligent tutoring systems \parencite{Choudhury2016,Freeman2016,Nguyen2014,Paassen2018JEDM,Rivers2015}.
As such, the TED has certainly stood the test of time and is still of great interest to a broad community. Unfortunately,
though, a detailed tutorial on the TED seems to lack, such that users tend to treat it as a black box.
This is unfortunate as the TED lends itself for straightforward adjustments to the application
domain at hand, and this potential remains under-utilized.

This contribution is an attempt to provide a comprehensive
tutorial to the TED, enabling users to implement it themselves, adjust it to their needs, and
compute not only the distance as such but also the optimal edits which transform $\ltree$ into $\rtree$.
Note that we focus here on the original version of the TED with a time complexity of
$\effic(\lnodelim^2 \cdot \rnodelim^2)$ and a space complexity of $\effic(\lnodelim \cdot \rnodelim)$,
where $\lnodelim$ and $\rnodelim$ are the number of nodes in $\ltree$
and $\rtree$, respectively \parencite{Zhang1989}.
Recent innovations have improved the worst-case time complexity to cubic time
\parencite{Pawlik2011,Pawlik2016}, but require deeper knowledge regarding tree decompositions. Furthermore,
the practical runtime complexity of the original TED algorithm is still
competitive for balanced trees, such that we regard it as a good choice in many practical scenarios
\parencite{Pawlik2011}.

Our tutorial roughly follows the structure of the original paper of \textcite{Zhang1989}, that is,
we start by first defining trees (section~\ref{sec:trees}) and edit scripts on trees
(section~\ref{sec:edits}), which are the basis for the TED. To make the TED more flexible, we
introduce generalized cost functions on edits (section~\ref{sec:costs}), which are a good interface
to adjust the TED for custom applications. We conclude the theory section by introducing
mappings between subtrees (section~\ref{sec:mappings}), which constitute the interface for an efficient
treatment of the TED.

These concepts form the basis for our key theorems, namely that the cheapest mapping between two
trees can be decomposed via recurrence equations, which in turn form the basis for Zhang and Shasha's
dynamic programming algorithm for the TED (section~\ref{sec:dp}). Finally, we conclude
this tutorial with a section on the backtracing for the TED, meaning
that we describe how to efficiently compute the cheapest edit script transforming one tree into
another (section~\ref{sec:coopts}).

A reference implementation of the algorithms presented in this work can be found at\\
\url{https://pypi.org/project/edist/}.

\section{Theory and Definitions}
\label{sec:ted}

We begin our description of the tree edit distance (TED) by defining trees, forests, and tree edits, which provides the
basis for our first definition of the TED. We will then revise this definition by permitting
customized costs for each edit, which yields a generalized version of the TED.
Finally, we will introduce the concept of a tree mapping, which will form the basis for the dynamic programming
algorithm.

\subsection{Trees}
\label{sec:trees}

\begin{dfn}[Alphabet, Tree, Label, Children, Leaf, Subtree, Parent, Ancestor, Forest]
We define an \emph{alphabet} as an arbitrary set $\alphabet$.

We define a \emph{tree} $\tree$ over the alphabet $\alphabet$ as an expression of the form
$\node(\tree_1, \ldots, \tree_\childlim)$, where $\node \in \alphabet$, and
$\tree_1, \ldots, \tree_\childlim$ is a (possibly empty) list of
trees over $\alphabet$.
We denote the set of all trees over $\alphabet$ as $\trees(\alphabet)$.

We call $\node$ the \emph{label} of $\ltree$, and we call
$\tree_1, \ldots, \tree_\childlim$ the \emph{children} of $\tree$.
If a tree has no children (i.e. $\childlim = 0$), we call it a \emph{leaf}.
In terms of notation, we will generally omit the brackets for leaves, i.e.\ $\node$ is a notational
shorthand for the tree $\node()$.

We define a \emph{subtree} of $\tree$ as either $\tree$ itself, or as a subtree of a child of
$\tree$.
Conversely, we call $\tree$ the \emph{parent} of $\rtree$ if $\rtree$ is a child of
$\tree$, and we call $\tree$ an \emph{ancestor} of $\rtree$ if $\tree$ is either the parent of
$\rtree$ or an ancestor of the parent of $\tree$. We call the multi-set of labels for all subtrees
of a tree the \emph{nodes} of the tree.

We call a list of trees $\tree_1, \ldots, \tree_\childlim$ from $\trees(\alphabet)$ a \emph{forest}
over $\alphabet$, and we denote the set of all possible forests over $\alphabet$ as
$\trees(\alphabet)^*$. We denote the empty forest as $\epsilon$.
\end{dfn}

As an example, consider the alphabet $\alphabet = \{\sym{a}, \sym{b}\}$. Some example trees
over $\alphabet$ are $\sym{a}$, $\sym{b}$, $\sym{a}(\sym{a})$, $\sym{a}(\sym{b})$,
$\sym{b}(\sym{a}, \sym{b})$, and $\sym{a}(\sym{b}(\sym{a}, \sym{b}), \sym{b})$.

An example forest over this alphabet is $\sym{a}, \sym{b}, \sym{b}(\sym{a}, \sym{b})$.
Note that each tree is also a forest. This is important as many of our proofs in this
paper will be concerned with forests, and these proofs apply to trees as well.

Now, consider the example tree $\ltree = \sym{a}(\sym{b}(\sym{c}, \sym{d}), \sym{e})$ from
Figure~\ref{fig:preorder} (left). $\sym{a}$ is the label of $\ltree$, and $\sym{b}(\sym{c}, \sym{d})$ as 
well as $\sym{e}$ are the children of $\ltree$. Conversely, $\ltree$ is the parent of
$\sym{b}(\sym{c}, \sym{d})$ and $\sym{e}$. The leaves of $\ltree$ are $\sym{c}$, $\sym{d}$, and
$\sym{e}$. The subtrees of $\ltree$ are $\ltree$, $\sym{b}(\sym{c}, \sym{d})$, $\sym{c}$,
$\sym{d}$, and $\sym{e}$. The nodes of $\ltree$ are $\sym{a}$, $\sym{b}$, $\sym{c}$, $\sym{d}$,
and $\sym{e}$.

\subsection{Tree Edits}
\label{sec:edits}

Next, we shall consider \emph{edits} on trees, that is, functions which \emph{change} trees
(or forests). In particular, we define:

\begin{dfn}[Tree Edit, Edit Script]
A tree edit over the alphabet $\alphabet$ is a function $\edit$
which maps a forest over $\alphabet$ to another forest over $\alphabet$, that is,
a tree edit $\edit$ over $\alphabet$ is any kind of function
$\edit : \trees(\alphabet)^* \to \trees(\alphabet)^*$.

In particular, we define a \emph{deletion} as the following function $\del$.
\begin{align*}
\del(\epsilon) &:= \epsilon \\
\del(\lnode(\rtree_1, \ldots, \rtree_\rnodelim), \ltree_2, \ldots, \ltree_\childlim) &:= \rtree_1, \ldots, \rtree_\rnodelim, \ltree_2, \ldots, \ltree_\childlim
\end{align*}
We define a \emph{replacement} with node $\rnode \in \alphabet$ as the following function $\rep_\rnode$.
\begin{align*}
\rep_\rnode(\epsilon) &:= \epsilon \\
\rep_\rnode(\lnode(\rtree_1, \ldots, \rtree_\rnodelim), \ltree_2, \ldots, \ltree_\childlim) &:=
\rnode(\rtree_1, \ldots, \rtree_\rnodelim), \ltree_2, \ldots, \ltree_\childlim
\end{align*}
And we define an \emph{insertion} of node $\rnode \in \alphabet$ as parent of
the trees $\lchildidx$ to $\rchildidx-1$ as the following function
$\ins_{\rnode, \lchildidx, \rchildidx}$.
\begin{align*}
\ins_{\rnode, \lchildidx, \rchildidx}(\ltree_1, \ldots, \ltree_\childlim) &:=
\begin{cases}
\ltree_1, \ldots, \ltree_\childlim & \text{if } \rchildidx > \childlim + 1, \lchildidx > \rchildidx, \text{ or } \lchildidx < 1 \\
\ltree_1, \ldots, \ltree_{\lchildidx-1}, \rnode, \ltree_\lchildidx, \ldots, \ltree_\childlim & \text{if } 1 \leq \lchildidx = \rchildidx \leq \childlim + 1 \\
\ltree_1, \ldots, \ltree_{\lchildidx-1}, \rnode(\ltree_\lchildidx, \ldots, \ltree_{\rchildidx-1}),
\ltree_\rchildidx, \ldots, \ltree_\childlim  & \text{if } 1 \leq \lchildidx < \rchildidx \leq \childlim + 1
\end{cases}
\end{align*}

We define an \emph{edit script} $\script$ as a list of tree edits $\edit_1, \ldots, \edit_\editlim$.
We define the application of an edit script $\script = \edit_1, \ldots, \edit_\editlim$ to
a tree $\ltree$ as the composition of all edits, that is: $\script(\ltree) := \edit_1 \circ \ldots \circ \edit_\editlim(\ltree)$,
where $\circ$ denotes the contravariant composition operator, i.e.\ $f \circ g(x) := g(f(x))$.

Let $\edits$ be a set of tree edits. We denote the set of all possible edit scripts using edits from
$\edits$ as $\edits^*$. We denote the empty script as $\epsilon$.
\end{dfn}

As an example, consider the alphabet $\alphabet = \{\sym{a}, \sym{b}\}$ and the edit $\rep_\sym{b}$,
which replaces the first node in a forest with a $\sym{b}$. If we apply this edit to the example
tree $\sym{a}(\sym{b}, \sym{a})$, we obtain $\rep_\sym{b}(\sym{a}(\sym{b}, \sym{a})) =
\sym{b}(\sym{b}, \sym{a})$. Now, consider the edit script $\script := \del, \rep_\sym{a}, \ins_{\sym{a}, 1, 3}$,
which yields the following result for the example tree $\sym{a}(\sym{b}, \sym{a})$.
\begin{align*}
\script(\sym{a}(\sym{b}, \sym{a})) &= \del \circ \rep_\sym{a} \circ \ins_{\sym{a}, 1, 3}(\sym{a}(\sym{b}, \sym{a})) \\
&= \rep_\sym{a} \circ \ins_{\sym{a}, 1, 3}(\sym{b}, \sym{a}) \\
&= \ins_{\sym{a}, 1, 3}(\sym{a}, \sym{a}) \\
&= \sym{a}(\sym{a}, \sym{a}) \\
\end{align*}

Note that tree edits are defined over forests, not only over trees. This is necessary
because, as in our example above, deletions may change trees into forests and need to be followed
up with insertions to obtain a tree again.

Based on edit scripts, we can define the TED.

\begin{dfn}[Edit Distance]
Let $\alphabet$ be an alphabet and $\edits$ be a set of tree edits over $\alphabet$. Then,
the TED according to $\edits$ is defined as the function

\begin{align}
&\dist_\edits : \trees(\alphabet) \times \trees(\alphabet) \to \N \\
&\dist_\edits(\ltree, \rtree) = \min_{\script \in \edits^*} \Big\{ |\script| \Big| \script(\ltree) = \rtree \Big\}
\end{align}
\end{dfn}

In other words, we define the TED between two trees $\ltree$ and $\rtree$
as the minimum number of edits we need to transform $\ltree$ to $\rtree$.

Our definition of tree edit is very broad and includes many edits which are not meaningful
in most tasks. Therefore, the standard TED of \textcite{Zhang1989} is restricted to the three kinds
of special edits listed above, namely deletions, which remove a single node from a forest,
insertions, which insert a single node into a forest,
and replacements, which replace a single node in a forest with another node.
Up to now, we have only defined versions of these edits which apply to the \emph{first} node in a forest.
We now go on to define variants which can be applied to \emph{any} node in a given forest. To this
end, we need a way to uniquely identify single nodes in a forest. We address this problem
via the concept of a \emph{pre-order} (sometimes called \emph{depth-first-search order}).
The pre-order just lists all subtrees of a forest recursively, starting with the first tree in its
forest, followed by the pre-order of its children and the pre-order of the remaining trees.
More precisely, we define the pre-order as follows.

\begin{dfn}[Pre-Order]
\label{dfn:preorder}
Let $\lforest = \lnode(\rtree_1, \ldots, \rtree_\rnodelim), \ltree_2, \ldots, \ltree_\childlim$ be a (non-empty)
forest over some alphabet $\alphabet$.
Then, we define the \emph{pre-order} of $\lforest$ as the list
\begin{equation}
\pre(\lforest) := \lnode(\rtree_1, \ldots, \rtree_\rnodelim) \concat \pre(\rtree_1, \ldots, \rtree_\rnodelim) \concat \pre(\ltree_2, \ldots, \ltree_\childlim),
\end{equation}
where $\concat$ denotes list concatenation.

We define the pre-order of the empty forest as $\pre(\epsilon) = \epsilon$.

As a shorthand, we denote the $\lnodeidx$th subtree $\pre(\lforest)_\lnodeidx$ as
$\ltree_\lnodeidx$, and we denote the label of $\ltree_\lnodeidx$ as $\lnode_\lnodeidx$.

We define the \emph{size} of the forest $\lforest$ as the length of the pre-order,
that is $|\lforest| := |\pre(\lforest)|$.

Further, we define $\paridx_\ltree(\lnodeidx)$ as the pre-order index of the parent of $\ltree_\lnodeidx$,
that is, $\ltree_{\paridx_\ltree(\lnodeidx)}$ is the parent of $\ltree_\lnodeidx$. If there is no
parent, we define $\paridx_\ltree(\lnodeidx) := 0$.

Finally, we define $\childidx_\ltree(\lnodeidx)$ as the child index of $\ltree_\lnodeidx$,
that is, $\ltree_\lnodeidx$ is the $\childidx_\ltree(\lnodeidx)$th child of
$\ltree_{\paridx_\ltree(\lnodeidx)}$.
If $\paridx_\ltree(\lnodeidx) = 0$, we define $\childidx_\ltree(\lnodeidx)$ as the index of $\ltree_\lnodeidx$ in
the forest, that is, $\ltree_\lnodeidx$ is the $\childidx_\ltree(\lnodeidx)$th tree in $\lforest$.
\end{dfn}

Consider the example of the tree $\ltree = \sym{a}(\sym{b}(\sym{c}, \sym{d}), \sym{e})$ from
Figure~\ref{fig:preorder} (left). Here, the pre-order is
$\pre(\ltree) = \sym{a}(\sym{b}(\sym{c}, \sym{d}), \sym{e})$,
$\sym{b}(\sym{c}, \sym{d})$, $\sym{c}$, $\sym{d}$, $\sym{e}$.
Figure~\ref{fig:preorder} (right) lists for all $\lnodeidx$ the subtrees $\ltree_\lnodeidx$,
the nodes $\lnode_\lnodeidx$, the parents $\paridx_\ltree(\lnodeidx)$, and the child indices $\childidx_\ltree(\lnodeidx)$.

\begin{figure}
\begin{center}
\begin{tikzpicture}[sibling distance=1cm, level distance=1.1cm, node distance=0.7cm]
\begin{scope}[shift={(-3,0)}]
\Tree [.\node (a) {$\sym{a}$};
	[.\node (b) {$\sym{b}$};
		\node (c) {$\sym{c}$};
		\node (d) {$\sym{d}$};
	]
	\node (e) {$\sym{e}$};
]

\node [left of=a] {1};
\node [left of=b] {2};
\node [left of=c] {3};
\node [right of=d] {4};
\node [right of=e] {5};

\end{scope}

\node[below, outer sep=-0.5cm] at (+4,0) {
\begin{tabular}{ccccc}
$\lnodeidx$ & $\ltree_\lnodeidx$ & $\lnode_\lnodeidx$ & $\paridx_\ltree(\lnodeidx)$ & $\childidx_\ltree(\lnodeidx)$ \\
\cmidrule(lr){1-1} \cmidrule(lr){2-5}
$1$ & $\sym{a}(\sym{b}(\sym{c}, \sym{d}), \sym{e})$ & $\sym{a}$ & $0$     & $1$ \\
$2$ & $\sym{b}(\sym{c}, \sym{d})$& $\sym{b}$          & $1$               & $1$ \\
$3$         & $\sym{c}$          & $\sym{c}$          & $2$               & $1$ \\
$4$         & $\sym{d}$          & $\sym{d}$          & $2$               & $2$ \\
$5$         & $\sym{e}$          & $\sym{e}$          & $1$               & $2$
\end{tabular}
};
\end{tikzpicture}
\end{center}
\caption{Left: The tree $\ltree = \sym{a}(\sym{b}(\sym{c}, \sym{d}), \sym{e})$
with pre-order indices drawn next to each node.
Right: A table listing the subtrees $\ltree_\lnodeidx$, the nodes $\lnode_\lnodeidx$, the parents $\paridx_\ltree(\lnodeidx)$,
and the child indices $\childidx_\ltree(\lnodeidx)$ for all pre-order indices $\lnodeidx \in \{1, \ldots, 5\}$
for the tree $\ltree = \sym{a}(\sym{b}(\sym{c}, \sym{d}), \sym{e})$.}
\label{fig:preorder}
\end{figure}

Based on the pre-order, we can specify replacements, deletions, and insertions as follows:

\begin{dfn}[Replacements, Deletions, Insertions]
Let $\alphabet$ be some alphabet, let $\ltree_1 =
\lnode(\rtree_1, \ldots, \rtree_\rnodelim)$ be a tree over $\alphabet$, and let
$\lforest = \ltree_1, \ldots, \ltree_\childlim$ be a (non-empty) forest over $\alphabet$.

We define a deletion of the $\lnodeidx$th node as the following
function $\del_\lnodeidx$.
\begin{equation*}
\del_\lnodeidx(\lforest) :=
\begin{cases}
\lforest & \text{if } \lnodeidx < 1 \\
\del(\lforest) & \text{if } \lnodeidx = 1 \\
\lnode\big(\del_{\lnodeidx-1}(\rtree_1, \ldots, \rtree_\rnodelim\big), \ltree_2, \ldots, \ltree_\childlim & \text{if } 1 < \lnodeidx \leq \siz{\ltree_1} \\
\ltree_1, \del_{\lnodeidx-\siz{\ltree_1}}(\ltree_2, \ldots, \ltree_\childlim) & \text{if } \lnodeidx > \siz{\ltree_1} \\
\end{cases}
\end{equation*}
We define $\del_\lnodeidx(\epsilon) = \epsilon$.

We define a replacement of the $\lnodeidx$th node with $\rnode \in \alphabet$ as the following
function $\rep_{\lnodeidx, \rnode}$.
\begin{equation*}
\rep_{\lnodeidx, \rnode}(\lforest) :=
\begin{cases}
\lforest & \text{if } \lnodeidx < 1 \\
\rep_\rnode(\lforest) & \text{if } \lnodeidx = 1 \\
\lnode\big(\rep_{\lnodeidx-1, \rnode}(\rtree_1, \ldots, \rtree_\rnodelim\big), \ltree_2, \ldots, \ltree_\childlim & \text{if } 1 < \lnodeidx \leq \siz{\ltree_1} \\
\ltree_1, \rep_{\lnodeidx-\siz{\ltree_1}, \rnode}(\ltree_2, \ldots, \ltree_\childlim) & \text{if } \lnodeidx > \siz{\ltree_1} \\
\end{cases}
\end{equation*}
We define $\rep_{\lnodeidx, \rnode}(\epsilon) = \epsilon$.

Finally, we define an \emph{insertion} of node $\rnode \in \alphabet$ as parent of
the children $\lchildidx$ to $\rchildidx-1$ of the $\lnodeidx$th node as the following function
$\ins_{\lnodeidx, \rnode, \lchildidx, \rchildidx}$.
\begin{align*}
\ins_{\lnodeidx, \rnode, \lchildidx, \rchildidx}(\lforest) &:=
\begin{cases}
\lforest & \text{if } \lnodeidx < 0 \\
\ins_{\rnode, \lchildidx, \rchildidx}(\lforest) & \text{if } \lnodeidx = 0 \\
\lnode\big(\ins_{\lnodeidx-1, \rnode, \lchildidx, \rchildidx}(\rtree_1, \ldots, \rtree_\rnodelim)\big) \ltree_2, \ldots, \ltree_\childlim & \text{if } 1 \leq \lnodeidx \leq \siz{\ltree_1} \\
\ltree_1, \ins_{\lnodeidx-\siz{\ltree_1}, \rnode, \lchildidx, \rchildidx}(\ltree_2, \ldots, \ltree_\childlim) & \text{if } \lnodeidx > \siz{\ltree_1} \\
\end{cases}
\end{align*}
We define $\ins_{\lnodeidx, \rnode, \lchildidx, \rchildidx}(\epsilon)$ as
$\ins_{\rnode, \lchildidx, \rchildidx}(\epsilon)$ if $\lnodeidx = 0$ and as $\epsilon$ otherwise.

We define the standard TED edit set $\edits_\alphabet$
for the alphabet $\alphabet$ as the following set:
$\edits_\alphabet := \{ \del_\lnodeidx  | \lnodeidx \in \N \}
\cup \{ \rep_{\lnodeidx, \rnode} | \lnodeidx \in \N, \rnode \in \alphabet \}
\cup \{ \ins_{\lnodeidx, \rnode, \lchildidx, \rchildidx} | \lnodeidx \in \N_0, \lchildidx, \rchildidx \in \N, \rnode \in \alphabet \}$.
\end{dfn}

Note that we leave the forest $\lforest$ unchanged if $\lnodeidx > |\lforest|$, if the edit is a
deletion or replacement and $\lnodeidx < 1$, and if the edit is an insertion and $\lnodeidx < 0$.

Consider the example of the script $\script = \rep_{1, \sym{f}}, \del_2, \del_2, \rep_{2, \sym{g}}, \del_3$ in Figure~\ref{fig:edits}. This script transforms the tree $\ltree = \sym{a}(\sym{b}(\sym{c}, \sym{d}), \sym{e})$
into the tree $\rtree = \sym{f}(\sym{g})$. Because there is no shorter script over $\edits_\alphabet$ that transforms $\ltree$ to $\rtree$, the standard tree
edit distance $\dist_{\edits_\alphabet}$ between $\ltree$ and $\rtree$ is $5$. Note that the deletion of $\sym{b}$ in the tree
$\sym{f}(\sym{b}(\sym{c}, \sym{d}), \sym{e})$ results in the tree $\sym{f}(\sym{c}, \sym{d}, \sym{e})$, meaning
that the children of $\sym{b}$, namely $\sym{d}$ and $\sym{e}$, are now children of the parent of $\sym{b}$,
namely $\sym{f}$. Deletions can also lead to trees becoming forests. In particular, a deletion of $\sym{f}$
in the tree $\sym{f}(\sym{c}, \sym{d}, \sym{e})$ would result in the forest $\sym{c}, \sym{d}, \sym{e}$.

Conversely, an insertion takes (some of) the children of a tree and uses them as children of the newly inserted
node. For example, the insertion of $\sym{b}$ in the tree $\sym{f}(\sym{c}, \sym{d}, \sym{e})$ in Figure~\ref{fig:edits}
uses the children $\sym{c}$ and $\sym{d}$ of $\sym{f}$ as children for the new node $\sym{b}$.
Insertions can also be used to transform forests to trees by inserting a new node at the root. For example,
the insertion $\ins_{0, \sym{f}, 1, 4}$ applied to the forest $\sym{c}, \sym{d}, \sym{e}$ would result in the
tree $\sym{f}(\sym{c}, \sym{d}, \sym{e})$.

\begin{figure}
\begin{center}
\begin{tikzpicture}[node distance=2cm]
\node (abcde) {%
\Tree [.$\sym{a}$
	[.$\sym{b}$ $\sym{c}$ $\sym{d}$ ]
	$\sym{e}$
]
};

\node (fbcde) [right of=abcde, node distance=2.5cm] {%
\Tree [.$\sym{f}$
	[.$\sym{b}$ $\sym{c}$ $\sym{d}$ ]
	$\sym{e}$
]
};

\node (fcde) [right of=fbcde, node distance=2.5cm] {%
\Tree [.$\sym{f}$ $\sym{c}$ $\sym{d}$ $\sym{e}$ ]
};

\node (fde) [right of=fcde, node distance=2.5cm] {%
\Tree [.$\sym{f}$ $\sym{d}$ $\sym{e}$ ]
};

\node (fge) [right of=fde] {%
\Tree [.$\sym{f}$ $\sym{g}$ $\sym{e}$ ]
};

\node (fg) [right of=fge] {%
\Tree [.$\sym{f}$ $\sym{g}$ ]
};

\path[edge]%
(abcde.north east) edge [bend left] node [above] {$\rep_{1, \sym{f}}$} (fbcde.north west)
(fbcde.north east) edge [bend left] node [above] {$\del_2$}          (fcde.north west)
(fcde.north east)  edge [bend left] node [above] {$\del_2$}          (fde.north west)
(fde.north east)   edge [bend left] node [above] {$\rep_{2, \sym{g}}$} (fge.north west)
(fge.north east)   edge [bend left] node [above] {$\del_3$}          (fg.north west)
(fg.south west)    edge [bend left] node [below] {$\ins_{1, \sym{e}, 2, 2}$} (fge.south east)
(fge.south west)   edge [bend left] node [below] {$\rep_{2, \sym{d}}$}       (fde.south east)
(fde.south west)   edge [bend left] node [below] {$\ins_{1, \sym{c}, 1, 1}$} (fcde.south east)
(fcde.south west)  edge [bend left] node [below=0.3cm] {$\ins_{1, \sym{b}, 1, 3}$} (fbcde.south east)
(fbcde.south west) edge [bend left] node [below] {$\rep_{1, \sym{a}}$}       (abcde.south east);

\end{tikzpicture}
\end{center}
\caption{An illustration of a shortest edit script transforming the tree $\ltree$ on the left
to the tree $\rtree$ on the right and a shortest edit script transforming the tree $\rtree$
on the right to the tree $\ltree$ on the left. The intermediate trees resulting from the
application of single edits are shown in the middle.}
\label{fig:edits}
\end{figure}

\subsection{Cost Functions}
\label{sec:costs}

Up until now we have defined the edit distance based on the \emph{length} of the script
required to transform $\ltree$ into $\rtree$. However, we may want to regard some edits as
more costly than others, because some elements may be easier to replace.
This is reflected in manually defined \emph{cost matrices}, such as the
PAM and BLOSUM matrices from bioinformatics \parencite{Henikoff1992}. In general,
we can express the cost of edits in terms of a \emph{cost function}.

\begin{dfn}[Cost function]
A \emph{cost function} $\cost$ over some alphabet $\alphabet$ with $\gap \notin \alphabet$
is defined as a function $\cost : (\alphabet \cup \{ \gap \}) \times (\alphabet \cup \{ \gap \}) \to \R$, where
$\gap$ is called the special \emph{gap symbol}.

Now, let $\cost$ be a cost function over $\alphabet$. Then, we define the cost $\Cost(\edit, \lforest)$
of an edit $\edit \in \edits_\alphabet$ as zero if $\edit(\lforest) = \lforest$,
i.e.\ if the edit does not change the input forest.

Otherwise, we define the cost of a replacement $\rep_{\lnodeidx, \rnode}$ with respect
to some input forest $\lforest$ as $\Cost(\rep_{\lnodeidx, \rnode}, \lforest) := \cost(\lnode_\lnodeidx, \rnode)$;
we define the cost of a deletion $\del_\lnodeidx$ with respect to some input forest
$\lforest$ as $\Cost(\del_\lnodeidx, \lforest) := \cost(\lnode_\lnodeidx, \gap)$;
and we define the cost of an insertion $\ins_{\lnodeidx, \rnode, \lchildidx, \rchildidx}$
with respect to some input forest $\lforest$ as
$\Cost(\ins_{\lnodeidx, \rnode, \lchildidx, \rchildidx}, \lforest) := \cost(\gap, \rnode)$.

Finally, we define the cost $\Cost(\script, \lforest)$ of an edit script $\script = \edit_1, \ldots, \edit_\editlim$
with respect to some input forest $\lforest$ recursively as $\Cost(\script, \lforest) :=
\Cost(\edit_1, \lforest) + \Cost((\edit_2, \ldots, \edit_\editlim), \edit_1(\lforest))$,
with the base case $\Cost(\epsilon, \lforest) = 0$ for the empty script.
\end{dfn}

Intuitively, the cost of an edit script is just the sum over the costs of any single edit in the script.
As an example, consider our example script in Figure~\ref{fig:edits}. For this script
we obtain the cost:
\begin{align*}
  &\Cost\Big(\rep_{1, \sym{f}}, \del_2, \del_2, \rep_{2, \sym{g}}, \del_3, \sym{a}(\sym{b}(\sym{c}, \sym{d}), \sym{e})\Big) \\
= &\cost(\sym{a}, \sym{f}) + \Cost\Big(\del_2, \del_2, \rep_{2, \sym{g}}, \del_3, \sym{f}(\sym{b}(\sym{c}, \sym{d}), \sym{e})\Big) \\
= &\cost(\sym{a}, \sym{f}) + \cost(\sym{b}, \gap) + \Cost\Big(\del_2, \rep_{2, \sym{g}}, \del_3, \sym{f}(\sym{c}, \sym{d}, \sym{e})\Big) \\
= &\cost(\sym{a}, \sym{f}) + \cost(\sym{b}, \gap) + \cost(\sym{c}, \gap) + \Cost\Big(\rep_{2, \sym{g}}, \del_3, \sym{f}(\sym{d}, \sym{e})\Big) \\
= &\cost(\sym{a}, \sym{f}) + \cost(\sym{b}, \gap) + \cost(\sym{c}, \gap) + \cost(\sym{d}, \sym{g}) + \Cost\Big(\del_3, \sym{f}(\sym{g}, \sym{e})\Big) \\
= &\cost(\sym{a}, \sym{f}) + \cost(\sym{b}, \gap) + \cost(\sym{c}, \gap) + \cost(\sym{d}, \sym{g}) + \cost(\sym{e}, \gap) + \Cost\Big(\epsilon, \sym{f}(\sym{g})\Big) \\
= &\cost(\sym{a}, \sym{f}) + \cost(\sym{b}, \gap) + \cost(\sym{c}, \gap) + \cost(\sym{d}, \sym{g}) + \cost(\sym{e}, \gap) +  0
\end{align*}

Based on the notion of cost, we can generalize the TED as follows.

\begin{dfn}[Generalized Tree Edit Distance]
Let $\alphabet$ be an alphabet, let $\edits_\alphabet$ be the standard TED edit set over $\alphabet$,
and let $\cost$ be a cost function over $\alphabet$. Then, the generalized TED over $\alphabet$ is
defined as the function

\begin{align}
&\dist_\cost : \trees(\alphabet) \times \trees(\alphabet) \to \R \\
&\dist_\cost(\ltree, \rtree) = \min_{\script \in \edits_\alphabet^*} \Big\{ \Cost(\script, \ltree) \Big| \script(\ltree) = \rtree \Big\}
\end{align}
\end{dfn}

As an example, consider the cost function $\cost(\lnode, \rnode) = 1$ if $\lnode \neq \rnode$ and $0$
if $\lnode = \rnode$. In that case, every edit (except for self-replacements) costs $1$, such that
the generalized edit distance corresponds to the length of the shortest edit script.
If we change this cost function to be $0$ for a replacement of $\sym{a}$ with $\sym{f}$,
our edit distance between the two example trees in Figure~\ref{fig:edits} decreases from $5$ to $4$.
If we set the cost $\cost(\sym{a}, \sym{a}) = -1$, the edit distance becomes ill-defined, because we
can always make an edit script cheaper by appending another self-replacement of $\sym{a}$ with $\sym{a}$.

This begs the question: Which properties does the cost function $\cost$ need to fulfill in order
to ensure a \enquote{reasonable} edit distance? To answer this question, we first define what it means for
a distance to be \enquote{reasonable}.
Here, we turn to the mathematical notion of a \emph{metric}.

\begin{dfn}[Metric]

Let $\alphabet$ be some set. A function $\dist : \alphabet \times \alphabet \to \R$ is called a \emph{metric} if
for all $x, y, z \in \alphabet$ it holds:

\begin{align*}
\dist(x, y) & \geq 0 & \text{(non-negativity)} \\
\dist(x, x) & = 0 & \text{(self-equality)} \\
\dist(x, y) & > 0 \text{ if } x \neq y & \text{(discernibility)} \\
\dist(x, y) & = \dist(y, x) & \text{(symmetry)} \\
\dist(x, z) + \dist(z, y) & \geq \dist(x, y) & \text{(triangular inequality)}
\end{align*}

\end{dfn}

All five of these properties make intuitive sense: We require a reasonable distance to not return negative values,
we require that every object should have a distance of $0$ to itself,
we require that no two different objects can occupy the same space,
we require that any object $x$ is as far from $y$ as $y$ is from $x$, and we require that the fastest route
from $x$ to $y$ is a straight line, that is, there is no point $z$ through which we could travel such that
we reach $y$ faster from $x$ compared to taking the direct distance.

Interestingly, it is relatively easy to show that the generalized TED is a metric if the cost function is a metric.

\begin{thm}
If $\cost$ is a metric over $\alphabet$, then the generalized TED $\dist_\cost$
is a metric over $\trees(\alphabet)$. More specifically:
\begin{enumerate}
\item If $\cost$ is non-negative, then $\dist_\cost$ is non-negative.
\item If $\cost$ is non-negative and self-equal, then $\dist_\cost$ is self-equal.
\item If $\cost$ is non-negative and discernible, then $\dist_\cost$ is discernible.
\item If $\cost$ is non-negative and symmetric, then $\dist_\cost$ is symmetric.
\item If $\cost$ is non-negative, $\dist_\cost$ conforms to the triangular inequality.
\end{enumerate}

\begin{proof}
Note that we require non-negativity as a pre-requisite for any of the metric conditions,
because negative cost function values may lead to an ill-defined distance, as in the example
above.

We now prove any of the four statements in turn:
\begin{description}
\item[Non-negativity:] The TED is a sum of outputs of $\cost$. Because $\cost$
is non-negative, $\dist_\cost$ is as well.
\item[Self-Equality:] The empty edit script $\epsilon$ transforms $\ltree$ to $\ltree$ and
has a cost of $0$. Because $\dist_\cost$ is non-negative, this is the cheapest edit sequence,
therefore $\dist_\cost(\ltree, \ltree) = 0$ for all $\ltree$.
\item[Discernibility:] Let $\ltree \neq \rtree$ be two different trees and let $\script = \edit_1, \ldots, \edit_\editlim$ be an edit
script such that $\script(\ltree) = \rtree$. We now define $\ltree_0 = \ltree$ and $\ltree_\editidx$
recursively as $\edit_\editidx(\ltree_{\editidx-1})$ for all $\editidx \in \{1, \ldots, \editlim\}$.
Accordingly, there must exist an $\editidx \in \{1, \ldots, \editlim\}$ such that
$\ltree_\editidx \neq \ltree_{\editidx-1}$, otherwise $\ltree = \rtree$. However, in that case, the
costs of $\edit_\editidx$ must be $\cost(\lnode, \rnode)$ for some $\lnode \neq \rnode$.
Because $\cost$ is discernible, $\cost(\lnode, \rnode) > 0$. Further, because $\cost$ is non-negative,
$\Cost(\script, \ltree)$ is a sum of non-negative contributions with at least one strictly positive
contribution, which means that $\Cost(\script, \ltree) > 0$. Because this reasoning applies for
any script $\script$ with $\script(\ltree) = \rtree$, it holds: $\dist_\cost(\ltree, \rtree) > 0$.
\item[Symmetry:] Let $\script = \edit_1, \ldots, \edit_\editlim$ be the cheapest edit script which
transforms $\ltree$ to $\rtree$. Now, we can inductively construct an \emph{inverse}
edit script as follows: If $\script$ is the empty script, then the empty script
also transforms $\rtree$ to $\ltree$. If $\script$ is not empty, consider the first
edit $\edit_1$:
\begin{itemize}
\item If $\edit_1 = \rep_{\lnodeidx, \rnode}$, we construct the edit
$\edit^{-1}_1 = \rep_{\lnodeidx, \lnode_\lnodeidx}$. For this edit it holds:
$\edit_1 \circ \edit^{-1}_1(\ltree) = \ltree$. Further, for the cost it holds:
$\Cost(\edit_1, \ltree) = \cost(\lnode_\lnodeidx, \rnode) = \cost(\rnode, \lnode_\lnodeidx)
= \Cost(\edit_1^{-1}, \edit_1(\ltree))$.
\item If $\edit_1 = \ins_{\lnodeidx, \rnode, \lchildidx, \rchildidx}$,
we construct the edit $\edit^{-1}_1 = \del_{\lnodeidx'}$ where $\lnodeidx'$
is the index of the newly inserted node in the forest $\edit_1(\ltree)$.
Therefore, we obtain $\edit_1 \circ \edit^{-1}_1(\ltree) = \ltree$.
Further, for the cost it holds: $\Cost(\edit_1, \ltree) = \cost(\gap, \rnode) = \cost(\rnode, \gap)
= \Cost(\edit_1^{-1}, \edit_1(\ltree))$.
\item If $\edit_1 = \del_\lnodeidx$, we construct the edit
$\edit_1^{-1} = \ins_{\paridx_\ltree(\lnodeidx), \lnode_\lnodeidx, \childidx_\ltree(\lnodeidx), \childidx_\ltree(\lnodeidx) + |\ltree_\lnodeidx|-1}$.
That is, we construct an insertion which re-inserts the node that has been deleted by $\edit_1$,
and uses all its prior children.
Therefore, we obtain $\edit_1 \circ \edit^{-1}_1(\ltree) = \ltree$.
Further, for the cost it holds: $\Cost(\edit_1, \ltree) = \cost(\lnode_\lnodeidx, \gap) =
\cost(\gap, \lnode_\lnodeidx) = \Cost(\edit_1^{-1}, \edit_1(\ltree))$.
\end{itemize}

It follows by induction that we can construct an entire script $\script^{-1}$, which
transforms $\rtree$ to $\ltree$, because $\ltree = \script\circ \script^{-1}(\ltree) = \script^{-1}(\rtree)$.
Further, this script costs the same as $\script$, because
$\Cost(\script^{-1}, \script(\ltree)) = \Cost(\script^{-1}, \rtree) = \Cost(\script, \ltree)$.

Because $\script$ was by definition a cheapest edit script which transforms $\ltree$ to $\rtree$
we obtain: $\dist_\cost(\rtree, \ltree) \leq \Cost(\script^{-1}, \rtree) = \Cost(\script, \ltree) = \dist_\cost(\ltree, \rtree)$.
It remains to show that $\dist_\cost(\rtree, \ltree) \geq \dist_\cost(\ltree, \rtree)$.

Assume that $\dist_\cost(\rtree, \ltree) < \dist_\cost(\ltree, \rtree)$. Then, there is an
edit script $\tilde \script = \edit_1, \ldots, \edit_{\editlim'}$ which transforms $\rtree$ to $\ltree$
and is cheaper than $\dist_\cost(\ltree, \rtree)$. However, using the same
argument as before, we can generate an inverse edit script $\tilde \script^{-1}$ with the same
cost as $\tilde \script$ that transforms $\ltree$ to $\rtree$, such that
$\dist_\cost(\ltree, \rtree) \leq \dist_\cost(\rtree, \ltree) < \dist_\cost(\ltree, \rtree)$,
which is a contradiction. Therefore $\dist_\cost(\rtree, \ltree) = \dist_\cost(\ltree, \rtree)$.
\item[Triangular Inequality:] Assume that there are three trees $\ltree$, $\rtree$, and
$\ztree$, such that $\dist_\cost(\ltree, \ztree) + \dist_\cost(\ztree, \rtree) < \dist_\cost(\ltree, \rtree)$.
Now, let $\script$ and $\script'$ be cheapest edit scripts which transform $\ltree$ to $\ztree$ and
$\ztree$ to $\rtree$ respectively, that is,
$\script(\ltree) = \ztree$, $\script'(\ztree) = \rtree$,
$\Cost(\script, \ltree) = \dist_\cost(\ltree, \ztree)$, and
$\Cost(\script', \ztree) = \dist_\cost(\ztree, \rtree)$.
The concatenation of both scripts $\script'' = \script \concat \script'$ is
per construction a script such that $\script''(\ltree) = \script \concat \script'(\ltree) = \rtree$
and $\Cost(\script'', \ltree) = \Cost(\script, \ltree) + \Cost(\script', \ztree)
= \dist_\cost(\ltree, \ztree) + \dist_\cost(\ztree, \rtree)$. It follows that
$\dist_\cost(\ltree, \ztree) + \dist_\cost(\ztree, \rtree) < \dist_\cost(\ltree, \rtree) \leq \dist_\cost(\ltree, \ztree) + \dist_\cost(\ztree, \rtree)$
which is a contradiction. Therefore, the triangular inequality holds.
\end{description}
\end{proof}
\end{thm}

As an example of the symmetry part of the proof, consider again Figure~\ref{fig:edits}.
Here, the inverse script for $\script = \rep_{1, \sym{f}}, \del_2, \del_2, \rep_{2, \sym{g}}, \del_3$
is $\script^{-1} = \ins_{1, \sym{e}, 2, 2}, \rep_{2, \sym{d}}, \ins_{1, \sym{c}, 1, 1}, \ins_{1, \sym{b}, 1, 3}, \rep_{1, \sym{a}}$.
For the cost we obtain:
\begin{align*}
&\Cost\Big(\ins_{1, \sym{e}, 2, 2}, \rep_{2, \sym{d}}, \ins_{1, \sym{c}, 1, 1}, \ins_{1, \sym{b}, 1, 3}, \rep_{1, \sym{a}}, \sym{f}(\sym{g})\Big) \\
= &\cost(\gap, \sym{e}) + \Cost\Big(\rep_{2, \sym{d}}, \ins_{1, \sym{c}, 1, 1}, \ins_{1, \sym{b}, 1, 3}, \rep_{1, \sym{a}}, \sym{f}(\sym{g}, \sym{e})\Big) \\
= &\cost(\gap, \sym{e}) + \cost(\sym{g}, \sym{d}) + \Cost\Big(\ins_{1, \sym{c}, 1, 1}, \ins_{1, \sym{b}, 1, 3}, \rep_{1, \sym{a}}, \sym{f}(\sym{d}, \sym{e})\Big) \\
= &\cost(\gap, \sym{e}) + \cost(\gap, \sym{d}) + \cost(\gap, \sym{c}) + \Cost\Big(\ins_{1, \sym{b}, 1, 3}, \rep_{1, \sym{a}}, \sym{f}(\sym{c}, \sym{d}, \sym{e})\Big) \\
= &\cost(\gap, \sym{e}) + \cost(\gap, \sym{d}) + \cost(\gap, \sym{c}) + \cost(\gap, \sym{b}) + \Cost\Big(\rep_{1, \sym{a}}, \sym{f}(\sym{b}(\sym{c}, \sym{d}), \sym{e})\Big) \\
= &\cost(\gap, \sym{e}) + \cost(\gap, \sym{d}) + \cost(\gap, \sym{c}) + \cost(\sym{g}, \sym{b}) + \cost(\sym{f}, \sym{a}) + \Cost\Big(\epsilon, \sym{a}(\sym{b}(\sym{c}, \sym{d}), \sym{e})\Big) \\
= &\cost(\gap, \sym{e}) + \cost(\gap, \sym{d}) + \cost(\gap, \sym{c}) + \cost(\sym{g}, \sym{b}) + \cost(\sym{f}, \sym{a}) + 0,
\end{align*}
which is exactly the same cost as for the script $\script$, if $\cost$ is symmetric.

\subsection{Mappings}
\label{sec:mappings}

While edit scripts capture the intuitive notion of editing a tree, they are not a viable representation to develop an
efficient algorithm. In particular, edit scripts are highly redundant, in the sense that there may be many different edit scripts
which transform a tree $\ltree$ into a tree $\rtree$ and have the same cost. For example, to transform
the tree $\ltree = \sym{a}(\sym{b}(\sym{c}, \sym{d}), \sym{e})$ to the tree $\rtree = \sym{f}(\sym{g})$,
we can not only use the edit script in Figure~\ref{fig:edits}, but we could also use the script
$\del_5, \del_3, \del_2, \rep_{2, \sym{g}}, \rep_{1, \sym{f}}$, which has the same cost, irrespective
of the cost function.
To avoid these redundancies, we need a representation which is invariant against changes in order
of the edits, and instead just counts which nodes are replaced, which nodes are deleted, and which nodes are inserted.
Such a representation is offered by \emph{tree mappings}.

\begin{dfn}[Tree Mapping]
Let $\lforest$ and $\rforest$ be forests over some alphabet $\alphabet$, and let
$\lnodelim = |\lforest|$ and $\rnodelim = |\rforest|$.

A \emph{tree mapping} between $\lforest$ and $\rforest$ is defined as a set of tuples
$\map \subseteq \{1, \ldots, \lnodelim\} \times \{1, \ldots, \rnodelim\}$, such that
for all $(\lnodeidx, \rnodeidx), (\lnodeidx', \rnodeidx') \in \map$ it holds:
\begin{enumerate}
\item Each node of $\lforest$ is assigned to at most one node in $\rforest$, i.e.\
$\lnodeidx = \lnodeidx' \Rightarrow \rnodeidx = \rnodeidx'$.
\item Each node of $\rforest$ is assigned to at most one node in $\lforest$, i.e.\
$\rnodeidx = \rnodeidx' \Rightarrow \lnodeidx = \lnodeidx'$.
\item The mapping preserves the pre-order of both trees, i.e.\
$\lnodeidx \geq \lnodeidx' \iff \rnodeidx \geq \rnodeidx'$.
\item The mapping preserves the ancestral ordering in both trees, that is: if
the subtree rooted at $\lnodeidx$ is an ancestor of the subtree rooted at $\lnodeidx'$,
then the subtree rooted at $\rnodeidx$ is also an ancestor of the subtree rooted at $\rnodeidx'$,
and vice versa.
\end{enumerate}
\end{dfn}

\begin{figure}
\begin{center}
\begin{tikzpicture}

\begin{scope}

\node[above] at (0.5, 1) {$\map = \{(1, 1), (4, 2) \}$};
\node[above] at (0.5, 0.5) {valid};

\Tree [.\node (a) {$\sym{a}$};
	[.\node (b) {$\sym{b}$};
		\node (c) {$\sym{c}$};
		\node (d) {$\sym{d}$};
	]
	\node (e) {$\sym{e}$};
]

\begin{scope}[shift={(1.5, 0)}]

\Tree [.\node (f) {$\sym{f}$};
	\node (g) {$\sym{g}$};
]

\end{scope}

\path[edge, densely dashed, semithick]%
(a) edge (f)
(d) edge (g);

\end{scope}

\begin{scope}[shift={(4, 0)}]

\node[above] at (0.5, 1) {$\map = \{(1, 1), (1, 2) \}$};
\node[above] at (0.5, 0.5) {invalid};

\Tree [.\node (a) {$\sym{a}$};
	[.\node (b) {$\sym{b}$};
		\node (c) {$\sym{c}$};
		\node (d) {$\sym{d}$};
	]
	\node (e) {$\sym{e}$};
]

\begin{scope}[shift={(1.5, 0)}]

\Tree [.\node (f) {$\sym{f}$};
	\node (g) {$\sym{g}$};
]

\end{scope}

\path[edge, densely dashed, semithick]%
(a) edge (f)
(a) edge (g);

\end{scope}

\begin{scope}[shift={(8, 0)}]

\node[above] at (0.5, 1) {$\map = \{(1, 1), (2, 1) \}$};
\node[above] at (0.5, 0.5) {invalid};

\Tree [.\node (a) {$\sym{a}$};
	[.\node (b) {$\sym{b}$};
		\node (c) {$\sym{c}$};
		\node (d) {$\sym{d}$};
	]
	\node (e) {$\sym{e}$};
]

\begin{scope}[shift={(1.5, 0)}]

\Tree [.\node (f) {$\sym{f}$};
	\node (g) {$\sym{g}$};
]

\end{scope}

\path[edge, densely dashed, semithick]%
(a) edge (f)
(b) edge (f);

\end{scope}

\begin{scope}[shift={(2, -4)}]

\node[above] at (0.5, 1) {$\map = \{(1, 2), (2, 1) \}$};
\node[above] at (0.5, 0.5) {invalid};

\Tree [.\node (a) {$\sym{a}$};
	[.\node (b) {$\sym{b}$};
		\node (c) {$\sym{c}$};
		\node (d) {$\sym{d}$};
	]
	\node (e) {$\sym{e}$};
]

\begin{scope}[shift={(1.5, 0)}]

\Tree [.\node (f) {$\sym{f}$};
	\node (g) {$\sym{g}$};
]

\end{scope}

\path[edge, densely dashed, semithick]%
(a) edge (g)
(b) edge (f);

\end{scope}

\begin{scope}[shift={(6, -4)}]

\node[above] at (0.5, 1) {$\map = \{(3, 1), (5, 2) \}$};
\node[above] at (0.5, 0.5) {invalid};

\Tree [.\node (a) {$\sym{a}$};
	[.\node (b) {$\sym{b}$};
		\node (c) {$\sym{c}$};
		\node (d) {$\sym{d}$};
	]
	\node (e) {$\sym{e}$};
]

\begin{scope}[shift={(1.5, 0)}]

\Tree [.\node (f) {$\sym{f}$};
	\node (g) {$\sym{g}$};
]

\end{scope}

\path[edge, densely dashed, semithick]%
(c) edge [bend left=15] (f)
(e) edge (g);

\end{scope}

\end{tikzpicture}
\end{center}
\caption{One example mapping between the trees $\ltree = \sym{a}(\sym{b}(\sym{c}, \sym{d}), \sym{e})$ and
$\rtree = \sym{f}(\sym{g})$ (top left) and four sets which are not valid mappings due to violations of one
of the four mapping constraints.}
\label{fig:mappings}
\end{figure}

Intuitively, a tuple $(\lnodeidx, \rnodeidx)$ in a tree mapping $\map$ expresses that node $\lnodeidx$
is replaced with node $\rnodeidx$. If an index does not occur in $\map$, it is deleted/inserted.
The four constraints in the definition of a tree mapping have the purpose to ensure that we
can find a corresponding edit script for each mapping.
As an example, consider again our two trees $\ltree = \sym{a}(\sym{b}(\sym{c}, \sym{d}), \sym{e})$
and $\rtree = \sym{f}(\sym{g})$ from Figure~\ref{fig:edits}. The mapping corresponding to the
edit script in this figure would be $\map = \{(1, 1), (4, 2)\}$ because node $\lnode_1 = \sym{a}$
is replaced with node $\rnode_1 = \sym{f}$ and node $\lnode_4 = \sym{d}$ is replaced with
node $\rnode_2 = \sym{g}$. All remaining nodes are deleted and inserted, respectively.
The empty mapping $\map = \{ \}$ would correspond to deleting all nodes in $\ltree$ and then inserting
all nodes in $\rtree$, which is also a valid mapping but would likely be more costly.

The set $\map = \{ (1, 1), (1, 2) \}$ would not be a valid mapping because the node $\lnode_1 = \sym{a}$
is assigned to multiple nodes in $\rtree$ and thus we can not construct an edit script corresponding
to this mapping. For such an edit script we would need a \enquote{copy} edit.
For the same reason, the set $\map = \{ (1, 1), (2, 1) \}$ is not a valid mapping. Here, the node
$\rnode_1 = \sym{f}$ is assigned to multiple nodes in $\ltree$.

$\map = \{ (1, 2), (2, 1) \}$ is an example of a set that is not a valid tree mapping because
of the third criterion. To construct an edit script corresponding to this mapping we would
need a \enquote{swap} edit, i.e.\ an edit which can exchange nodes
$\lnode_1 = \sym{a}$ and $\lnode_2 = \sym{b}$ in $\ltree$.
Finally, the set $\map = \{ (3, 1), (5, 2) \}$ is not a valid mapping due to the fourth
criterion. In particular, the subtree $\rtree_1 = \sym{f}(\sym{g})$ is an ancestor
of the subtree $\rtree_2 = \sym{g}$ in $\rtree$, but the subtree $\ltree_3 = \sym{c}$
is \emph{not} an ancestor of the subtree $\ltree_5 = \sym{e}$. This last criterion is more subtle,
but you will find that each edit we can apply - be it replacement, deletion, or insertion -
preserves the ancestral order in the tree. Conversely, this means that we can not make
a node an ancestor of another node if it was not before. This also makes intuitive
sense because it means that nodes can not be mapped to nodes in completely distinct
subtrees.

Now that we have considered some examples, it remains to show that we can construct a
corresponding edit script for each mapping in general.

\begin{thm}\label{thm:map_to_script}
Let $\lforest$ and $\rforest$ be forests over some alphabet $\alphabet$ and let $\map$ be a
tree mapping between $\lforest$ and $\rforest$. Then, the output of Algorithm~\ref{alg:map_to_script}
for $\lforest$, $\rforest$ and $\map$ is an edit script $\script_\map$
over $\edits_\alphabet$, such that $\script_\map(\lforest) = \rforest$
and
\begin{enumerate}
\item If $(\lnodeidx, \rnodeidx) \in \map$, then the edit $\rep_{\lnodeidx, \rnode_\rnodeidx}$
is part of the script.
\item For all $\lnodeidx$ which are \emph{not} part of the mapping - i.e.\ $\nexists \rnodeidx : (\lnodeidx, \rnodeidx) \in \map$ -
the edit $\del_\lnodeidx$ is part of the script.
\item For all $\rnodeidx$ which are \emph{not} part of the mapping - i.e.\ $\nexists \lnodeidx : (\lnodeidx, \rnodeidx) \in \map$ -
an edit $\ins_{\paridx_\rforest(\rnodeidx), \rnode_\rnodeidx, \childidx_\rnodeidx, \childidx_\rnodeidx + \childlim_\rnodeidx}$ for some $\childlim_\rnodeidx$ is part of the script.
\end{enumerate}
Further, no other edits are part of $\script_\map$ than the edits mentioned above.
Algorithm~\ref{alg:map_to_script} runs in $\effic(\lnodelim + \rnodelim)$ worst-case time.

\begin{proof}
The three constraints are fulfilled because we iterate over all entries $(\lnodeidx, \rnodeidx)$
and create one replacement per such entry, we iterate over all $\lnodeidx \in \unmappedLeft$ and
create one deletion per such $\lnodeidx$, and we iterate over all $\rnodeidx \in \unmappedRight$
and we create one insertion per such entry. It is also clear that $\effic(\lnodelim + \rnodelim)$
because we iterate over all $\lnodeidx \in \{1, \ldots, \lnodelim\}$ and over all
$\rnodeidx \in \{1, \ldots, \rnodelim\}$. Assuming that $I$ and $J$ permit insertion
as well as containment tests in constant time, and that the list concatenations in
\emph{num-descendants} are possible in constant time, this leaves us with $\effic(\lnodelim + \rnodelim)$.

It is less obvious that $\script_\map(\lforest) = \rforest$. We show this by an induction
over the cardinality of $\map$. First, consider $\map = \emptyset$. In that case, we obtain
$\unmappedLeft = \{1, \ldots, \lnodelim\}$, $\unmappedRight = \{1, \ldots, \rnodelim\}$, and
$\childlim_0 = \ldots = \childlim_\rnodelim = 0$. Therefore, the resulting script is
$\script_\map = \del_\lnodelim, \ldots, \del_1, \ins_{\paridx_\rforest(1), \rnode_1, \childidx_1, \childidx_1},
\ldots, \ins_{\paridx_\rforest(\rnodelim), \rnode_\rnodelim, \childidx_\rnodelim, \childidx_\rnodelim}$.
This script obviously first deletes all nodes in $\lforest$ and then inserts all nodes
from $\rforest$ in the correct configuration.

Now assume that the theorem holds for all mappings $\map$ between $\lforest$ and
$\rforest$ with $|\map| \leq k$,
and consider a mapping $\map$ between $\lforest$ and $\rforest$ with $|\map| = k + 1$.

Let $(\lnodeidx, \rnodeidx)$ be the entry of $\map$ with smallest $\rnodeidx$,
let $\map' = \map \setminus \{ (\lnodeidx, \rnodeidx) \}$ and let $\edit_{\map'}$
be the corresponding edit script for $\map'$ according to Algorithm~\ref{alg:map_to_script}.
Per induction, $\edit_{\map'}(\lforest) = \rforest$.

Now, let $\unmappedLeft = \{ \lnodeidx | \nexists \rnodeidx : (\lnodeidx, \rnodeidx) \in \map \}$,
$\unmappedLeft' = \{ \lnodeidx | \nexists \rnodeidx : (\lnodeidx, \rnodeidx) \in \map' \}$,
$\unmappedRight = \{ \rnodeidx | \nexists \lnodeidx : (\lnodeidx, \rnodeidx) \in \map \}$,
and $\unmappedRight' = \{ \rnodeidx | \nexists \lnodeidx : (\lnodeidx, \rnodeidx) \in \map' \}$.
We observe that $\unmappedLeft' = \unmappedLeft \cup \{ \lnodeidx\}$ and $\unmappedRight' = \unmappedRight \cup \{\rnodeidx\}$,
so our resulting script $\script_\map$ will not delete node $\lnodeidx$ and
not insert node $\rnodeidx$, but otherwise contain all deletions and insertions of script
$\script_{\map'}$. We also know that node $\lnode_\lnodeidx$ will be replaced with node
$\rnode_\rnodeidx$, such that all nodes of $\rforest$ are contained after applying
$\script_\map$. It remains to show that node $\rnode_\rnodeidx$ is positioned correctly
in $\script_\map(\lforest)$, such that $\script_\map(\lforest) = \rforest$.

Let $P_\unmappedRight(\rnodeidx)$ be a set that is recursively defined as $P_\unmappedRight(\rnodeidx') = \emptyset$
if $\rnodeidx' \notin \unmappedRight$, and $P_\unmappedRight(\rnodeidx') = \{ \rnodeidx' \} \cup P_\unmappedRight(\paridx_\rtree(\rnodeidx'))$
if $\rnodeidx' \in \unmappedRight$. In other words, $P_\unmappedRight(\rnodeidx')$ contains all ancestors of $\rnodeidx'$, until
we find an ancestor that is not inserted. Now, consider all inserted ancestors of $\rnodeidx$, that is, $P_\unmappedRight(\paridx_\rtree(\rnodeidx))$.
Further, let $(\rnodelim, \childlim_0, \ldots, \childlim_\rnodelim) = \text{num-descendants}(\rforest, 0, \unmappedRight)$
and $(\rnodelim, \childlim'_0, \ldots, \childlim'_\rnodelim) = \text{num-descendants}(\rforest, 0, \unmappedRight')$.
For all elements $\rnodeidx' \in P_\unmappedRight(\paridx_\rtree(\rnodeidx))$ we obtain $\childlim_{\rnodeidx'} = \childlim'_{\rnodeidx'} + 1$,
and for all other nodes we obtain $\childlim_{\rnodeidx'} = \childlim'_{\rnodeidx'}$.
In other words, all ancestors of $\rnodeidx$ which are inserted use one more child compared to before, but no other node will.
This additional child is $\rnodeidx$, such that the ancestral structure is preserved
and we obtain $\script_\map(\lforest) = \rforest$.
\end{proof}
\end{thm}

\begin{algorithm}
\caption{An algorithm to transform a mapping $\map$ into a corresponding edit script $\script_\map$
according to Theorem~\ref{thm:map_to_script}.}
\label{alg:map_to_script}
\begin{algorithmic}
\Function{map-to-script}{Two forests $\lforest$ and $\rforest$, a tree mapping $\map$ between $\lforest$
and $\rforest$.}
\State $I \gets \{ \lnodeidx | \nexists \rnodeidx : (\lnodeidx, \rnodeidx) \in \map \}$.
\State $J \gets \{ \rnodeidx | \nexists \lnodeidx : (\lnodeidx, \rnodeidx) \in \map \}$.
\State Initialize $\script$ as empty.
\For{$(\lnodeidx, \rnodeidx) \in \map$}
	\State $\script \gets \script \concat \rep_{\lnodeidx, \rnode_\rnodeidx}$. \Comment replacements
\EndFor
\For{$\lnodeidx \in I$ in descending order}
	\State $\script \gets \script \concat \del_\lnodeidx$. \Comment deletions
\EndFor
\State $(\rnodelim, \childlim_0, \ldots, \childlim_\rnodelim) \gets$ \Call{num-descendants}{$\rforest$, $0$, $J$}. \Comment number of children for each inserted node
\For{$\rnodeidx \in \unmappedRight$ in ascending order}
	\State $\script \gets \script \concat \ins_{\paridx_\rforest(\rnodeidx), \rnode_\rnodeidx, \childidx_\rnodeidx, \childidx_\rnodeidx + \childlim_\rnodeidx}$, \Comment{insertions}
\EndFor
\State \Return $\script$.
\EndFunction
\end{algorithmic}
\begin{algorithmic}
\Function{num-descendants}{Forest $\rforest = \rnode(\bar{z}_1, \ldots, \bar{z}_k), \rtree_2, \ldots, \rtree_\childlim$, index $\rnodeidx$, index set $J$}
\State $\bar \childlim \gets \epsilon$.
\State $\tilde \childlim \gets 0$. \Comment The number of mapped descendants of this forest
\For{$\childidx \gets 1, \ldots, \childlim$}
	\State $\rnodeidx \gets \rnodeidx + 1$.
	\State $(\rnodeidx', \tilde \childlim_\rnodeidx, \ldots, \tilde \childlim_{\rnodeidx'}) \gets $
		\Call{num-descendants}{$\bar{z}_1, \ldots, \bar{z}_k$, $\rnodeidx$, $J$}.
	\State $\bar \childlim \gets \bar \childlim \concat \tilde \childlim_\rnodeidx, \ldots, \tilde \childlim_{\rnodeidx'}$.
	\If{$\rnodeidx \notin \unmappedRight$}
		\State $\tilde \childlim \gets \tilde \childlim + 1$.
	\Else
		\State $\tilde \childlim \gets \tilde \childlim + \tilde \childlim_\rnodeidx$.
	\EndIf
	\State $\rnodeidx \gets \rnodeidx'$.
\EndFor
\State \Return ($\rnodeidx$, $\tilde \childlim \concat \bar \childlim$).
\EndFunction
\end{algorithmic}
\end{algorithm}

As an example, consider again the mapping $\map = \{(1, 1), (4, 2)\}$ between the trees
$\ltree = \sym{a}(\sym{b}(\sym{c}, \sym{d}), \sym{e})$ and $\rtree = \sym{f}(\sym{g})$ from Figure~\ref{fig:edits}.
Here we have the non-mapped nodes $\unmappedLeft = \{ 2, 3, 5\}$ and $\unmappedRight = \{ \}$. 
Therefore, Algorithm~\ref{alg:map_to_script} returns the script $\rep_{1, \sym{f}}, \rep_{4, \sym{g}}, \del_5, \del_3, \del_2$.
Note that deletions are done in descending order to ensure that the pre-order indices in the tree do not
change for intermediate trees.

For the inverse mapping $\map = \{(1,1), (2, 4) \}$ between $\rtree$ and $\ltree$ we have
$\unmappedLeft = \{ \}$ and $\unmappedRight = \{2, 3, 5\}$. Further, the output of
\emph{num-descendants} is $(\rnodelim = 5, \childlim_0 = 1, \childlim_1 = 1, \childlim_2 = 1, \childlim_3 = 0, \childlim_4 = 0, \childlim_5 = 0)$.
Therefore, we obtain the script $\rep_{1, \sym{a}}, \rep_{2, \sym{d}},
\ins_{1, \sym{b}, 1, 2}, \ins_{2, \sym{c}, 1, 1}, \ins_{1, \sym{e}, 2, 2}$.

Our next task is to demonstrate that the inverse direction is also possible, that is, we can find a corresponding
mapping for each script.

\begin{thm}\label{thm:script_to_map}
Let $\lforest$ and $\rforest$ be forests over some alphabet $\alphabet$, and let $\script$ be
an edit script such that $\script(\lforest) = \rforest$. Then, the following, recursively
defined set $\map_{\script}$, is a mapping between $\lforest$ and $\rforest$:

\begin{align*}
\map_\epsilon &:= \{ (1, 1), \ldots, (\lnodelim, \lnodelim) \} \\
\map_{\edit_1, \ldots, \edit_\editlim} &:=
\begin{cases}
\quad \map_{\edit_1, \ldots, \edit_{\editlim-1}} & \text{if } \edit_\editlim = \rep_{\rnodeidx, \rnode_\rnodeidx} \\
\begin{array}{l}
\{ (\lnodeidx, \rnodeidx') | (\lnodeidx, \rnodeidx') \in \map_{\edit_1, \ldots, \edit_{\editlim-1}} , \rnodeidx' < \rnodeidx \} \quad \cup \\
\{ (\lnodeidx, \rnodeidx' - 1) | (\lnodeidx, \rnodeidx') \in \map_{\edit_1, \ldots, \edit_{\editlim-1}} , \rnodeidx' > \rnodeidx \}
\end{array}
& \text{if } \edit_\editlim = \del_\rnodeidx \\
\begin{array}{l}
\{ (\lnodeidx, \rnodeidx') | (\lnodeidx, \rnodeidx') \in \map_{\edit_1, \ldots, \edit_{\editlim-1}} , \rnodeidx' < \rnodeidx \} \quad \cup \\
\{ (\lnodeidx, \rnodeidx' + 1) | (\lnodeidx, \rnodeidx') \in \map_{\edit_1, \ldots, \edit_{\editlim-1}} , \rnodeidx' \geq \rnodeidx \}
\end{array}
& \text{if } \edit_\editlim = \ins_{\paridx_\rforest(\rnodeidx), \rnode_\rnodeidx, \childidx_\rnodeidx, \childidx_\rnodeidx + \childlim_\rnodeidx}
\end{cases}
\end{align*}
where $\childlim_\rnodeidx$ is the number of children of $\tree_\rnodeidx$.

\begin{proof}
We prove the claim via induction over the length of $\script$. $\map_\epsilon$ obviously conforms to all mapping constraints.

Now, assume that the claim is true for all scripts $\script$ with $|\script| \leq \editlim$ and consider a script
$\script = \edit_1, \ldots, \edit_{\editlim+1}$. Let $\script' = \edit_1, \ldots, \edit_\editlim$.
Due to induction, we know that $\map_{\script'}$ is a valid mapping between $\lforest$ and $\script'(\lforest)$.
Now, consider the last edit $\edit_{\editlim+1}$.

First, we observe that, if $\map_{\script'}$ fulfills the first three criteria of a mapping, $\map_{\script}$
does as well, because we never introduce many-to-one mappings and respect the pre-order. The only criterion left
in question is the fourth, namely whether $\map_{\script}$ respects the ancestral ordering of $\rforest$.

If $\edit_{\editlim+1}$ is a replacement, the tree structure of $\edit(\lforest)$ is the same as for
$\script'(\lforest)$. Therefore, $\map_{\script} = \map_{\script'}$ is also a valid mapping between $\lforest$ and $\rforest$.

If $\edit_{\editlim+1}$ is a deletion $\del_{\rnodeidx}$, then node $\rnode_\rnodeidx$ in
$\script'(\lforest)$ is missing from $\rforest$ and all subtrees with pre-order indices higher than
$\rnodeidx$ decrease their index by one, which is reflected by $\map_{\script}$. Further,
$\map_{\script}$ only removes a tuple, but does not add a tuple, such that all ancestral relationships
present in $\map_{\script}$ were also present in $\map_{\script'}$. Finally, a deletion does not
break any of the ancestral relationships because any ancestor of $\rtree_\rnodeidx$
remains an ancestor of all children of $\rtree_\rnodeidx$ in $\rforest$.
Therefore, $\map_{\script}$ is a valid mapping between $\lforest$ and $\rforest$.

If $\edit_{\editlim+1}$ is an insertion $\ins_{\paridx_\rforest(\rnodeidx), \rnode_\rnodeidx, \childidx_\rnodeidx, \childidx_\rnodeidx + \childlim_\rnodeidx}$,
then $\rnode_\rnodeidx$ is a new node in $\rforest$ and all subtrees with pre-order indices as high or higher
than $\rnodeidx$ in $\script'(\lforest)$ increase their index by one, which is reflected by $\map_{\script}$.
Further, $\map_{\script}$ leaves all tuples intact, such that all ancestral relationships of $\map_{\script'}$
are preserved. Finally, an insertion does not break any ancestral relationships because $\rtree_{\paridx_\rforest(\rnodeidx)}$
is still an ancestor of all nodes it was before, except that there is now a new node $\rnode_\rnodeidx$
in between. Therefore, $\map_{\script}$ is a valid mapping between $\lforest$ and $\rforest$.
\end{proof}
\end{thm}

\begin{figure}
\begin{center}
\begin{tikzpicture}

\begin{scope}

\node at (1, 1) {$\rep_{1, \sym{f}}$};

\begin{scope}
\Tree [.\node (al) {$\sym{a}$};
	[.\node (bl) {$\sym{b}$}; \node (cl) {$\sym{c}$}; \node (dl) {$\sym{d}$}; ]
	\node (el) {$\sym{e}$};
]
\end{scope}

\begin{scope}[shift={(2, 0)}]
\Tree [.\node (fr) {$\sym{f}$};
	[.\node (br) {$\sym{b}$}; \node (cr) {$\sym{c}$}; \node (dr) {$\sym{d}$}; ]
	\node (er) {$\sym{e}$};
]
\end{scope}

\path[edge, densely dashed, semithick]%
(al) edge[bend left] (fr)
(bl) edge[bend left] (br)
(cl) edge[bend left] (cr)
(dl) edge[bend left] (dr)
(el) edge[bend left] (er);

\end{scope}

\begin{scope}[shift={(4, 0)}]

\node at (1, 1) {$\del_2$};

\begin{scope}
\Tree [.\node (al) {$\sym{a}$};
	[.\node (bl) {$\sym{b}$}; \node (cl) {$\sym{c}$}; \node (dl) {$\sym{d}$}; ]
	\node (el) {$\sym{e}$};
]
\end{scope}

\begin{scope}[shift={(2, 0)}]
\Tree [.\node (fr) {$\sym{f}$}; \node (cr) {$\sym{c}$}; \node (dr) {$\sym{d}$}; \node (er) {$\sym{e}$}; ]
\end{scope}

\path[edge, densely dashed, semithick]%
(al) edge[bend left] (fr)
(cl) edge (cr)
(dl) edge (dr)
(el) edge[bend left] (er);

\end{scope}

\begin{scope}[shift={(8, 0)}]

\node at (1, 1) {$\ins_{2, \sym{e}, 2, 2}$};

\begin{scope}
\Tree [.\node (fl) {$\sym{f}$}; \node (gl) {$\sym{g}$}; ]
\end{scope}

\begin{scope}[shift={(2, 0)}]
\Tree [.\node (fr) {$\sym{f}$}; \node (gr) {$\sym{g}$}; \node (er) {$\sym{e}$}; ]
\end{scope}

\path[edge, densely dashed, semithick]%
(fl) edge[bend left] (fr)
(gl) edge[bend left] (gr);

\end{scope}

\end{tikzpicture}
\end{center}
\caption{An illustration of the recursive construction of the corresponding mapping $\map_{\script}$
for three edits from Figure~\ref{fig:edits}.
For each edit of the script, the mapping is updated to be consistent with all edits up until now.
In particular, the mapping starts as a one-to-one mapping, is left unchanged for all replacements,
and is shifted for all deletions and insertions}
\label{fig:script_to_map}
\end{figure}

As an example, consider the edit scripts shown in Figure~\ref{fig:edits}. For the script
$\script = \rep_{1, \sym{f}}, \del_2, \del_2, \rep_{2, \sym{g}},\allowbreak \del_3$, which transforms the
tree $\ltree = \sym{a}(\sym{b}(\sym{c}, \sym{d}), \sym{e})$ into the tree $\rtree = \sym{f}(\sym{g})$,
we obtain the following mappings $\map_\editidx$ after the $\editidx$th edit:
\begin{align*}
\map_0 &= \{ (1, 1), (2, 2), (3, 3), (4, 4), (5, 5) \} & \text{initial}\\
\map_1 &= \{ (1, 1), (2, 2), (3, 3), (4, 4), (5, 5) \} & \rep_{1, \sym{f}}\\
\map_2 &= \{ (1, 1), (3, 2), (4, 3), (5, 4) \} & \rep_{1, \sym{f}}, \del_2 \\
\map_3 &= \{ (1, 1), (4, 2), (5, 3) \} & \rep_{1, \sym{f}}, \del_2, \del_2 \\
\map_4 &= \{ (1, 1), (4, 2), (5, 3) \} & \rep_{1, \sym{f}}, \del_2, \del_2, \rep_{2, \sym{g}} \\
\map_5 &= \{ (1, 1), (4, 2) \} & \rep_{1, \sym{f}}, \del_2, \del_2, \rep_{2, \sym{g}}, \del_3 \\
\end{align*}

Conversely, for the script $\script^{-1} = \ins_{1, \sym{e}, 2, 2}, \rep_{2, \sym{d}}, \ins_{1, \sym{c}, 1, 1}, \ins_{1, \sym{b}, 1, 3}, \rep_{1, \sym{a}}$,
which transforms $\rtree$ into $\ltree$, we obtain the following mappings.
\begin{align*}
\map_0 &= \{ (1, 1), (2, 2) \} & \text{initial}\\
\map_1 &= \{ (1, 1), (2, 2) \} & \ins_{1, \sym{e}, 2, 2}\\
\map_2 &= \{ (1, 1), (2, 2) \} & \ins_{1, \sym{e}, 2, 2}, \rep_{2, \sym{d}} \\
\map_3 &= \{ (1, 1), (2, 3) \} & \ins_{1, \sym{e}, 2, 2}, \rep_{2, \sym{d}}, \ins_{1, \sym{c}, 1, 1} \\
\map_4 &= \{ (1, 1), (2, 4) \} & \ins_{1, \sym{e}, 2, 2}, \rep_{2, \sym{d}}, \ins_{1, \sym{c}, 1, 1}, \ins_{1, \sym{b}, 1, 3} \\
\map_5 &= \{ (1, 1), (2, 4) \} & \ins_{1, \sym{e}, 2, 2}, \rep_{2, \sym{d}}, \ins_{1, \sym{c}, 1, 1}, \ins_{1, \sym{b}, 1, 3}, \rep_{1, \sym{a}} \\
\end{align*}

The influence of the different kinds of edits on the mapping is also illustrated in Figure~\ref{fig:script_to_map}.

Now that we have shown that edit scripts and mappings can be related on a structural level,
it remains to show that they are also related in terms of \emph{cost}. To that end, we need to
define the cost of a mapping:

\begin{dfn}[Mapping cost]
Let $\lforest$ and $\rforest$ be forests over some alphabet $\alphabet$, and let $\cost$ be a cost function
over $\alphabet$. Further, let $\map$ be a mapping between $\lforest$ and $\rforest$,
let $\unmappedLeft = \{ \lnodeidx \in \{1, \ldots, |\lforest|\} | \nexists \rnodeidx : (\lnodeidx, \rnodeidx) \in \map \}$, and let
$\unmappedRight = \{ \rnodeidx \in \{1, \ldots, |\rforest|\} | \nexists \lnodeidx : (\lnodeidx, \rnodeidx) \in \map \}$

The \emph{cost} of the mapping $\map$ is defined as:
\begin{equation}
\Cost(\map, \lforest, \rforest) = \sum_{(\lnodeidx, \rnodeidx) \in \map} \cost(\lnode_\lnodeidx, \rnode_\rnodeidx)
+ \sum_{\lnodeidx \in \unmappedLeft} \cost(\lnode_\lnodeidx, \gap) + \sum_{\rnodeidx \in \unmappedRight} \cost(\gap, \rnode_\rnodeidx)
\end{equation}
\end{dfn}

For example, consider the mapping $\map = \{ (1, 1), (4, 2) \}$ between the trees in Figure~\ref{fig:edits}.
This mapping has cost
\begin{equation*}
\Cost\Big(\{(1, 1), (4, 2)\}, \sym{a}(\sym{b}(\sym{c}, \sym{d}), \sym{e}), \sym{f}(\sym{g})\Big)
= \cost(\sym{a}, \sym{f}) + \cost(\sym{d}, \sym{g}) + \cost(\sym{b}, \gap) + \cost(\sym{c}, \gap) + \cost(\sym{e}, \gap)
\end{equation*}
Note that this is equivalent to the cost of the edit script $\script = \rep_{1, \sym{f}}, \del_2, \del_2, \rep_{2, \sym{g}}, \del_3$.
However, the cost of an edit script is not always equal to the cost of its corresponding mapping. 
For example, consider the two
trees $\ltree = \sym{a}$ and $\rtree = \sym{b}$ and the script $\script = \rep_{1, \sym{c}}, \rep_{1, \sym{b}}$,
which transforms $\ltree$ to $\rtree$. Here, the corresponding mapping is $\map = \{(1, 1)\}$ with the
cost $\Cost(\map, \ltree, \rtree) = \cost(\sym{a}, \sym{b})$. However, the cost of the edit script
is $\Cost(\script, \ltree) = \cost(\sym{a}, \sym{c}) + \cost(\sym{c}, \sym{b})$, which will be
at least as expensive if the cost function conforms to the triangular inequality.

In general, we can show that mappings are at most as expensive as scripts if $\cost$ is
non-negative, self-equal, and conforms to the triangular inequality.

\begin{thm} \label{thm:cost_equivalence}
Let $\lforest$ and $\rforest$ be forests over some alphabet $\alphabet$ and $\cost$ be a cost function over
$\alphabet$. Further, let $\script$ be an edit script over $\edits_\alphabet$
with $\script(\lforest) = \rforest$, and let $\map$ be a mapping between $\lforest$ and
$\rforest$. Then it holds:
\begin{enumerate}
\item The corresponding script $\script_\map$ for $\map$ according to Algorithm~\ref{alg:map_to_script}
has the same cost as $\map$, that is: $\Cost(\map, \lforest, \rforest) = \Cost(\script_\map, \lforest)$.
\item If $\cost$ is non-negative, self-equal, and conforms to the triangular inequality,
the corresponding mapping $\map_{\script}$ for $\script$ according to
Theorem~\ref{thm:script_to_map} is at most as expensive as $\script$, that is:
$\Cost(\map_{\script}, \lforest, \rforest) \leq \Cost(\script, \lforest)$.
\end{enumerate}

\begin{proof}
Let $\lnodelim = |\lforest|$ and $\rnodelim = |\rforest|$, let
$\unmappedLeft = \{ \lnodeidx \in \{1, \ldots, \lnodelim\} | \nexists \rnodeidx : (\lnodeidx, \rnodeidx) \in \map \}$, and let
$\unmappedRight = \{ \rnodeidx \in \{1, \ldots, \rnodelim\} | \nexists \lnodeidx : (\lnodeidx, \rnodeidx) \in \map \}$

\begin{enumerate}
\item Due to Theorem~\ref{thm:map_to_script} we know that the script $\script_\map$ for $\map$
contains exactly one replacement $\rep_{\lnodeidx, \rnode_\rnodeidx}$ per entry $(\lnodeidx, \rnodeidx) \in \map$,
exactly one deletion $\del_\lnodeidx$ per unmapped index $\lnodeidx \in \unmappedLeft$ and exactly one
insertion $\ins_{\paridx_\rforest(\rnodeidx), \rnode_\rnodeidx, \childidx_\rnodeidx, \childidx + \childlim_\rnodeidx}$ per
unmapped index $\rnodeidx \in \unmappedRight$. Therefore, the cost of $\script_\map$ is:
\begin{equation}
\Cost(\script_\map, \lforest) = \sum_{(\lnodeidx, \rnodeidx) \in \map} \cost(\lnode_\lnodeidx, \rnode_\rnodeidx)
+ \sum_{\lnodeidx \in \unmappedLeft} \cost(\lnode_\lnodeidx, \gap)
+ \sum_{\rnodeidx \in \unmappedRight} \cost(\gap, \rnode_\rnodeidx)
\end{equation}
which is per definition equal to $\Cost(\map, \lforest, \rforest)$.
\item We show this claim via induction over the length of $\script$.
First, consider the case $\script = \epsilon$. Then, $\lforest = \rforest$ and
$\map_{\script} = \{(1, 1), \ldots, (\lnodelim, \lnodelim)\}$.
Because $\cost$ is self-equal, we obtain for the cost of $\map_{\script}$:
\begin{equation*}
\Cost(\map_{\script}, \lforest, \rforest)
= \sum_{(\lnodeidx, \rnodeidx) \in \map} \cost(\lnode_\lnodeidx, \rnode_\rnodeidx)
= \sum_{\lnodeidx = 1}^{\lnodelim} \cost(\lnode_\lnodeidx, \lnode_\lnodeidx)
= 0 = \cost(\epsilon, \lforest)
\end{equation*}

Now, assume that the claim holds for all $\script'$ with $|\script'| \leq \editlim$,
and consider a script $\script = \edit_1, \ldots, \edit_{\editlim + 1}$.
Let $\script' = \edit_1, \ldots, \edit_\editlim$ and let $\rforest' = \script'(\lforest)$.
Then, we consider the last edit $\edit_{\editlim + 1}$ and distinguish the following cases:

If $\edit_{\editlim + 1} = \rep_{\rnodeidx, \rnode}$ we have $\map_{\script} = \map_{\script'}$.
Further, if there is an $\lnodeidx \in \{1, \ldots, \lnodelim\}$ such that
$(\lnodeidx, \rnodeidx) \in \map_{\script'}$, we obtain for the cost:
\begin{align*}
\Cost(\script, \lforest) &= \Cost(\script', \lforest) + \cost(\rnode'_\rnodeidx, \rnode_\rnodeidx)
\stackrel{\text{Induction}}{\geq} \Cost(\map_{\script'}, \lforest, \rforest') + \cost(\rnode'_\rnodeidx, \rnode_\rnodeidx) \\
&= \Cost(\map_{\script}, \lforest, \rforest) - \cost(\lnode_\lnodeidx, \rnode_\rnodeidx)
+ \cost(\lnode_\lnodeidx, \rnode'_\rnodeidx)
+ \cost(\rnode'_\rnodeidx, \rnode_\rnodeidx) \\
&\geq \Cost(\map_{\script}, \lforest, \rforest) - \cost(\lnode_\lnodeidx, \rnode_\rnodeidx)
+ \cost(\lnode_\lnodeidx, \rnode_\rnodeidx) = \Cost(\map_{\script}, \lforest, \rforest)
\end{align*}
In case there is no $\lnodeidx \in \{1, \ldots, \lnodelim\}$ such that
$(\lnodeidx, \rnodeidx) \in \map_{\script'}$, we obtain for the cost:
\begin{align*}
\Cost(\script, \lforest) &= \Cost(\script', \lforest) + \cost(\rnode'_\rnodeidx, \rnode_\rnodeidx)
\stackrel{\text{Induction}}{\geq} \Cost(\map_{\script'}, \lforest, \rforest') + \cost(\rnode'_\rnodeidx, \rnode_\rnodeidx) \\
&= \Cost(\map_{\script}, \lforest, \rforest) - \cost(\gap, \rnode_\rnodeidx)
+ \cost(\gap, \rnode'_\rnodeidx)
+ \cost(\rnode'_\rnodeidx, \rnode_\rnodeidx) \\
&\geq \Cost(\map_{\script}, \lforest, \rforest) - \cost(\gap, \rnode_\rnodeidx)
+ \cost(\gap, \rnode_\rnodeidx) = \Cost(\map_{\script}, \lforest, \rforest)
\end{align*}

If $\edit_{\editlim +1} = \del_{\rnodeidx}$, consider first the case that there
exists some $\lnodeidx$ such that $(\lnodeidx, \rnodeidx) \in \map_{\script'}$.
Then, we obtain for the cost:
\begin{align*}
\Cost(\script, \lforest) &= \Cost(\script', \lforest) + \cost(\rnode'_\rnodeidx, \gap)
\stackrel{\text{Induction}}{\geq} \Cost(\map_{\script'}, \lforest, \rforest') +  \cost(\rnode'_\rnodeidx, \gap) \\
&= \Cost(\map_{\script}, \lforest, \rforest) - \cost(\lnode_\lnodeidx, \gap)
+ \cost(\lnode_\lnodeidx, \rnode'_\rnodeidx)
+ \cost(\rnode'_\rnodeidx, \gap) \\
&\geq \Cost(\map_{\script}, \lforest, \rforest) - \cost(\lnode_\lnodeidx, \gap)
+ \cost(\lnode_\lnodeidx, \gap) = \Cost(\map_{\script}, \lforest, \rforest)
\end{align*}
If there exists no such $\lnodeidx$, we obtain for the cost:
\begin{align*}
\Cost(\script, \lforest) &= \Cost(\script', \lforest) + \cost(\rnode'_\rnodeidx, \gap)
\stackrel{\text{Induction}}{\geq} \Cost(\map_{\script'}, \lforest, \rforest') +  \cost(\rnode'_\rnodeidx, \gap) \\
&= \Cost(\map_{\script}, \lforest, \rforest) + \cost(\gap, \rnode'_\rnodeidx)
+ \cost(\rnode'_\rnodeidx, \gap) \geq \Cost(\map_{\script}, \lforest, \rforest)
\end{align*}

Finally, if $\edit_{\editlim + 1} = \ins_{\paridx_\rforest(\rnodeidx), \rnode_\rnodeidx, \childidx_\rnodeidx, \childidx_\rnodeidx + \childlim_\rnodeidx}$,
we obtain for the cost
\begin{align*}
\Cost(\script, \lforest) &= \Cost(\script', \lforest) + \cost(\gap, \rnode_\rnodeidx)
\stackrel{\text{Induction}}{\geq} \Cost(\map_{\script'}, \lforest, \rforest') + \cost(\gap, \rnode_\rnodeidx) \\
&= \Cost(\map_{\script}, \lforest, \rforest) - \cost(\gap, \rnode_\rnodeidx)
+ \cost(\gap, \rnode_\rnodeidx)
= \Cost(\map_{\script}, \lforest, \rforest)
\end{align*}

This concludes our proof by induction.
\end{enumerate}

\end{proof}
\end{thm}

It follows directly that we can compute the TED by computing the cheapest mapping instead of the cheapest
edit script.

\begin{thm}
Let $\lforest$ and $\rforest$ be forests over some alphabet $\alphabet$ and $\cost$ be a cost function over
$\alphabet$ that is non-negative, self-equal, and conforms to the triangular inequality. Then it holds:

\begin{align}
&\min_{\script \in \edits_\alphabet^*} \{ \Cost(\script, \lforest) | \script(\lforest) = \rforest \} = \notag \\
&\min \{ \Cost(\map, \lforest, \rforest) | \map \text{ is a tree mapping between $\lforest$ and $\rforest$} \}
\end{align}

\begin{proof}
First, we define two abbreviations for the minima, namely:
\begin{align*}
\dist_\cost^\text{script}(\lforest, \rforest) &:= \min_{\script \in \edits_\alphabet^*} \{ \Cost(\script, \lforest) | \script(\lforest) = \rforest \} \\
\dist_\cost^\text{map}(\lforest, \rforest)    &:= \min \{ \Cost(\map, \lforest, \rforest) | \map \text{ is a tree mapping between $\lforest$ and $\rforest$} \}
\end{align*}

Let $\script$ be an edit script such that $\script(\lforest) = \rforest$ and $\Cost(\script, \lforest) = \dist_\cost^\text{script}(\lforest, \rforest)$.
Then, we know due to Theorem~\ref{thm:cost_equivalence} that the corresponding mapping $\map_{\script}$ is at most as expensive
as $\script$, i.e.\ $\Cost(\map_{\script}, \lforest, \rforest) \leq \Cost(\script, \lforest) = \dist_\cost^\text{script}(\lforest, \rforest)$. This
implies: $\dist_\cost^\text{map}(\lforest, \rforest) \leq \dist_\cost^\text{script}(\lforest, \rforest)$.

Conversely, let $\map$ be a tree mapping between $\lforest$ and $\rforest$, such that $\Cost(\map, \lforest, \rforest)
= \dist_\cost^\text{map}(\lforest, \rforest)$.
Then, we know due to Theorem~\ref{thm:cost_equivalence} that the corresponding edit script $\script_\map$ has
the same cost as $\map$, i.e.\ $\Cost(\script_\map, \lforest) = \Cost(\map, \lforest, \rforest)$. This implies:
$\dist_\cost^\text{script}(\lforest, \rforest) \leq \dist_\cost^\text{map}(\lforest, \rforest)$.
\end{proof}
\end{thm}

This concludes our theory on edit scripts, cost functions, and mappings. We have now laid enough groundwork to
efficiently compute the TED.

\section{The Dynamic Programming Algorithm}
\label{sec:dp}

To compute the TED between two trees $\ltree$ and $\rtree$ efficiently, we require a way
to decompose the TED into parts, such that we can compute the distance between subtrees of
$\ltree$ and $\rtree$ and combine those partial TEDs to an overall TED.
In order to do that, we need to define what we mean by \enquote{partial trees}.

\begin{dfn}[subforest]
Let $\lforest$ be a forest of size $\lnodelim = |\lforest|$.
Further, let $\lnodeidx, \rnodeidx \in \{1, \ldots, \lnodelim\}$ with $\lnodeidx \leq \rnodeidx$.
We define the \emph{subforest} of $\lforest$ from $\lnodeidx$ to $\rnodeidx$, denoted as
$\lforest[\lnodeidx, \rnodeidx]$, as the first output of Algorithm~\ref{alg:subforest} for the input
$\lforest$, $\lnodeidx$, $\rnodeidx$, and $0$.
\end{dfn}

\begin{algorithm}
\caption{An algorithm to retrieve the subforest from $\lnodeidx$ to $\rnodeidx$ of a forest $\lforest$.}
\label{alg:subforest}
\begin{algorithmic}
\Function{subforest}{Forest $\lforest = \ltree_1, \ldots, \ltree_\childlim$, start index $\lnodeidx$, end index $\rnodeidx$, current index $k$}
\State $\rforest \gets \epsilon$.
\For{$\childidx = 1, \ldots, \childlim$}
	\State $k \gets k + 1$.
	\State Let $x(\rtree_1, \ldots, \rtree_\rnodelim) \gets \ltree_\childidx$.
	\If{$k > \rnodeidx$}
		\State \Return $(\rforest, k)$.
	\ElsIf{$k \geq \lnodeidx$}
		\State $(\rforest', k) \gets $ \Call{subforest}{$\rtree_1, \ldots, \rtree_\rnodelim$, $\lnodeidx$, $\rnodeidx$, $k$}.
		\State $\rforest \gets \rforest \concat \lnode(\rforest')$.
	\Else
		\State $(\rforest', k) \gets $ \Call{subforest}{$\rtree_1, \ldots, \rtree_\rnodelim$, $\lnodeidx$, $\rnodeidx$, $k$}.
		\State $\rforest \gets \rforest'$.
	\EndIf
\EndFor
\State \Return $(\rforest, k)$.
\EndFunction
\end{algorithmic}
\end{algorithm}

As an example, consider the left tree $\ltree = \sym{a}(\sym{b}(\sym{c}, \sym{d}), \sym{e})$ from
Figure~\ref{fig:preorder} (left). For this example, we find:
$\lforest[1, 1] = \sym{a}$, $\lforest[2, 4] = \sym{b}(\sym{c}, \sym{d})$, $\lforest[3, 5] = \sym{c}, \sym{d}, \sym{e}$,
and $\lforest[2, 1] = \epsilon$.

Note that $\lforest[2, 4] = \ltree_2$, that is: The subforest of $\ltree$
from $2$ to $4$ is exactly the subtree rooted at $2$. In general, subforests which correspond
to subtrees are important special cases, which we can characterize in terms of
\emph{outermost right leafs}.

\begin{dfn}[outermost right leaf]
Let $\lforest$ be a forest of size $\lnodelim = |\lforest|$.
Further, let $\lnodeidx \in \{1, \ldots, \lnodelim\}$. We define the \emph{outermost right leaf}
of $\lnodeidx$ as
\begin{equation}
\rleaf_\lforest(\lnodeidx) = \lnodeidx + |\ltree_\lnodeidx|-1.
\end{equation}
\end{dfn}

Again, consider the tree $\ltree = \sym{a}(\sym{b}(\sym{c}, \sym{d}), \sym{e})$ from Figure~\ref{fig:edits}.
For this tree, we have $\rleaf_{\ltree}(1) = 5$, $\rleaf_{\ltree}(2) = 4$, $\rleaf_{\ltree}(2) = 4$,
$\rleaf_{\ltree}(3) = 3$, $\rleaf_{\ltree}(4) = 4$, $\rleaf_{\ltree}(5) = 5$.

More generally, we can show that the subforest from $i$ to its outermost right leaf is always the
subtree rooted at $i$.

\begin{thm} \label{thm:subtrees}
Let $\lforest$ be a forest. For any $\lnodeidx \in \{1, \ldots, |\lforest|\}$
it holds:
\begin{equation}
\lforest[\lnodeidx, \rleaf_\lforest(\lnodeidx)] = \ltree_\lnodeidx
\end{equation}

\begin{proof}
First, note that the pre-order algorithm (see definition~\ref{dfn:preorder}) visits parents before children and
left children before right children. Therefore, the largest index within a subtree must be the
outermost right leaf.
The claim follows because the subforest Algorithm~\ref{alg:subforest} visits nodes in
the same order as the pre-order algorithm and therefore
$\lforest[\lnodeidx, \rleaf_\lforest(\lnodeidx)] = \lforest[\lnodeidx, \lnodeidx + |\ltree_\lnodeidx|-1]
= \tree_\lnodeidx$.
\end{proof}
\end{thm}

Now, we can define the edit distance between partial trees, which we call the \emph{subforest edit distance}:

\begin{dfn}[Subforest edit distance]
Let $\lforest$ and $\rforest$ be forests over some alphabet $\alphabet$, let
$\cost$ be a cost function over $\alphabet$, let $\ltree_\lkeyrootidx$
be an ancestor of $\ltree_\lnodeidx$ in $\lforest$, and let $\rtree_\rkeyrootidx$ be an ancestor
of $\rtree_\rnodeidx$ in $\rforest$. Then, we define the \emph{subforest edit distance} between
the subforests $\lforest[\lnodeidx, \rleaf_\lforest(\lkeyrootidx)]$ and $\rforest[\rnodeidx, \rleaf_\rforest(\rkeyrootidx)]$
as
\begin{equation}
\fdist_\cost(\lforest[\lnodeidx, \rleaf_\lforest(\lkeyrootidx)], \rforest[\rnodeidx, \rleaf_\rforest(\rkeyrootidx)]) :=
\min_{\script \in \edits_\alphabet^*} \Big\{ \Cost(\script, \lforest[\lnodeidx, \rleaf_\lforest(\lkeyrootidx)]) \Big| \script(\lforest[\lnodeidx, \rleaf_\lforest(\lkeyrootidx)]) = \rforest[\rnodeidx, \rleaf_\rforest(\rkeyrootidx)] \Big\}
\end{equation}
\end{dfn}

It directly follows that:

\begin{thm}
Let $\lforest$ and $\rforest$ be trees over some alphabet $\alphabet$ of size $\lnodelim = |\lforest|$
and $\rnodelim = |\rforest|$ respectively. For every $\lnodeidx \in \{1, \ldots, \lnodelim\}$ and
$\rnodeidx \in \{1, \ldots, \rnodelim\}$ we have:
\begin{equation}
\fdist(\lforest[\lnodeidx, \rleaf_\lforest(\lnodeidx)], \rforest[\rnodeidx, \rleaf_\rforest(\rnodeidx)]) = \dist_\cost(\ltree_\lnodeidx, \rtree_\rnodeidx)
\end{equation}

\begin{proof}
From Theorem~\ref{thm:subtrees} we know that $\lforest[\lnodeidx, \rleaf_\lforest(\lnodeidx)] = \ltree_\lnodeidx$
and $\rforest[\rnodeidx, \rleaf_\rforest(\rnodeidx)] = \rtree_\rnodeidx$. Therefore, we have
\begin{equation*}
\fdist(\lforest[\lnodeidx, \rleaf_\lforest(\lnodeidx)], \rforest[\rnodeidx, \rleaf_\rforest(\rnodeidx)]) :=
\min_{\script \in \edits_\alphabet^*} \Big\{ \Cost(\script, \ltree_\lnodeidx) \Big| \script(\ltree_\lnodeidx) = \rtree_\rnodeidx \Big\}
\end{equation*}
which corresponds exactly to the definition of $\dist_\cost(\ltree_\lnodeidx, \rtree_\rnodeidx)$.
\end{proof}
\end{thm}

Finally, we can go on to prove the arguably most important theorem for the TED,
namely the recursive decomposition of the subforest edit distance:

\begin{thm} \label{thm:decompose}
Let $\lforest$ and $\rforest$ be non-empty forests over some alphabet $\alphabet$, let
$\cost$ be a cost function over $\alphabet$ that is non-negative, self-equal, and conforms to the triangular inequality, let $\ltree_\lkeyrootidx$
be an ancestor of $\ltree_\lnodeidx$ in $\lforest$, and let $\rtree_\rkeyrootidx$ be an ancestor
of $\rtree_\rnodeidx$ in $\rforest$. Then it holds:

\begin{align} 
\fdist_\cost(\lforest[\lnodeidx, &\rleaf_\lforest(\lkeyrootidx)], \rforest[\rnodeidx, \rleaf_\rforest(\rkeyrootidx)]) = 
\min\Big\{ \label{eq:fdist_decompose} \\
&\cost(\lnode_\lnodeidx, \gap) + \fdist_\cost(\lforest[\lnodeidx + 1, \rleaf_\lforest(\lkeyrootidx)], \rforest[\rnodeidx, \rleaf_\rforest(\rkeyrootidx)]), \notag \\
&\cost(\gap, \rnode_\rnodeidx) + \fdist_\cost(\lforest[\lnodeidx, \rleaf_\lforest(\lkeyrootidx)], \rforest[\rnodeidx + 1, \rleaf_\rforest(\rkeyrootidx)]), \notag \\
&\dist_\cost(\ltree_\lnodeidx, \rtree_\rnodeidx) + \fdist_\cost(\lforest[\rleaf_\lforest(\lnodeidx) + 1, \rleaf_\lforest(\lkeyrootidx)], \rforest[\rleaf_\rforest(\rnodeidx) + 1, \rleaf_\rforest(\rkeyrootidx)]) \Big\} \notag
\end{align}

Further it holds:

\begin{align}
\dist_\cost(\ltree_\lnodeidx, \rtree_\rnodeidx) = \min \{ 
&\cost(\lnode_\lnodeidx, \gap) + \fdist_\cost(\lforest[\lnodeidx + 1, \rleaf_\lforest(\lnodeidx)], \rforest[\rnodeidx, \rleaf_\rforest(\rnodeidx)]),  \label{eq:dist_decompose} \\
&\cost(\gap, \rnode_\rnodeidx) + \fdist_\cost(\lforest[\lnodeidx, \rleaf_\lforest(\lnodeidx)], \rforest[\rnodeidx + 1, \rleaf_\rforest(\rnodeidx)]), \notag \\
&\cost(\lnode_\lnodeidx, \rnode_\rnodeidx) + \fdist_\cost(\lforest[\lnodeidx + 1, \rleaf_\lforest(\lnodeidx)], \rforest[\rnodeidx + 1, \rleaf_\rforest(\rnodeidx)]) \Big\} \notag
\end{align}

\begin{proof}
We first show that an intermediate decomposition holds. In particular, we show that:
\begin{align} \label{eq:lemma_decompose}
\fdist_\cost(\lforest[\lnodeidx, &\rleaf_\lforest(\lkeyrootidx)], \rforest[\rnodeidx, \rleaf_\rforest(\rkeyrootidx)]) = 
\min\Big\{ \\
&\cost(\lnode_\lnodeidx, \gap) + \fdist_\cost(\lforest[\lnodeidx + 1, \rleaf_\lforest(\lkeyrootidx)], \rforest[\rnodeidx, \rleaf_\rforest(\rkeyrootidx)]), \notag \\
&\cost(\gap, \rnode_\rnodeidx) + \fdist_\cost(\lforest[\lnodeidx, \rleaf_\lforest(\lkeyrootidx)], \rforest[\rnodeidx + 1, \rleaf_\rforest(\rkeyrootidx)]), \notag \\
&\cost(\lnode_\lnodeidx, \rnode_\rnodeidx) + \fdist_\cost(\lforest[\lnodeidx + 1, \rleaf_\lforest(\lnodeidx)], \rforest[\rnodeidx + 1, \rleaf_\rforest(\rnodeidx)]) + \notag \\
&\fdist_\cost(\lforest[\rleaf_\lforest(\lnodeidx) + 1, \rleaf_\lforest(\lkeyrootidx)], \rforest[\rleaf_\rforest(\rnodeidx) + 1, \rleaf_\rforest(\rkeyrootidx)]) \Big\} \notag
\end{align}

Now, because we require that $\cost$ is non-negative, self-equal, and conforms to the triangular inequality, we know
that Theorem~\ref{thm:cost_equivalence} holds, that is, we know that we can replace the cost of
a cheapest edit script with the cost of a cheapest mapping. Let $\map$ be a cheapest mapping
between the subtrees $\lforest[\lnodeidx, \rleaf_\lforest(\lkeyrootidx)]$ and
$\rforest[\rnodeidx, \rleaf_\rforest(\rkeyrootidx)]$. Regarding $\lnodeidx$ and $\rnodeidx$, only
the following cases can occur:
\begin{enumerate}
\item $\lnodeidx$ is not part of the mapping. In that case, $\lnode_\lnodeidx$ is deleted and
we have $\fdist_\cost(\lforest[\lnodeidx, \rleaf_\lforest(\lkeyrootidx)], \rforest[\rnodeidx, \rleaf_\rforest(\rkeyrootidx)]) \allowbreak=
\cost(\lnode_\lnodeidx, \gap) + \fdist_\cost(\lforest[\lnodeidx + 1, \rleaf_\lforest(\lkeyrootidx)], \rforest[\rnodeidx, \rleaf_\rforest(\rkeyrootidx)])$.
\item $\rnodeidx$ is not part of the mapping. In that case, $\rnode_\rnodeidx$ is inserted and we
have $\cost(\gap, \rnode_\rnodeidx) + \fdist_\cost(\lforest[\lnodeidx, \rleaf_\lforest(\lkeyrootidx)], \allowbreak \rforest[\rnodeidx + 1, \rleaf_\rforest(\rkeyrootidx)])$.
\item Both $\lnodeidx$ and $\rnodeidx$ are part of the mapping. Let $\rnodeidx'$ be the index
$\lnodeidx$ is mapped to and let $\lnodeidx'$ be the index that is mapped to $\rnodeidx$, that is,
$(\lnodeidx, \rnodeidx') \in \map$ and $(\lnodeidx', \rnodeidx) \in \map$. Because of the third
constraint on mappings we know that $\lnodeidx \geq \lnodeidx' \iff \rnodeidx' \geq \rnodeidx$
and $\lnodeidx \leq \lnodeidx' \iff \rnodeidx' \leq \rnodeidx$. Now, consider the case that
$\lnodeidx' > \lnodeidx$. In that case we know that $\rnodeidx' < \rnodeidx$.
However, in that case, $\rnodeidx'$ is not part of the subforest
$\rforest[\rnodeidx, \rleaf_\rforest(\rkeyrootidx)]$, because $\rnodeidx$ is per definition the
smallest index within the subforest. Therefore, $(\lnodeidx, \rnodeidx')$ can not be part of
a cheapest mapping between our two considered subforests.

Conversely, consider the case $\lnodeidx' < \lnodeidx$. This is also not possible because in that
case $\lnodeidx'$ is not part of the subforest $\lforest[\lnodeidx, \rleaf_\lforest(\lkeyrootidx)]$.
Therefore, it must hold that $\lnodeidx' = \lnodeidx$. However, this implies by the first
constraint on mappings that $\rnodeidx' = \rnodeidx$. Therefore, $(\lnodeidx, \rnodeidx) \in \map$.
So we know that $\lnode_\lnodeidx$ is replaced with $\rnode_\rnodeidx$, which implies that
$\fdist_\cost(\lforest[\lnodeidx, \rleaf_\lforest(\lkeyrootidx)], \rforest[\rnodeidx, \rleaf_\rforest(\rkeyrootidx)]) = \cost(\lnode_\lnodeidx, \rnode_\rnodeidx) + \fdist_\cost(\lforest[\lnodeidx + 1, \rleaf_\lforest(\lkeyrootidx)], \rforest[\rnodeidx + 1, \rleaf_\rforest(\rkeyrootidx)])$.

However, if $(\lnodeidx, \rnodeidx) \in \map$, the fourth constraint on mappings implies that all
descendants of $\ltree_\lnodeidx$ are mapped to descendants of $\rtree_\rnodeidx$. More specifically,
for any $(\lnodeidx', \rnodeidx') \in \map$ where $\ltree_\lnodeidx$ is an ancestor of $\ltree_{\lnodeidx'}$
it must hold that $\rtree_\rnodeidx$ is also an ancestor of $\rtree_{\rnodeidx'}$.
Therefore, the subforest edit distance further decomposes into:
$\fdist_\cost(\lforest[\lnodeidx, \rleaf_\lforest(\lkeyrootidx)], \rforest[\rnodeidx, \rleaf_\rforest(\rkeyrootidx)]) =
\cost(\lnode_\lnodeidx, \rnode_\rnodeidx) + \fdist_\cost(\lforest[\lnodeidx + 1, \rleaf_\lforest(\lnodeidx)], \rforest[\rnodeidx + 1, \rleaf_\rforest(\rnodeidx)]) +
\fdist_\cost(\lforest[\rleaf_\lforest(\lnodeidx) + 1, \rleaf_\lforest(\lkeyrootidx)], \rforest[\rleaf_\rforest(\rnodeidx) + 1, \rleaf_\rforest(\rkeyrootidx)])$.
\end{enumerate}
Because we required that $\map$ is a cheapest mapping, the minimum of these three options must
be the case, which yields Equation~\ref{eq:lemma_decompose}.

Using this intermediate result, we can now go on to prove Equations~\ref{eq:fdist_decompose} and
\ref{eq:dist_decompose}. In particular, Equation~\ref{eq:dist_decompose} holds because we find that:
\begin{align*}
\dist_\cost(\ltree_\lnodeidx, \rtree_\rnodeidx) = &\fdist_\cost(\lforest[\lnodeidx, \rleaf_\lforest(\lnodeidx)], \rforest[\rnodeidx, \rleaf_\rforest(\rnodeidx)]) \\
= &\min\Big\{ \cost(\lnode_\lnodeidx, \gap) + \fdist_\cost(\lforest[\lnodeidx + 1, \rleaf_\lforest(\lnodeidx)], \rforest[\rnodeidx, \rleaf_\rforest(\rnodeidx)]), \\
&\cost(\gap, \rnode_\rnodeidx) + \fdist_\cost(\lforest[\lnodeidx, \rleaf_\lforest(\lnodeidx)], \rforest[\rnodeidx + 1, \rleaf_\rforest(\rnodeidx)]),  \\
&\cost(\lnode_\lnodeidx, \rnode_\rnodeidx) + \fdist_\cost(\lforest[\lnodeidx + 1, \rleaf_\lforest(\lnodeidx)], \rforest[\rnodeidx + 1, \rleaf_\rforest(\rnodeidx)]) +  \\
&\fdist_\cost(\lforest[\rleaf_\lforest(\lnodeidx) + 1, \rleaf_\lforest(\lnodeidx)], \rforest[\rleaf_\rforest(\rnodeidx) + 1, \rleaf_\rforest(\rnodeidx)]) \Big\} 
\end{align*}
where $\fdist_\cost(\lforest[\rleaf_\lforest(\lnodeidx) + 1, \rleaf_\lforest(\lnodeidx)], \rforest[\rleaf_\rforest(\rnodeidx) + 1, \rleaf_\rforest(\rnodeidx)]) = 0$
because the two input forests are empty.

With respect to Equation~\ref{eq:fdist_decompose}, we observe that one way to transform the subforest $\lforest[\lnodeidx, \rleaf_\lforest(\lkeyrootidx)]$
into the subforest $\rforest[\rnodeidx, \rleaf_\rforest(\rkeyrootidx)]$ is to transform the subforest
$\lforest[\lnodeidx, \rleaf_\lforest(\lnodeidx)]$ into the subforest $\rforest[\rnodeidx, \rleaf_\rforest(\rnodeidx)]$, and then the
subforest $\lforest[\rleaf_\lforest(\lnodeidx) + 1, \rleaf_\lforest(\lkeyrootidx)]$ into the subforest $\rforest[\rleaf_\rforest(\rnodeidx) + 1, \rleaf_\rforest(\rkeyrootidx)]$.
Therefore, we obtain:
\begin{align}
&\fdist_\cost(\lforest[\lnodeidx, \rleaf_\lforest(\lkeyrootidx)], \rforest[\rnodeidx, \rleaf_\rforest(\rkeyrootidx)]) \label{eq:observation1} \\
\leq & \fdist_\cost(\lforest[\lnodeidx, \rleaf_\lforest(\lnodeidx)], \rforest[\rnodeidx, \rleaf_\rforest(\rnodeidx)]) +
\fdist_\cost(\lforest[\rleaf_\lforest(\lnodeidx) + 1, \rleaf_\lforest(\lkeyrootidx)], \rforest[\rleaf_\rforest(\rnodeidx) + 1, \rleaf_\rforest(\rkeyrootidx)]) \notag \\
= & \dist_\cost(\ltree_\lnodeidx, \rtree_\rnodeidx) + 
\fdist_\cost(\lforest[\rleaf_\lforest(\lnodeidx) + 1, \rleaf_\lforest(\lkeyrootidx)], \rforest[\rleaf_\rforest(\rnodeidx) + 1, \rleaf_\rforest(\rkeyrootidx)]), \notag
\end{align}
Further, we observe that 
\begin{equation}
\dist_\cost(\ltree_\lnodeidx, \rtree_\rnodeidx) \leq \cost(\lnode_\lnodeidx, \rnode_\rnodeidx) + \fdist_\cost(\lforest[\lnodeidx + 1, \rleaf_\lforest(\lnodeidx)], \rforest[\rnodeidx + 1, \rleaf_\rforest(\rnodeidx)]), \label{eq:observation2}
\end{equation}
because this is only one of the three cases in Equation~\ref{eq:dist_decompose}.

Now, note that the first two cases in Equations~\ref{eq:fdist_decompose} and~\ref{eq:lemma_decompose}
are the same. Finally, consider that the last case of Equation~\ref{eq:lemma_decompose}. In that case,
we can conclude that:
\begin{align*}
&\dist_\cost(\ltree_\lnodeidx, \rtree_\rnodeidx) + \fdist_\cost(\lforest[\rleaf_\lforest(\lnodeidx) + 1, \rleaf_\lforest(\lkeyrootidx)], \rforest[\rleaf_\rforest(\rnodeidx) + 1, \rleaf_\rforest(\rkeyrootidx)]) \\
\stackrel{\ref{eq:observation2}}{\leq} & \cost(\lnode_\lnodeidx, \rnode_\rnodeidx) + \fdist_\cost(\lforest[\lnodeidx + 1, \rleaf_\lforest(\lnodeidx)], \rforest[\rnodeidx + 1, \rleaf_\rforest(\rnodeidx)]) + \\
& \fdist_\cost(\lforest[\rleaf_\lforest(\lnodeidx) + 1, \rleaf_\lforest(\lkeyrootidx)], \rforest[\rleaf_\rforest(\rnodeidx) + 1, \rleaf_\rforest(\rkeyrootidx)]) \\
\stackrel{\ref{eq:lemma_decompose}}{=} &\fdist_\cost(\lforest[\lnodeidx, \rleaf_\lforest(\lkeyrootidx)], \rforest[\rnodeidx, \rleaf_\rforest(\rkeyrootidx)]) \\
\stackrel{\ref{eq:observation1}}{\leq} & \dist_\cost(\ltree_\lnodeidx, \rtree_\rnodeidx) + \fdist_\cost(\lforest[\rleaf_\lforest(\lnodeidx) + 1, \rleaf_\lforest(\lkeyrootidx)], \rforest[\rleaf_\rforest(\rnodeidx) + 1, \rleaf_\rforest(\rkeyrootidx)])
\end{align*}
which implies that Equation~\ref{eq:fdist_decompose} holds.
\end{proof}
\end{thm}

As an example, consider the trees $\ltree = \sym{a}(\sym{b}(\sym{c}, \sym{d}), \sym{e})$ and
$\rtree = \sym{f}(\sym{g})$ from Figure~\ref{fig:edits} and the subforest edit distance
$\fdist_\cost(\ltree[2,5], \rtree[1,2]) = \fdist_\cost\big((\sym{b}(\sym{c}, \sym{d}), \sym{e}), \sym{f}(\sym{g})\big)$.
According to Equation~\ref{eq:fdist_decompose}, we can decompose this in three ways, corresponding
to the options to delete $\sym{b}$, replace $\sym{b}$ with $\sym{f}$, and insert $\sym{f}$.
In particular, we obtain the distance
$\cost(\sym{b}, \gap) + \fdist_\cost\big((\sym{c}, \sym{d}, \sym{e}), \sym{f}(\sym{g})\big)$ for the deletion,
$\dist_\cost\big(\sym{b}(\sym{c}, \sym{d}), \sym{f}(\sym{g})\big) + \fdist_\cost(\sym{e}, \epsilon)$ for the replacement,
and $\cost(\gap, \sym{f}) + \fdist_\cost\big((\sym{b}(\sym{c}, \sym{d}), \sym{e}), \sym{g}\big)$ for the insertion
(also refer to Figure~\ref{fig:decompose}).

Now, consider the replacement option. According to Equation~\ref{eq:dist_decompose}, we can
decompose the TED $\dist_\cost\big(\sym{b}(\sym{c}, \sym{d}), \sym{f}(\sym{g})\big)$
in three ways, corresponding to the options to delete $\sym{b}$, replace $\sym{b}$ with $\sym{f}$,
and insert $\sym{f}$. In particular, we obtain the distance
$\cost(\sym{b}, \gap) + \fdist_\cost\big((\sym{d}, \sym{e}), \sym{f}(\sym{g})\big)$ for the deletion,
$\cost(\sym{b}, \sym{f}) + \fdist_\cost\big((\sym{d}, \sym{e}), \sym{g}\big)$ for the replacement,
and $\cost(\gap, \sym{f}) + \fdist_\cost\big(\sym{b}(\sym{c}, \sym{d}), \sym{g}\big)$ for the insertion
(also refer to Figure~\ref{fig:decompose}).

\begin{figure}
\begin{center}
\begin{tikzpicture}
\node (initial) {$\fdist_\cost\Big($ \tikz[baseline=0cm]{\Tree [.$\sym{b}$ $\sym{c}$ $\sym{d}$ ]} $\sym{e}$, \tikz[baseline=0cm]{\Tree [.$\sym{f}$ $\sym{g}$ ]} $\Big)$};

\node (del)  at(-4,-1) {$\cost(\sym{b}, \gap) + \fdist_\cost\Big(\sym{c}\, \sym{d}\, \sym{e}$, \tikz[baseline=0cm]{\Tree [.$\sym{f}$ $\sym{g}$ ]} $\Big)$};
\node (rep)  at( 0,-3) {$\dist_\cost\Big($ \tikz[baseline=0cm]{\Tree [.$\sym{b}$ $\sym{c}$ $\sym{d}$ ]}, \tikz[baseline=0cm]{\Tree [.$\sym{f}$ $\sym{g}$ ]} $\Big) + \fdist_\cost(\sym{e}, \epsilon)$};
\node (ins)  at( 4,-1) {$\cost(\gap, \sym{f}) + \fdist_\cost\Big($ \tikz[baseline=0cm]{\Tree [.$\sym{b}$ $\sym{c}$ $\sym{d}$ ]} $\sym{e}$, $\sym{g} \Big)$};

\node (del2) at(-4,-4) {$\cost(\sym{b}, \gap) + \fdist_\cost\Big(\sym{c} \, \sym{d}$, \tikz[baseline=0cm]{\Tree [.$\sym{f}$ $\sym{g}$ ]} $\Big)$};
\node (rep2) at( 0,-6) {$\cost(\sym{b}, \sym{f}) + \fdist_\cost\Big(\sym{c}\, \sym{d}, \sym{g}\Big)$};
\node (ins2) at( 4,-4) {$\cost(\gap, \sym{f}) + \fdist_\cost\Big($ \tikz[baseline=0cm]{\Tree [.$\sym{b}$ $\sym{c}$ $\sym{d}$ ]}, $\sym{g} \Big)$};

\path[edge, semithick]%
(-1.5,0.55)  edge[out=180, in=90] node[above left]  {delete $\sym{b}$} (del.north)
(initial)    edge                 node[right]       {replace $\sym{b}$ with $\sym{f}$} (rep)
( 1.5,0.55)  edge[out=0, in=90]   node[above right] {insert $\sym{f}$} (ins.north)
(-2.1,-2.45) edge[out=180, in=90] node[above left]  {delete $\sym{b}$} (del2.north)
(rep)        edge                 node[right]       {replace $\sym{b}$ with $\sym{f}$}  (rep2)
( 2.1,-2.45) edge[out=0, in=90]   node[above right] {insert $\sym{f}$} (ins2.north);
\end{tikzpicture}
\end{center}
\caption{An illustration of the decompositions in Equation~\ref{eq:fdist_decompose} and~\ref{eq:dist_decompose}.}
\label{fig:decompose}
\end{figure}

For an efficient algorithm for the TED, we are missing only one last ingredient,
namely a valid base case for empty forests. This is easy enough to obtain:

\begin{thm}
Let $\lforest$ and $\rforest$ be forests over some alphabet $\alphabet$, let
$\cost$ be a cost function $\alphabet$ that is non-negative, self-equal, and conforms to the
triangular inequality, let $\ltree_\lkeyrootidx$
be an ancestor of $\ltree_\lnodeidx$ in $\lforest$, and let $\rtree_\rkeyrootidx$ be an ancestor
of $\rtree_\rnodeidx$ in $\rforest$. Then it holds:
\begin{align}
\fdist_\cost(\epsilon, \epsilon) &= 0 \label{eq:empty} \\
\fdist_\cost(\lforest[\lnodeidx, \rleaf_\lforest(\lkeyrootidx)], \epsilon) &= \cost(\lnode_\lnodeidx, \gap) + \fdist_\cost(\lforest[\lnodeidx + 1, \rleaf_\lforest(\lkeyrootidx)], \epsilon) \label{eq:del} \\
\fdist_\cost(\epsilon, \rforest[\rnodeidx, \rleaf_\rforest(\rkeyrootidx)]) &= \cost(\gap, \rnode_\rnodeidx) + \fdist_\cost(\epsilon, \rforest[\rnodeidx + 1, \rleaf_\rforest(\rkeyrootidx)]) \label{eq:ins} 
\end{align}

\begin{proof}
Because $\cost$ non-negative and self-equal, the cheapest script to transform an empty forest into an empty forest is
to do nothing. Further, because $\cost$ conforms to the trinagular inequality, the cheapest script
to transform a non-empty forest into an empty forest is to delete all nodes.
Finally, the cheapest script to transform an empty forest into a non-empty one is to insert all nodes.
\end{proof}
\end{thm}

Now, we are able to construct an efficient algorithm for the TED. We just need to
iterate over all possible pairs of subtrees in both input trees and compute the TED
between these subtrees. For this, we require the subforest edit distance for all pairs of subforests
in these subtrees. We store intermediate results for the subforest edit distance in an array
$\mat \fdist$ and intermediate results for the subtree edit distance in an array $\mat \dist$.
Finally, the edit distance between the whole input trees will be in the first entry of $\mat \dist$.

\begin{thm}
The output of Algorithm~\ref{alg:ted_proto} is the TED. Further, Algorithm~\ref{alg:ted_proto}
runs in $\effic(\lnodelim^2 \cdot \rnodelim^2)$ time and in $\effic(\lnodelim \cdot \rnodelim)$ space
complexity where $\lnodelim$ and $\rnodelim$ are the sizes of the input trees.

\begin{proof}
Each computational step is justified by one of the equations proven before (refer to the comments
in the pseudo-code). Therefore, the output of the algorithm is correct. Finally, the algorithm
runs in $\effic(\lnodelim^2 \cdot \rnodelim^2)$ because two of the nested for-loops run at most
$\lnodelim$ times and two of the loops run at most $\rnodelim$ times. Note that this bound is tight
because the worst case does occur for the trees shown in Figure~\ref{fig:worst_case}.
Regarding space complexity, we note that we maintain two matrices, $\mat \dist$ and $\mat \fdist$, each
with $\effic(\lnodelim \cdot \rnodelim)$ entries.
\end{proof}
\end{thm}

\begin{algorithm}
\caption{An efficient algorithm for the TED. Note that this algorithm is not yet
the most efficient one, but a proto-version of the actual TED algorithm of
\textcite{Zhang1989} which is shown later as Algorithm~\ref{alg:ted}.
The algorithm iterates over all subtrees of $\ltree$ and $\rtree$ and computes the tree edit
distance for them based on the forest edit distances between all subforests of the respective
subtrees.}
\label{alg:ted_proto}
\begin{algorithmic}
\Function{tree-edit-distance}{Two input trees $\ltree$ and $\rtree$, a cost function $\cost$.}
\State $\lnodelim \gets |\ltree|$, $\rnodelim \gets |\rtree|$.
\State $\mat \dist \gets \lnodelim \times \rnodelim$ matrix of zeros. \Comment $\mat \dist_{\lnodeidx, \rnodeidx} = \dist_\cost(\ltree_\lnodeidx, \rtree_\rnodeidx)$.
\State $\mat \fdist \gets (\lnodelim + 1) \times (\rnodelim + 1)$ matrix of zeros.
\State \Comment $\mat \fdist_{\lnodeidx, \rnodeidx} = \fdist_\cost(\lforest[\lnodeidx, \rleaf_{\ltree}(\lkeyrootidx)], \rforest[\rnodeidx, \rleaf_{\rtree}(\rkeyrootidx)])$.
\For{$\lkeyrootidx \gets \lnodelim, \ldots, 1$}
\For{$\rkeyrootidx \gets \rnodelim, \ldots, 1$}
\State $\mat \fdist_{\rleaf_\lforest(\lkeyrootidx) + 1, \rleaf_\rforest(\rkeyrootidx) + 1} \gets 0$. \Comment Equation~\ref{eq:empty}
\For{$\lnodeidx \gets \rleaf_\lforest(\lkeyrootidx), \ldots, \lkeyrootidx$}
	\State $\mat \fdist_{\lnodeidx, \rleaf_\rforest(\rkeyrootidx) + 1} \gets \mat \fdist_{\lnodeidx + 1, \rleaf_\rforest(\rkeyrootidx) + 1} + \cost(\lnode_{\lnodeidx}, \gap)$.
	\Comment Equation~\ref{eq:del}
\EndFor
\For{$\rnodeidx \gets \rleaf_\rforest(\rkeyrootidx), \ldots, \rkeyrootidx$}
	\State $\mat \fdist_{\rleaf_\lforest(\lkeyrootidx) + 1, \rnodeidx} \gets \mat \fdist_{\rleaf_\lforest(\lkeyrootidx) + 1, \rnodeidx + 1} + \cost(\gap, \rnode_{\rnodeidx})$.
	\Comment Equation~\ref{eq:ins}
\EndFor
\For{$\lnodeidx \gets \rleaf_\lforest(\lkeyrootidx), \ldots, \lkeyrootidx$}
	\For{$\rnodeidx \gets \rleaf_\rforest(\rkeyrootidx), \ldots, \rkeyrootidx$}
		\If{$\rleaf_{\ltree}(\lnodeidx) = \rleaf_{\ltree}(\lkeyrootidx) \wedge \rleaf_{\rtree}(\rnodeidx) = \rleaf_{\rtree}(\rkeyrootidx)$}
			\State $\mat \fdist_{\lnodeidx, \rnodeidx} \gets \min \{
			              \mat \fdist_{\lnodeidx + 1, \rnodeidx    } + \cost(\lnode_{\lnodeidx}, \gap),$
			\State $\quad \mat \fdist_{\lnodeidx    , \rnodeidx + 1} + \cost(\gap, \rnode_{\rnodeidx}),$
			\State $\quad \mat \fdist_{\lnodeidx + 1, \rnodeidx + 1} + \cost(\lnode_{\lnodeidx}, \rnode_{\rnodeidx}) \}$. \Comment Equation~\ref{eq:dist_decompose}
			\State $\mat \dist_{\lnodeidx, \rnodeidx} \gets \mat \fdist_{\lnodeidx, \rnodeidx}$.
		\Else
			\State $\mat \fdist_{\lnodeidx, \rnodeidx} \gets \min \{
			\mat \fdist_{\lnodeidx + 1, \rnodeidx    } + \cost(\lnode_{\lnodeidx}, \gap),$
			\State $\quad \mat \fdist_{\lnodeidx    , \rnodeidx + 1} + \cost(\gap, \rnode_{\rnodeidx}),$
			\State $\quad \mat \fdist_{\rleaf_{\ltree}(\lnodeidx) + 1, \rleaf_{\rtree}(\rnodeidx) + 1} + \mat \dist_{\lnodeidx, \rnodeidx} \}$. \Comment Equation~\ref{eq:fdist_decompose}
		\EndIf
	\EndFor
\EndFor
\EndFor
\EndFor
\State \Return $\mat \dist_{1, 1}$.
\EndFunction
\end{algorithmic}
\end{algorithm}

\begin{figure}
\begin{center}
\begin{tikzpicture}
\Tree[. $\lnode_1$ $\lnode_2$
	[. $\lnode_3$ $\lnode_4$
		[. $\ldots$ $\node_{\lnodelim - 3}$
			[. $\lnode_{\lnodelim - 2}$ $\lnode_{\lnodelim - 1}$ $\lnode_\lnodelim$ ]
		]
	]
]
\end{tikzpicture}
\end{center}
\caption{An example tree structure which yields the worst-case runtime of $\effic(\lnodelim^2 \cdot \rnodelim^2)$
in Algorithm~\ref{alg:ted_proto} as well as Algorithm~\ref{alg:ted}.}
\label{fig:worst_case}
\end{figure}

As \textcite{Zhang1989} point out, we can be even more efficient in our algorithm if we re-use
already computed subforest edit distances. In particular, we can re-use the subforest edit distance
whenever the outermost right leaf is equal:

\begin{thm} \label{thm:keyroots}
Let $\lforest$ and $\rforest$ be forests over some alphabet $\alphabet$, let
$\cost$ be a cost function over $\alphabet$, let $\ltree_\lkeyrootidx$
be an ancestor of $\ltree_{\lkeyrootidx'}$ in $\lforest$ such that $\rleaf_\lforest(\lkeyrootidx) = \rleaf_\lforest(\lkeyrootidx')$, 
and let $\rtree_\rkeyrootidx$ be an ancestor of $\rtree_{\rkeyrootidx'}$ such that $\rleaf_\rforest(\rkeyrootidx) = \rleaf_\rforest(\rkeyrootidx')$.
Then it holds for all $\lnodeidx$ such that $\ltree_{\lkeyrootidx'}$ is an ancestor of $\ltree_\lnodeidx$
and all $\rnodeidx$ such that $\rtree_{\rkeyrootidx'}$ is an ancestor of $\rtree_\rnodeidx$:

\begin{equation}
\fdist_\cost(\lforest[\lnodeidx, \rleaf_\lforest(\lkeyrootidx)], \rforest[\rnodeidx, \rleaf_\rforest(\rkeyrootidx)])
= \fdist_\cost(\lforest[\lnodeidx, \rleaf_\lforest(\lkeyrootidx')], \rforest[\rnodeidx, \rleaf_\rforest(\rkeyrootidx')])
\end{equation}

\begin{proof}
Because we required that $\rleaf_\lforest(\lkeyrootidx) = \rleaf_\lforest(\lkeyrootidx')$ and
$\rleaf_\rforest(\rkeyrootidx) = \rleaf_\rforest(\rkeyrootidx')$, this follows directly.
\end{proof}
\end{thm}

Therefore, we can make our algorithm faster by letting the two outer loops only run over nodes
for which we can \emph{not} re-use the subforest edit distance. Those nodes are the so-called
\emph{keyroots} of our input trees.

\begin{dfn}[keyroots]
Let $\lforest$ be a forest over some alphabet $\alphabet$ and let $\ltree_\lnodeidx$ be a leaf
in $\lforest$. We define the \emph{keyroot} of $\ltree_\lnodeidx$ as
\begin{equation}
\keyroot_\lforest(\lnodeidx) = \min \{ \lkeyrootidx | \rleaf_\lforest(\lkeyrootidx) = \lnodeidx \}
\end{equation}

We define the keyroots of $\lforest$, denoted as $\keyroots(\lforest)$ as the set of keyroots for
all leaves of $\lforest$.
\end{dfn}

For example, if we inspect tree $\ltree = \sym{a}(\sym{b}(\sym{c}, \sym{d}), \sym{e})$ from Figure~\ref{fig:edits}
our leaves are $\ltree_3 = \sym{c}$, $\ltree_4 = \sym{d}$, and $\ltree_5 = \sym{e}$. The corresponding
key roots are $\keyroot_{\ltree}(3) = 3$, $\keyroot_{\ltree}(4) = 2$, and $\keyroot_{\ltree}(5) = 1$. Accordingly, the
set of keyroots $\keyroots(\ltree)$ is $\{1, 2, 3\}$.

Computing the keyroots is possible using Algorithm~\ref{alg:keyroots}.

\begin{algorithm}
\caption{An algorithm to compute the key roots of a forest.}
\label{alg:keyroots}
\begin{algorithmic}
\Function{keyroots}{A forest $\lforest$.}
\State $\lnodelim \gets |\lforest|$.
\State $R \gets \emptyset$.
\State $\keyroots \gets \epsilon$.
\For{$\lnodeidx \gets 1, \ldots, \lnodelim$}
	\State $\rleaf \gets \rleaf_\lforest(\lnodeidx)$.
	\If{$\rleaf \notin R$}
		\State $R \gets R \cup \{ \rleaf \}$.
		\State $\keyroots \gets \keyroots \concat \lnodeidx$.
	\EndIf
\EndFor
\State \Return $\keyroots$.
\EndFunction
\end{algorithmic}
\end{algorithm}

This yields the TED Algorithm~\ref{alg:ted} of \textcite{Zhang1989}.

\begin{thm}
Algorithm~\ref{alg:ted} computes the TED. Further, Algorithm~\ref{alg:ted} runs
in $\effic(\lnodelim^2 \cdot \rnodelim^2)$ and has $\effic(\lnodelim \cdot \rnodelim)$ space
complexity.

\begin{proof}
The runtime proof is simple: Because we use a subset of outer loop iterations compared to Algorithm~\ref{alg:ted_proto},
we are at most as slow. Still, the runtime bound is tight, because the set of keyroots is per definition
as large as the set of leaves of a tree, and the number of leaves of a tree can grow linearly with
the size of a tree, as is the case in Figure~\ref{fig:worst_case}. The space requirements are
the same as for Algorithm~\ref{alg:ted_proto}.

Further, Algorithm~\ref{alg:ted} still computes the same result as Algorithm~\ref{alg:ted_proto},
because according to Theorem~\ref{thm:keyroots} the same subforest edit distances are computed as
before.
\end{proof}
\end{thm}

\begin{algorithm}
\caption{The $\effic(\lnodelim^2 \cdot \rnodelim^2)$ TED algorithm of \textcite{Zhang1989}.
The algorithm iterates over all subtrees of $\ltree$ and $\rtree$ rooted at key roots and computes
the TED for them based on the forest edit distances between all subforests of the
respective subtrees.
Refer to our \href{https://gitlab.ub.uni-bielefeld.de/bpaassen/python-edit-distances/-/blob/master/edist/ted.pyx}{project web site}
for a reference implementation.}
\label{alg:ted}
\begin{algorithmic}
\Function{tree-edit-distance}{Two input trees $\ltree$ and $\rtree$, a cost function $\cost$.}
\State $\lnodelim \gets |\ltree|$, $\rnodelim \gets |\rtree|$.
\State $\keyroots(\ltree) \gets $ \Call{keyroots}{$\ltree$}.
\State $\keyroots(\rtree) \gets $ \Call{keyroots}{$\rtree$}.
\State $\mat \dist \gets \lnodelim \times \rnodelim$ matrix of zeros. \Comment $\mat \dist_{\lnodeidx, \rnodeidx} = \dist_\cost(\ltree_\lnodeidx, \rtree_\rnodeidx)$.
\State $\mat \fdist \gets (\lnodelim + 1) \times (\rnodelim + 1)$ matrix of zeros.
\State \Comment $\mat \fdist_{\lnodeidx, \rnodeidx} = \fdist_\cost(\lforest[\lnodeidx, \rleaf_{\ltree}(\lkeyrootidx)], \rforest[\rnodeidx, \rleaf_{\rtree}(\rkeyrootidx)])$.
\For{$\lkeyrootidx \in \keyroots(\ltree)$ in descending order}
\For{$\rkeyrootidx \in \keyroots(\rtree)$ in descending order}
\State $\mat \fdist_{\rleaf_\lforest(\lkeyrootidx) + 1, \rleaf_\rforest(\rkeyrootidx) + 1} \gets 0$. \Comment Equation~\ref{eq:empty}
\For{$\lnodeidx \gets \rleaf_\lforest(\lkeyrootidx), \ldots, \lkeyrootidx$}
	\State $\mat \fdist_{\lnodeidx, \rleaf_\rforest(\rkeyrootidx) + 1} \gets \mat \fdist_{\lnodeidx + 1, \rleaf_\rforest(\rkeyrootidx) + 1} + \cost(\lnode_{\lnodeidx}, \gap)$.
	\Comment Equation~\ref{eq:del}
\EndFor
\For{$\rnodeidx \gets \rleaf_\rforest(\rkeyrootidx), \ldots, \rkeyrootidx$}
	\State $\mat \fdist_{\rleaf_\lforest(\lkeyrootidx) + 1, \rnodeidx} \gets \mat \fdist_{\rleaf_\lforest(\lkeyrootidx) + 1, \rnodeidx + 1} + \cost(\gap, \rnode_{\rnodeidx})$.
	\Comment Equation~\ref{eq:ins}
\EndFor
\For{$\lnodeidx \gets \rleaf_\lforest(\lkeyrootidx), \ldots, \lkeyrootidx$}
	\For{$\rnodeidx \gets \rleaf_\rforest(\rkeyrootidx), \ldots, \rkeyrootidx$}
		\If{$\rleaf_{\ltree}(\lnodeidx) = \rleaf_{\ltree}(\lkeyrootidx) \wedge \rleaf_{\rtree}(\rnodeidx) = \rleaf_{\rtree}(\rkeyrootidx)$}
			\State $\mat \fdist_{\lnodeidx, \rnodeidx} \gets \min \{
			              \mat \fdist_{\lnodeidx + 1, \rnodeidx    } + \cost(\lnode_{\lnodeidx}, \gap),$
			\State $\quad \mat \fdist_{\lnodeidx    , \rnodeidx + 1} + \cost(\gap, \rnode_{\rnodeidx}),$
			\State $\quad \mat \fdist_{\lnodeidx + 1, \rnodeidx + 1} + \cost(\lnode_{\lnodeidx}, \rnode_{\rnodeidx}) \}$. \Comment Equation~\ref{eq:dist_decompose}
			\State $\mat \dist_{\lnodeidx, \rnodeidx} \gets \mat \fdist_{\lnodeidx, \rnodeidx}$.
		\Else
			\State $\mat \fdist_{\lnodeidx, \rnodeidx} \gets \min \{
			\mat \fdist_{\lnodeidx + 1, \rnodeidx    } + \cost(\lnode_{\lnodeidx}, \gap),$
			\State $\quad \mat \fdist_{\lnodeidx    , \rnodeidx + 1} + \cost(\gap, \rnode_{\rnodeidx}),$
			\State $\quad \mat \fdist_{\rleaf_{\ltree}(\lnodeidx) + 1, \rleaf_{\rtree}(\rnodeidx) + 1} + \mat \dist_{\lnodeidx, \rnodeidx} \}$. \Comment Equation~\ref{eq:fdist_decompose}
		\EndIf
	\EndFor
\EndFor
\EndFor
\EndFor
\State \Return $\mat \dist_{1, 1}$.
\EndFunction
\end{algorithmic}
\end{algorithm}

\newcommand{\dpc}[3]{\node[#2] (#1) {#3}; }

\begin{figure}
\begin{center}
\begin{tikzpicture}

\begin{scope}
\Tree%
[. \node[class0] {$\sym{a}$};
	[. \node[class1] {$\sym{b}$};
		\node[class2] {$\sym{c}$};
		\node[class1] {$\sym{d}$};
	]
	\node[class0] {$\sym{e}$};
]
\end{scope}

\begin{scope}[shift={(3, 0)}]
\Tree%
[. \node[class0] {$\sym{f}$};
	\node[class0] {$\sym{g}$};
]
\end{scope}


\matrix(d)[below,matrix of nodes,nodes={align=center},row 2/.style={anchor=south}] at (7, 0.5) {
%
%
$\dist_\cost(\ltree_\lnodeidx, \rtree_\rnodeidx)$ & $\rnodeidx$    & $1$                                     & $2$ \\
$\lnodeidx$ & $\ltree_\lnodeidx$ \textbackslash $\rtree_\rnodeidx$ & {\color{aluminium6} $\sym{f}(\sym{g})$} & {\color{aluminium6} $\sym{g}$} \\
$1$         & {\color{aluminium6} $\sym{a}(${\color{skyblue3}$\sym{b}(${\color{orange3}$\sym{c}$}$, \sym{d})$}$, \sym{e})$} & \dpc{d11}{class0color}{$5$} & \dpc{d12}{class0color}{$5$} \\
$2$         & {\color{skyblue3}$\sym{b}(${\color{orange3}$\sym{c}$}$, \sym{d})$} & \dpc{d21}{class1color}{$3$} & \dpc{d22}{class1color}{$3$} \\
$3$         & {\color{orange3}$\sym{c}$}                           & \dpc{d31}{class2color}{$2$}             & \dpc{d32}{class2color}{$1$} \\
$4$         & {\color{skyblue3}$\sym{d}$}                          & \dpc{d41}{class1color}{$2$}             & \dpc{d42}{class1color}{$1$} \\
$5$         & {\color{aluminium6}$\sym{e}$}                        & \dpc{d51}{class0color}{$2$}             & \dpc{d52}{class0color}{$1$} \\
};

\draw(d-2-1.south west)--(d-2-1.south east);
\draw(d-2-2.south west)--(d-2-2.south east);
\draw(d-2-3.south west)--(d-2-4.south east);



\matrix(D31)[below,matrix of nodes,nodes={align=center},row 2/.style={anchor=south}] at (4.5, -3.5) {
$\fdist_\cost(\ltree[\lnodeidx, 3], \rtree[\rnodeidx, 2])$ & $\rnodeidx$   & $1$ & $2$ & $3$ \\
$\lnodeidx$ & $\ltree[\lnodeidx, 3]$ \textbackslash $\rtree[\rnodeidx, 2]$ & {\color{aluminium6} $\sym{f}(\sym{g})$} & {\color{aluminium6} $\sym{g}$} & $\epsilon$ \\
$3$           & {\color{orange3}$\sym{c}$} & \dpc{D3312}{class2color}{$2$} & \dpc{D3322}{class2color}{$1$} & $1$ \\
$4$           & $\epsilon$                 & $2$                           & \dpc{D4322}{}{$1$}            & \dpc{D4332}{}{$0$} \\
};

\draw(D31-2-1.south west)--(D31-2-1.south east);
\draw(D31-2-2.south west)--(D31-2-2.south east);
\draw(D31-2-3.south west)--(D31-2-5.south east);


\path%
(D3312) edge (D4322) edge (D3322)
(D3322) edge (D4332)
(D4322) edge (D4332);


\matrix(D21)[below,matrix of nodes,nodes={align=center},row 2/.style={anchor=south}] at (4.5, -6.5) {
$\fdist_\cost(\ltree[\lnodeidx, 4], \rtree[\rnodeidx, 2])$ & $\rnodeidx$   & $1$ & $2$ & $3$ \\
$\lnodeidx$ & $\ltree[\lnodeidx, 4]$ \textbackslash $\rtree[\rnodeidx, 2]$ & {\color{aluminium6} $\sym{f}(\sym{g})$} & {\color{aluminium6} $\sym{g}$} & $\epsilon$ \\
$2$           & {\color{skyblue3}$\sym{b}(${\color{orange3}$\sym{c}$}$, \sym{d})$}   & \dpc{D2412}{class1color}{$3$}           & \dpc{D2422}{class1color}{$3$}  & $3$ \\
$3$           & {\color{orange3}$\sym{c}$}, {\color{skyblue3}$\sym{d}$}              & $3$                                     & \dpc{D3422}{}{$2$}             & \dpc{D3432}{}{$2$}  \\
$4$           & {\color{skyblue3}$\sym{d}$}                                          & \dpc{D4412}{class1color}{$2$}           & \dpc{D4422}{class1color}{$1$}  & \dpc{D4432}{}{$1$} \\
$5$           & $\epsilon$ & $2$ & \dpc{D5422}{}{$1$} & \dpc{D5432}{}{$0$} \\
};

\draw(D21-2-1.south west)--(D21-2-1.south east);
\draw(D21-2-2.south west)--(D21-2-2.south east);
\draw(D21-2-3.south west)--(D21-2-5.south east);


\path%
(D2412) edge (D3422)
(D2422) edge[shorten <=-2pt, shorten >=-2pt] (D3422) edge (D3432)
(D3422) edge[shorten <=-2pt, shorten >=-2pt] (D4422) edge[class2, semithick] (D4432)
(D3432) edge[shorten <=-2pt, shorten >=-2pt] (D4432)
(D4412) edge (D4422) edge (D5422)
(D4422) edge (D5432)
(D4432) edge[shorten <=-2pt, shorten >=-2pt] (D5432)
(D5422) edge (D5432);


\matrix(D11)[below,matrix of nodes,nodes={align=center},row 2/.style={anchor=south}] at (4.5, -10.5) {
$\fdist_\cost(\ltree[\lnodeidx, 5], \rtree[\rnodeidx, 2])$ & $\rnodeidx$   & $1$ & $2$ & $3$ \\
$\lnodeidx$ & $\ltree[\lnodeidx, 5]$ \textbackslash $\rtree[\rnodeidx, 2]$ & {\color{aluminium6} $\sym{f}(\sym{g})$} & {\color{aluminium6} $\sym{g}$} & $\epsilon$ \\
$1$           & {\color{aluminium6} $\sym{a}(${\color{skyblue3}$\sym{b}(${\color{orange3}$\sym{c}$}$, \sym{d})$}$, \sym{e}$}
                                                                                     & \dpc{D1512}{class0color}{$5$}           & \dpc{D1522}{class0color}{$5$} & $5$ \\
$2$           & {\color{skyblue3}$\sym{b}(${\color{orange3}$\sym{c}$}$, \sym{d})$}, {\color{aluminium6} $\sym{e}$}
                                                                                     & \dpc{D2512}{}{$4$}                      & \dpc{D2522}{}{$4$}            & \dpc{D2532}{}{$4$} \\
$3$           & {\color{orange3}$\sym{c}$}, {\color{skyblue3} $\sym{d}$}, {\color{aluminium6} $\sym{e}$}
                                                                                     & $4$                                     & \dpc{D3522}{}{$3$}            & \dpc{D3532}{}{$3$} \\
$4$           & {\color{skyblue3}$\sym{d}$}, {\color{aluminium6} $\sym{e}$}          & $3$                                     & \dpc{D4522}{}{$2$}            & \dpc{D4532}{}{$2$} \\
$5$           & {\color{aluminium6}$\sym{e}$}                                        & \dpc{D5512}{class0color}{$2$}           & \dpc{D5522}{class0color}{$1$} & \dpc{D5532}{}{$1$} \\
$6$           & $\epsilon$                                                           & $2$                                     & \dpc{D6522}{}{$1$}            & \dpc{D6532}{}{$0$} \\
};

\draw(D11-2-1.south west)--(D11-2-1.south east);
\draw(D11-2-2.south west)--(D11-2-2.south east);
\draw(D11-2-3.south west)--(D11-2-5.south east);


\path%
(D1512) edge[shorten <=-2pt, shorten >=-2pt] (D2512) edge (D2522)
(D1522) edge[shorten <=-2pt, shorten >=-2pt] (D2522) edge (D2532)
(D2512) edge[class1, semithick] (D5532)
(D2522) edge[shorten <=-2pt, shorten >=-2pt] (D3522) edge[class1, semithick] (D5532)
(D2532) edge[shorten <=-2pt, shorten >=-2pt] (D3532)
(D3522) edge[shorten <=-2pt, shorten >=-2pt] (D4522) edge[class2, semithick] (D4532)
(D3532) edge[shorten <=-2pt, shorten >=-2pt] (D4532)
(D4522) edge[shorten <=-2pt, shorten >=-2pt] (D5522) edge[class1, semithick] (D5532)
(D4532) edge[shorten <=-2pt, shorten >=-2pt] (D5532)
(D5512) edge (D5522) edge (D6522)
(D5522) edge (D6532)
(D5532) edge[shorten <=-2pt, shorten >=-2pt] (D6532)
(D6522) edge (D6532);

\begin{pgfonlayer}{bg}
\draw[class0, rounded corners] (d11.north west) rectangle (d12.south east);
\draw[class1, rounded corners] (d21.north west) rectangle (d22.south east);
\draw[class2, rounded corners] (d31.north west) rectangle (d32.south east);
\draw[class1, rounded corners] (d41.north west) rectangle (d42.south east);
\draw[class0, rounded corners] (d51.north west) rectangle (d52.south east);

\draw[class2, rounded corners] (D3312.north west) rectangle (D3322.south east);
\draw[edge, class2, fill=none, semithick, densely dashed, out=0, in=0] (D3322.east) to (d32.east);

\draw[class1, rounded corners] (D2412.north west) rectangle (D2422.south east);
\draw[edge, class1, fill=none, semithick, densely dashed, out=0, in=0] (D2422.east) to (d22.east);

\draw[class1, rounded corners] (D4412.north west) rectangle (D4422.south east);
\draw[edge, class1, fill=none, semithick, densely dashed, out=0, in=0] (D4422.east) to (d42.east);

\draw[class0, rounded corners] (D1512.north west) rectangle (D1522.south east);
\draw[edge, class0, fill=none, semithick, densely dashed, out=0, in=0] (D1522.east) to (d12.east);

\draw[class0, rounded corners] (D5512.north west) rectangle (D5522.south east);
\draw[edge, class0, fill=none, semithick, densely dashed, out=0, in=0] (D5522.east) to (d52.east);

\end{pgfonlayer}

\end{tikzpicture}
\end{center}
\caption{An illustration of the TED Algorithm~\ref{alg:ted} of \textcite{Zhang1989}
for the two input trees from Figure~\ref{fig:edits}. Nodes with the same right outermost leaf
are shown in the same color. For these nodes, the subforest edit distance is re-used.
Top right: The TED between all subtrees of the input trees.
Bottom: The subforest edit distances for all key root pairs. All entries which correspond
to a subtree edit distance are highlighted in color and linked with dashed arrows to the
corresponding entries in the subtree edit distance table at the top right.
Co-optimal mappings are indicated by solid lines linking the entries of the dynamic
programming table to the distances they are decomposed into.}
\label{fig:ted}
\end{figure}

As an example, consider the trees $\ltree = \sym{a}(\sym{b}(\sym{c}, \sym{d}), \sym{e})$ and
$\rtree = \sym{f}(\sym{g})$ from Figure~\ref{fig:edits} and the trivial cost function
$\cost(\lnode, \rnode) = 0$ if $\lnode = \rnode$ and $1$ otherwise. The TED algorithm first
considers the edit distance between the subtrees $\ltree_3 = \sym{c}$ and $\rtree_2 = \sym{f}(\sym{g})$,
which can be computed based on the subforest edit distances
$\fdist_\cost(\ltree[4, 3], \rtree[3, 2]) = \fdist_\cost(\epsilon, \epsilon) = 0$,
$\fdist_\cost(\ltree[4, 3], \rtree[2, 2]) = \fdist_\cost(\epsilon, \sym{g}) = 1$,
$\fdist_\cost(\ltree[4, 3], \rtree[1, 2]) = \fdist_\cost(\epsilon, \sym{f}(\sym{g})) = 2$,
$\fdist_\cost(\ltree[3, 3], \rtree[3, 2]) = \fdist_\cost(\sym{c}, \epsilon) = 1$, and
$\fdist_\cost(\ltree[3, 3], \rtree[2, 2]) = \fdist_\cost(\sym{c}, \sym{g}) = 1$. Based on these
intermediate results, we can infer that $\dist_\cost(\ltree_3, \rtree_1) =
\fdist_\cost(\ltree[3, 3], \rtree[1, 2]) = 2$. Note that this calculation also yields
the edit distance between the subtrees $\ltree_3 = \sym{c}$ and $\rtree_2 = \sym{g}$ as an
intermediate result (also refer to Figure~\ref{fig:ted}, middle).
Next, we compute the subtree edit distance between $\ltree_2 = \sym{b}(\sym{c}, \sym{d})$ and
$\rtree_1 = \sym{f}(\sym{g})$, which also yields the subtree edit distances
$\dist_\cost(\ltree_4, \rtree_2)$, $\dist_\cost(\ltree_4, \rtree_1)$, and
$\dist_\cost(\ltree_2, \rtree_2)$ as intermediate results (see Figure~\ref{fig:ted}, middle).
Finally, we can compute the subtree edit distance between $\ltree_1 = \ltree$ and $\rtree_1 = \rtree$
(see Figure~\ref{fig:ted}, bottom), which turns out to be $5$.

This concludes our description of the edit distance itself. However, in many situations it is not
only interesting to know the tree edit distance, but also which mapping (and which edit
script) corresponds to the edit distance. This is the topic of \emph{backtracing}.

\section{Backtracing and Co-Optimal Mappings}
\label{sec:coopts}

\begin{figure}
\begin{center}
\begin{tikzpicture}

\begin{scope}

\node[above] at (0.5,0.5) {$\map = \{(1, 1), (2, 2) \}$};

\Tree [.\node (a) {$\sym{a}$};
	[.\node (b) {$\sym{b}$};
		\node (c) {$\sym{c}$};
		\node (d) {$\sym{d}$};
	]
	\node (e) {$\sym{e}$};
]

\begin{scope}[shift={(1.5, 0)}]

\Tree [.\node (f) {$\sym{f}$};
	\node (g) {$\sym{g}$};
]

\end{scope}

\path[edge, densely dashed, semithick]%
(a) edge (f)
(b) edge[bend left] (g);

\end{scope}

\begin{scope}[shift={(4, 0)}]

\node[above] at (0.5,0.5) {$\map = \{(1, 1), (3, 2) \}$};

\Tree [.\node (a) {$\sym{a}$};
	[.\node (b) {$\sym{b}$};
		\node (c) {$\sym{c}$};
		\node (d) {$\sym{d}$};
	]
	\node (e) {$\sym{e}$};
]

\begin{scope}[shift={(1.5, 0)}]

\Tree [.\node (f) {$\sym{f}$};
	\node (g) {$\sym{g}$};
]

\end{scope}

\path[edge, densely dashed, semithick]%
(a) edge (f)
(c) edge (g);

\end{scope}

\begin{scope}[shift={(8, 0)}]

\node[above] at (0.5,0.5) {$\map = \{(1, 1), (4, 2) \}$};

\Tree [.\node (a) {$\sym{a}$};
	[.\node (b) {$\sym{b}$};
		\node (c) {$\sym{c}$};
		\node (d) {$\sym{d}$};
	]
	\node (e) {$\sym{e}$};
]

\begin{scope}[shift={(1.5, 0)}]

\Tree [.\node (f) {$\sym{f}$};
	\node (g) {$\sym{g}$};
]

\end{scope}

\path[edge, densely dashed, semithick]%
(a) edge (f)
(d) edge (g);

\end{scope}

\begin{scope}[shift={(0,-4)}]

\node[above] at (0.5,0.5) {$\map = \{(1, 1), (5, 2) \}$};

\Tree [.\node (a) {$\sym{a}$};
	[.\node (b) {$\sym{b}$};
		\node (c) {$\sym{c}$};
		\node (d) {$\sym{d}$};
	]
	\node (e) {$\sym{e}$};
]

\begin{scope}[shift={(1.5, 0)}]

\Tree [.\node (f) {$\sym{f}$};
	\node (g) {$\sym{g}$};
]

\end{scope}

\path[edge, densely dashed, semithick]%
(a) edge (f)
(e) edge (g);

\end{scope}

\begin{scope}[shift={(4,-4)}]

\node[above] at (0.5,0.5) {$\map = \{(2, 1), (3, 2) \}$};

\Tree [.\node (a) {$\sym{a}$};
	[.\node (b) {$\sym{b}$};
		\node (c) {$\sym{c}$};
		\node (d) {$\sym{d}$};
	]
	\node (e) {$\sym{e}$};
]

\begin{scope}[shift={(1.5, 0)}]

\Tree [.\node (f) {$\sym{f}$};
	\node (g) {$\sym{g}$};
]

\end{scope}

\path[edge, densely dashed, semithick]%
(b) edge (f)
(c) edge (g);

\end{scope}

\begin{scope}[shift={(8,-4)}]

\node[above] at (0.5,0.5) {$\map = \{(2, 2), (4, 2) \}$};

\Tree [.\node (a) {$\sym{a}$};
	[.\node (b) {$\sym{b}$};
		\node (c) {$\sym{c}$};
		\node (d) {$\sym{d}$};
	]
	\node (e) {$\sym{e}$};
]

\begin{scope}[shift={(1.5, 0)}]

\Tree [.\node (f) {$\sym{f}$};
	\node (g) {$\sym{g}$};
]

\end{scope}

\path[edge, densely dashed, semithick]%
(b) edge (f)
(d) edge (g);

\end{scope}

\end{tikzpicture}
\end{center}
\caption{All six co-optimal mappings between the trees $\ltree = \sym{a}(\sym{b}(\sym{c}, \sym{d}), \sym{e})$
and $\rtree = \sym{f}(\sym{g})$ from Figure~\ref{fig:ted}.}
\label{fig:coopt_mappings}
\end{figure}

Now that we have computed the TED, the next question is: Which edit script
corresponds to the TED? We have answered this question in part in Algorithm~\ref{alg:map_to_script},
which transforms a mapping into an edit script with the same cost. Therefore, we can
re-phrase the question: Which \emph{mapping} corresponds to the TED?
\textcite{Zhang1989} only hint at an answer in their own paper. Here, we shall analyze
this question in detail. We start by phrasing more precisely what we are looking for:

\begin{dfn}[Co-Optimal Mappings]
Let $\lforest$ and $\rforest$ be forests over some alphabet $\alphabet$ and let $\cost$ be a cost
function over $\alphabet$. We define a \emph{co-optimal mapping} $\map$ as a tree mapping
between $\lforest$ and $\rforest$ such that $\Cost(\map, \lforest, \rforest) = \min_{\map'} \Cost(\map', \lforest, \rforest)$.
\end{dfn}

For example, all co-optimal mappings for the trees in Figure~\ref{fig:ted} are shown in Figure~\ref{fig:coopt_mappings}.
Unfortunately, listing all co-optimal mappings is infeasible in general, as the following theorem
demonstrates:

\begin{thm} \label{thm:num_coopts}
Let $\bar{\sym{a}}(1)$ be the tree $\sym{a}$ over the alphabet $\alphabet = \{ \sym{a} \}$ and let
$\bar{\sym{a}}(\lnodelim) = \sym{a}\big(\bar{\sym{a}}(\lnodelim - 1) \big)$. Then, for any
metric cost function $\cost$ over $\alphabet$ and any $\lnodelim \in \N$ which is
divisible by $2$ it holds: There are $\lnodelim! / [(\lnodelim / 2)!]^2$ co-optimal mappings between
the trees $\ltree = \bar{\sym{a}}(\lnodelim)$ and $\rtree = \bar{\sym{a}}(\lnodelim / 2)$.
Further, this number is larger than $\frac{\sqrt{2\pi}}{e^2} \cdot \frac{2^{\lnodelim+1}}{\sqrt{\lnodelim}}$.

\begin{proof}
Because $\cost$ is metric, we have $\cost(\sym{a}, \sym{a}) = 0$ and $\cost(\sym{a}, \gap) > 0$.
Therefore, we want to replace as many $\sym{a}$ with $\sym{a}$ as possible to reduce the cost.
At most, we can replace $\lnodelim / 2$ $\sym{a}$ with $\sym{a}$, because $|\bar{\sym{a}}(\lnodelim / 2)| =
\lnodelim / 2$. Therefore, this corresponds to choosing $\lnodelim / 2$ nodes from $\ltree$
which are mapped to the $\lnodelim / 2$ nodes from $\rtree$. As we know from combinatorics, there are
\begin{equation*}
\begin{pmatrix} \lnodelim \\ \frac{\lnodelim}{2} \end{pmatrix} = \frac{\lnodelim!}{(\frac{\lnodelim}{2}!)^2}
\end{equation*}
ways to choose $\lnodelim / 2$ from $\lnodelim$ elements. Using \href{https://en.wikipedia.org/wiki/Stirling\%27s_approximation}{Stirling's approximation},
we then obtain the following lower bound:
\begin{equation*}
\frac{\lnodelim!}{(\frac{\lnodelim}{2}!)^2}
\geq \frac{\sqrt{2\pi} \cdot \lnodelim^{\lnodelim + \frac{1}{2}} \cdot e^{-\lnodelim}}{(e \cdot (\frac{\lnodelim}{2})^{\frac{\lnodelim}{2}+\frac{1}{2}} \cdot e^{-\frac{\lnodelim}{2}})^2} 
= \frac{\sqrt{2\pi}}{e^2} \cdot \frac{\lnodelim^{\lnodelim + \frac{1}{2}}}{(\frac{\lnodelim}{2})^{\lnodelim+\frac{1}{2}} \cdot \sqrt{\frac{\lnodelim}{2}}}
\cdot \frac{e^{-\lnodelim}}{e^{-\lnodelim}} 
= \frac{\sqrt{2\pi}}{e^2} \cdot \frac{2^{\lnodelim + \frac{1}{2}}}{\sqrt{\frac{\lnodelim}{2}}} = \frac{\sqrt{2\pi}}{e^2} \cdot \frac{2^{\lnodelim + 1}}{\sqrt{\lnodelim}}
\end{equation*}
\end{proof}
\end{thm}

While it is therefore infeasible to list \emph{all} co-optimal mappings, it is
still possible to return \emph{some} of the co-optimal mappings. In particular, it turns out that
constructing a co-optimal mapping corresponds to finding a path in a graph which we call the
\emph{co-optimal edit graph}. First, we define a general graph as follows:

\begin{dfn}[Directed Acyclic Graph (DAG)]
Let $\nodes$ be some set, and let $\edges \subseteq \nodes \times \nodes$.
Then we call $\graph = (\nodes, \edges)$ a \emph{graph}, $\nodes$ the \emph{nodes} of $\graph$ and
$\edges$ the \emph{edges} of $\graph$. We call $\graph$ a \emph{directed acyclic graph} (DAG)
if there exists a total ordering relation $<$ on $\nodes$, such that for all edges $(\lgnode, \rgnode) \in \edges$
it holds: $\lgnode < \rgnode$, i.e.\ edges occur only from lower nodes to higher nodes in the ordering.
\end{dfn}

We then define our co-optimal edit graph as follows.

\begin{dfn}[Co-optimal Edit Graph] \label{dfn:coopt_graph}
Let $\lforest$ and $\rforest$ be forests over some alphabet $\alphabet$
and let $\cost$ be a cost function over $\alphabet$.
Then, we define the \emph{co-optimal edit graph} between $\lforest$ and $\rforest$
according to $\cost$ as the graph $\graph_{\cost, \lforest, \rforest}
= (\nodes, \edges)$ with nodes $\nodes$ and edges $\edges$ as follows.

If $\lforest = \epsilon$ and $\rforest = \epsilon$ we define $\nodes := \{ (1, 1, 1, 1) \}$ and
$\edges := \emptyset$.

If $\lforest = \epsilon$ but $\rforest \neq \epsilon$ we define $\nodes := \big\{ (1, 1, 1, \rnodeidx) \big| \rnodeidx \in \{1, \ldots, \siz{\rforest} + 1 \} \big\}$
and $\edges := \big\{ \big((1, 1, 1, \rnodeidx), (1, 1, 1, \rnodeidx+1)\big) \big| \rnodeidx \in \{1, \ldots, \siz{\rforest} \} \big\}$.

If $\lforest \neq \epsilon$ but $\rforest = \epsilon$ we define $\nodes := \big\{ (1, \lnodeidx, 1, 1) \big| \lnodeidx \in \{1, \ldots, \siz{\lforest} + 1 \} \big\}$
and $\edges := \big\{ \big((1, \lnodeidx, 1, 1), (1, \lnodeidx+1, 1, 1)\big) \big| \lnodeidx \in \{1, \ldots, \siz{\lforest} \} \big\}$.

If neither forest is empty, we define:
\begin{align}
\nodes :=  &\Big\{(\lkeyrootidx, \lnodeidx, \rkeyrootidx, \rnodeidx) \Big|
\lkeyrootidx \in \keyroots(\lforest), \lnodeidx \in \{\lkeyrootidx, \ldots, \rleaf_\lforest(\lkeyrootidx)+1\},
\rkeyrootidx \in \keyroots(\rforest), \rnodeidx \in \{\rkeyrootidx, \ldots, \rleaf_\rforest(\rkeyrootidx)+1\} \Big\} \\
\edges :=
&\Big\{\big((\lkeyrootidx, \lnodeidx, \rkeyrootidx, \rnodeidx), (\lkeyrootidx, \lnodeidx+1, \rkeyrootidx, \rnodeidx)\big) \Big|
	\fdist_\cost\big(\lforest[\lnodeidx, \rleaf_\lforest(\lkeyrootidx)], \rforest[\rnodeidx, \rleaf_\rforest(\rkeyrootidx)]\big) \notag \\
& \qquad \qquad = \cost(\lnode_\lnodeidx, \gap) +
	\fdist_\cost\big(\lforest[\lnodeidx+1, \rleaf_\lforest(\lkeyrootidx)], \rforest[\rnodeidx, \rleaf_\rforest(\rkeyrootidx)]\big) \Big\} \cup \label{eq:coopt_graph_del} \\
&\Big\{\big((\lkeyrootidx, \lnodeidx, \rkeyrootidx, \rnodeidx), (\lkeyrootidx, \lnodeidx, \rkeyrootidx, \rnodeidx+1)\big) \Big|
	\fdist_\cost\big(\lforest[\lnodeidx, \rleaf_\lforest(\lkeyrootidx)], \rforest[\rnodeidx, \rleaf_\rforest(\rkeyrootidx)]\big) \notag \\
& \qquad \qquad = \cost(\gap, \rnode_\rnodeidx) +
	\fdist_\cost\big(\lforest[\lnodeidx, \rleaf_\lforest(\lkeyrootidx)], \rforest[\rnodeidx+1, \rleaf_\rforest(\rkeyrootidx)]\big) \Big\} \cup \label{eq:coopt_graph_ins} \\
&\Big\{\big((\lkeyrootidx, \lnodeidx, \rkeyrootidx, \rnodeidx), (\lkeyrootidx, \lnodeidx+1, \rkeyrootidx, \rnodeidx+1)\big) \Big|
	\fdist_\cost\big(\lforest[\lnodeidx, \rleaf_\lforest(\lkeyrootidx)], \rforest[\rnodeidx, \rleaf_\rforest(\rkeyrootidx)]\big) \notag \\
& \qquad \qquad = \cost(\lnode_\lnodeidx, \rnode_\rnodeidx) +
	\fdist_\cost\big(\lforest[\lnodeidx+1, \rleaf_\lforest(\lkeyrootidx)], \rforest[\rnodeidx+1, \rleaf_\rforest(\rkeyrootidx)]\big) \notag \\
& \qquad \qquad \wedge \rleaf_{\lforest}(\lnodeidx) = \rleaf_{\lforest}(\lkeyrootidx) \wedge \rleaf_{\rforest}(\rnodeidx) = \rleaf_{\rforest}(\rkeyrootidx)
\Big\} \label{eq:coopt_graph_rep1} \cup \\
&\Big\{\big((\lkeyrootidx, \lnodeidx, \rkeyrootidx, \rnodeidx), (\lkeyrootidx, \lnodeidx+1, \rkeyrootidx, \rnodeidx+1)\big) \Big|
	\fdist_\cost\big(\lforest[\lnodeidx, \rleaf_\lforest(\lkeyrootidx)], \rforest[\rnodeidx, \rleaf_\rforest(\rkeyrootidx)]\big) \notag \\
& \qquad \qquad = \cost(\lnode_\lnodeidx, \rnode_\rnodeidx) +
	\fdist_\cost\big(\lforest[\lnodeidx+1, \rleaf_\lforest(\lkeyrootidx)], \rforest[\rnodeidx+1, \rleaf_\rforest(\rkeyrootidx)]\big) \notag \\
& \qquad \qquad \wedge \cost(\lnode_\lnodeidx, \rnode_\rnodeidx) = \cost(\lnode_\lnodeidx, \gap) + \cost(\gap, \rnode_\rnodeidx)
\Big\} \cup \label{eq:coopt_graph_rep2} \\
&\Big\{\big((\lkeyrootidx, \lnodeidx, \rkeyrootidx, \rnodeidx), (\keyroot_\lforest(\lnodeidx), \lnodeidx+1, \keyroot_\rforest(\rnodeidx), \rnodeidx+1)\big) \Big|
	\fdist_\cost\big(\lforest[\lnodeidx, \rleaf_\lforest(\lkeyrootidx)], \rforest[\rnodeidx, \rleaf_\rforest(\rkeyrootidx)]\big) \notag \\
& \qquad \qquad = \fdist_\cost(\ltree^\lnodeidx, \rtree^\rnodeidx) +
	\fdist_\cost\big(\lforest[\rleaf_\lforest(\lnodeidx)+1, \rleaf_\lforest(\lkeyrootidx)], \rforest[\rleaf_\rforest(\rnodeidx)+1, \rleaf_\rforest(\rkeyrootidx)]\big) \notag \\
& \qquad \qquad \wedge \big(\rleaf_{\lforest}(\lnodeidx) \neq \rleaf_{\lforest}(\lkeyrootidx) \vee \rleaf_{\rforest}(\rnodeidx) \neq \rleaf_{\rforest}(\rkeyrootidx)\big) 
\wedge \cost(\lnode_\lnodeidx, \rnode_\rnodeidx) < \cost(\lnode_\lnodeidx, \gap) + \cost(\gap, \rnode_\rnodeidx)
\Big\} \cup \label{eq:coopt_graph_rep3} \\
&\Big\{\big((\lkeyrootidx, \rleaf_\lforest(\lkeyrootidx)+1, \rkeyrootidx, \rleaf_\rforest(\rkeyrootidx)+1), (\keyroot_\lforest(\rleaf_\lforest(\lkeyrootidx)+1), \rleaf_\lforest(\lkeyrootidx)+1, \keyroot_\rforest(\rleaf_\rforest(\rkeyrootidx)+1), \rleaf_\rforest(\rkeyrootidx)+1)\big) \Big | \notag \\
& \qquad \qquad \rleaf_\lforest(\lkeyrootidx)+1 \leq \siz{\lforest} \wedge
\rleaf_\rforest(\rkeyrootidx)+1 \leq \siz{\rforest}
\Big\} \cup \label{eq:coopt_graph_up1} \\
&\Big\{\big((\lkeyrootidx, \rleaf_\lforest(\lkeyrootidx)+1, \rkeyrootidx, \siz{\rforest}+1),
(\keyroot_\lforest(\rleaf_\lforest(\lkeyrootidx)+1), \rleaf_\lforest(\lkeyrootidx)+1, 1, \siz{\rforest}+1)\big)
\Big | \rleaf_\lforest(\lkeyrootidx)+1 \leq \siz{\lforest} \Big\} \cup \label{eq:coopt_graph_up2} \\
&\Big\{\big((\lkeyrootidx, \siz{\lforest}+1, \rkeyrootidx, \rleaf_\rforest(\rkeyrootidx)+1),
(1, \siz{\lforest}+1, \keyroot_\rforest(\rleaf_\rforest(\rkeyrootidx)+1), \rleaf_\rforest(\rkeyrootidx)+1)\big) 
\Big | \rleaf_\rforest(\rkeyrootidx)+1 \leq \siz{\rforest} \Big\}  \label{eq:coopt_graph_up3}
\end{align}
\end{dfn}

As this definition is quite extensive, we shall explain it in a bit more detail. The nodes of the
co-optimal edit graph are, essentially, the entries of the dynamic programming matrix $\mat \fdist$
of Algorithm~\ref{alg:ted}. Given that this matrix needs to be computed for every combination
of keyroots $(\lkeyrootidx, \rkeyrootidx) \in \keyroots(\lforest) \times \keyroots(\rforest)$,
we need four indices to identify a position in the dynamic programming matrix $\mat \fdist$ uniquely,
namely the keyroot indices and the matrix indices, leading to a quartuple $(\lkeyrootidx, \lnodeidx, \rkeyrootidx, \rnodeidx)$.
Now, with respect to the edges, Equation~\ref{eq:coopt_graph_del} defines the edges corresponding
to deletions (the first case in Equations~\ref{eq:fdist_decompose} and~\ref{eq:dist_decompose}),
Equation~\ref{eq:coopt_graph_ins} defines the edges corresponding to insertions (the second case in Equations~\ref{eq:fdist_decompose} and~\ref{eq:dist_decompose}),
Equation~\ref{eq:coopt_graph_rep1} defines the edges corresponding to replacements within a
subtree (the third case in Equation~\ref{eq:dist_decompose}), and
Equation~\ref{eq:coopt_graph_rep3} defines the edges corresponding to replacements of entire subtrees
(the third case in Equation~\ref{eq:fdist_decompose}).

The remaining edges cover special cases. In particular, Equation~\ref{eq:coopt_graph_rep2} covers
the case where all options, deletion, insertion, and replacement are co-optimal, and
Equations~\ref{eq:coopt_graph_up1}, \ref{eq:coopt_graph_up2}, and~\ref{eq:coopt_graph_up3} cover
cases where we are at the end of the dynamic programming matrix for a subtree and need to continue
the computation in the dynamic programming matrix for a larger subtree which includes the current
subtree.

As an example, consider the co-optimal edit graph between the trees $\ltree = \sym{a}(\sym{b}(\sym{c}, \sym{d}), \sym{e})$
and $\rtree = \sym{f}(\sym{g})$ from Figure~\ref{fig:edits}. An excerpt of this graph is shown in Figure~\ref{fig:coopt_graph}.

\begin{figure}
\begin{center}
\begin{tikzpicture}[xscale=2.8,yscale=1.8]

\node[align=center] (abcde_vs_fg) at (0, 0) {$(1, 1, 1, 1)$\\$\fdist_\cost\big( \sym{a}(\sym{b}(\sym{c}, \sym{d}), \sym{e}) , \sym{f}(\sym{g}) \big)$};
\node[align=center] (bcde_vs_fg)  at (0,-1) {$(1, 2, 1, 1)$\\$\fdist_\cost\big( (\sym{b}(\sym{c}, \sym{d}), \sym{e}) , \sym{f}(\sym{g}) \big)$};
\node[align=center] (bcde_vs_g)   at (2,-1) {$(1, 2, 1, 2)$\\$\fdist_\cost\big( (\sym{b}(\sym{c}, \sym{d}), \sym{e}) , \sym{g} \big)$};
\node[align=center] (cde_vs_g)    at (2,-2) {$(1, 3, 1, 2)$\\$\fdist_\cost\big( (\sym{c}, \sym{d}, \sym{e}) , \sym{g} \big)$};
\node[align=center] (de_vs_g)     at (2,-3) {$(1, 4, 1, 2)$\\$\fdist_\cost\big( (\sym{d}, \sym{e}) , \sym{g} \big)$};
\node[align=center] (e_vs_g)      at (2,-4) {$(1, 5, 1, 2)$\\$\fdist_\cost\big( \sym{e} , \sym{g} \big)$};
\node[align=center] (e_vs_eps)    at (3,-4) {$(1, 5, 1, 3)$\\$\fdist_\cost\big( \sym{e} , \epsilon \big)$};
\node[align=center] (eps_vs_eps)  at (3,-5) {$(1, 6, 1, 3)$\\$\fdist_\cost\big( \epsilon , \epsilon \big)$};

\begin{scope}[shift={(-1,-1)}]

\node[align=center] (cd_vs_g)     at (1,-1) {$(2, 3, 1, 2)$\\$\fdist_\cost\big( (\sym{c}, \sym{d}) , \sym{g} \big)$};
\node[align=center] (cd_vs_eps)   at (2,-1) {$(2, 3, 1, 3)$\\$\fdist_\cost\big( (\sym{c}, \sym{d}) , \epsilon \big)$};
\node[align=center] (d_vs_g)      at (1,-2) {$(2, 4, 1, 2)$\\$\fdist_\cost\big( \sym{d} , \sym{g} \big)$};
\node[align=center] (d_vs_eps)    at (2,-2) {$(2, 4, 1, 3)$\\$\fdist_\cost\big( \sym{d} , \epsilon \big)$};
\node[align=center] (eps_vs_eps2) at (2,-3) {$(2, 5, 1, 3)$\\$\fdist_\cost\big( \epsilon , \epsilon \big)$};

\end{scope}

\node[align=center] (eps_vs_eps3) at (3,-3) {$(3, 4, 1, 3)$\\$\fdist_\cost\big( \epsilon , \epsilon \big)$};

\node[skyblue3] [above left of=abcde_vs_fg] {1};
\node[skyblue3] [above left of=bcde_vs_fg]  {2};
\node[skyblue3] [above right of=bcde_vs_g]  {3};
\node[skyblue3] [above left of=cd_vs_g]     {4};
\node[skyblue3] [above right of=cde_vs_g]   {5};
\node[skyblue3] [left of=cd_vs_eps]         {6};
\node[skyblue3] [above left of=d_vs_g]      {7};
\node[skyblue3] [right of=de_vs_g]          {8};
\node[skyblue3] [right of=eps_vs_eps3]      {9};
\node[skyblue3] [left of=d_vs_eps]          {10};
\node[skyblue3] [left of=e_vs_g]            {11};
\node[skyblue3] [left of=eps_vs_eps2]       {12};
\node[skyblue3] [right of=e_vs_eps]         {13};
\node[skyblue3] [right of=eps_vs_eps]       {14};

\path[edge, semithick]%
(abcde_vs_fg) edge (bcde_vs_fg) edge (bcde_vs_g)
(bcde_vs_fg)  edge (cd_vs_g)
(bcde_vs_g)   edge (cde_vs_g) edge (cd_vs_eps)
(cde_vs_g)    edge (de_vs_g) edge (eps_vs_eps3)
(de_vs_g)     edge (e_vs_g) edge (e_vs_eps)
(e_vs_g)      edge (eps_vs_eps)
(e_vs_eps)    edge (eps_vs_eps)
(cd_vs_g)     edge (d_vs_g) edge[out=315,in=135] (eps_vs_eps3)
(cd_vs_eps)   edge (d_vs_eps)
(d_vs_g)      edge (eps_vs_eps2)
(d_vs_eps)    edge (eps_vs_eps2)
(eps_vs_eps2) edge[bend right=35] (e_vs_eps)
(eps_vs_eps3) edge[bend right] (d_vs_eps);

\end{tikzpicture}
\end{center}
\caption{An excerpt of the co-optimal edit graph between the trees $\ltree$ and $\rtree$ from
Figure~\ref{fig:edits}. The figure only shows nodes which are
reachable from $(1, 1, 1, 1)$. Further, to support clarity, the nodes are labelled
with indices \emph{and} with the corresponding subforest edit distance. The indices in blue
mark the order according to the ordering relationship $<$ as defined in Theorem~\ref{thm:coopt_graph_order}}
\label{fig:coopt_graph}
\end{figure}

An important insight regarding the co-optimal edit graph is that it is acyclic.

\begin{thm}\label{thm:coopt_graph_order}
Let $\lforest$ and $\rforest$ be forests over some alphabet $\alphabet$, let $\cost$ be a cost
function over $\alphabet$, and let $\graph_{\cost, \lforest, \rforest}$ be the co-optimal edit
graph with respect to $\lforest$, $\rforest$, and $\cost$. Then, $\graph_{\cost, \lforest, \rforest}$
is a directed acyclic graph with the ordering relation $(\lkeyrootidx, \lnodeidx, \rkeyrootidx, \rnodeidx) <
(\lkeyrootidx', \lnodeidx', \rkeyrootidx', \rnodeidx')$ if and only if $\lnodeidx < \lnodeidx'$,
or $\lnodeidx = \lnodeidx'$ and $\rnodeidx < \rnodeidx'$,
or $\lnodeidx = \lnodeidx'$ and $\rnodeidx = \rnodeidx'$ and $\lkeyrootidx > \lkeyrootidx'$,
or $\lnodeidx = \lnodeidx'$ and $\rnodeidx = \rnodeidx'$ and $\lkeyrootidx = \lkeyrootidx'$ and
$\rkeyrootidx < \rkeyrootidx'$.

\begin{proof}
Follows directly from the definition of the edges.
\end{proof}
\end{thm}

The ordering for the example graph in Figure~\ref{fig:coopt_graph} is displayed as blue indices.

Each edge in the co-optimal edit graph corresponds to an edit which could be used in a co-optimal
edit script. Accordingly, we should be able to join edges in the co-optimal edit graph together,
such that we obtain a complete, co-optimal edit script. This notion of joining edges is captured
by the notion of a \emph{path}.

\begin{dfn}[path]
Let $\graph = (\nodes, \edges)$ be a graph. A \emph{path} $\pth$ from $\lgnode \in \nodes$ to
$\rgnode \in \nodes$ is defined as a sequence of nodes $\pth = \rgnode_0, \ldots, \rgnode_\pathlim$,
such that $\rgnode_0 = \lgnode$, $\rgnode_\pathlim = \rgnode$, and for all
$\pathidx \in \{1, \ldots, \pathlim \} : (\rgnode_{\pathidx - 1}, \rgnode_\pathidx) \in \edges$.

If a path from $\lgnode$ to $\rgnode$ exists, we call $\rgnode$ \emph{reachable} from $\lgnode$.
\end{dfn}

Note that our definition permits trivial paths of length $\pathlim = 0$ from any node to itself.
Next, we define the corresponding mapping to a path in the co-optimal edit graph:

\begin{dfn}[corresponding mapping]
Let $\lforest$ and $\rforest$ be forests over some alphabet $\alphabet$, let $\cost$ be a 
cost function over $\alphabet$, and let $\graph_{\cost, \lforest, \rforest} = (\nodes, \edges)$ be the co-optimal edit
graph with respect to $\lforest$, $\rforest$, and $\cost$. Further, let $\rgnode_0, \ldots, \rgnode_\pathlim$
be a path from $1$ to $|\nodes|$ in $\graph_{\cost, \lforest, \rforest}$. Then, we define the
\emph{corresponding mapping} $\map_{\lforest, \rforest}(\rgnode_0, \ldots, \rgnode_\pathlim)$ for path
$\rgnode_0, \ldots, \rgnode_\pathlim$ as $\map = \emptyset$ if $\lforest$ or $\rforest$ are empty and
as follows otherwise:
\begin{equation}
\map_{\lforest, \rforest}(\rgnode_0, \ldots, \rgnode_\pathlim) = \{ (\lnodeidx, \rnodeidx) | \rgnode_\pathidx = (\lkeyrootidx, \lnodeidx, \rkeyrootidx, \rnodeidx),
\rgnode_{\pathidx + 1} = (\lkeyrootidx', \lnodeidx + 1, \rkeyrootidx', \rnodeidx +1) \text{ for any }
\lkeyrootidx, \rkeyrootidx, \lkeyrootidx', \rkeyrootidx', \pathidx \}
\end{equation}
\end{dfn}

Consider the example path $\pth = (1, 1, 1, 1), (1, 2, 1, 2), (1, 3, 1, 2), (3, 4, 1, 3), (2, 4, 1, 3), (2, 5, 1, 3),\allowbreak (1, 5, 1, 3), (1, 6, 1, 3)$
in Figure~\ref{fig:coopt_graph}. The corresponding
mapping $\map_{a(b(c, d), e), f(g)}(\pth)$ would be $\{ (1, 1), (3, 2) \}$,
corresponding to the replacement edges on the path.

Given these definitions, we can go on to show our key theorem for backtracing, namely that 
the corresponding mapping for every path through the co-optimal edit graph is a co-optimal mapping,
and that every co-optimal mapping corresponds to a path through the co-optimal edit graph.

\begin{thm} \label{thm:coopt_path}
Let $\lforest$ and $\rforest$ be forests over some alphabet $\alphabet$, let $\cost$ be a 
cost function over $\alphabet$ that is non-negative, self-equal, and conforms to the trinagular
inequality, and let $\graph_{\cost, \lforest, \rforest}$ be the co-optimal edit
graph with respect to $\lforest$, $\rforest$, and $\cost$. Then it holds:
\begin{enumerate}
\item For all paths $\pth$ from $(1, 1, 1, 1)$ to $(1, |\lforest|+1, 1, |\rforest|+1)$ in
$\graph_{\cost, \lforest, \rforest}$, the corresponding mapping $\map_{\lforest, \rforest}(\pth)$
is a co-optimal mapping between $\lforest$ and $\rforest$.
\item For all co-optimal mappings $\map$ between $\lforest$ and $\rforest$, there exists at least one
path $\pth$ from $(1, 1, 1, 1)$ to $(1, |\lforest|+1, 1, |\rforest|+1)$ in $\graph_{\cost, \lforest, \rforest}$
such that $\map_{\lforest, \rforest}(\pth) = \map$.
\end{enumerate}
\begin{proof}
As a shorthand, we will call a path from $(1, 1, 1, 1)$ to $(1, |\lforest|+1, 1, |\rforest|+1)$
in a co-optimal edit graph $\graph_{\cost, \lforest, \rforest}$ a path \emph{through} that graph.

We start by considering the trivial cases of empty forests. If $\lforest = \epsilon$ or $\rforest = \epsilon$,
the only co-optimal mapping is $\map = \emptyset$. It remains to show that, in these cases, the
co-optimal edit graph contains only paths which correspond to this mapping.

\begin{description}
\item[$\lforest = \epsilon$ and $\rforest = \epsilon$:] In this case, we obtain
$\nodes = \{(1, 1, 1, 1)\}$ and $\edges = \emptyset$. Accordingly,
the trivial path $\pth = (1, 1, 1, 1)$ is the only possible path through $\graph_{\cost, \lforest, \rforest}$
and it does indeed hold $\map_\pth = \emptyset$.
\item[$\lforest = \epsilon$ and $\rforest \neq \epsilon$:] In this case, we obtain
$\nodes = \big\{ (1, 1, 1, \rnodeidx) \big| \rnodeidx \in \{1, \ldots, \siz{\rforest} + 1 \} \big\}$
and
\begin{equation*}
\edges = \big\{ \big((1, 1, 1, \rnodeidx), (1, 1, 1, \rnodeidx+1)\big) \big| \rnodeidx \in \{1, \ldots, \siz{\rforest} \} \big\}.
\end{equation*}
Accordingly, the only possible path through $\graph_{\cost, \lforest, \rforest}$ is
$\pth = (1, 1, 1, 1), (1, 1, 1, 2), \ldots, (1, 1, 1, \siz{\rforest}+1)$.
And indeed it holds $\map_\pth = \emptyset$.
\item[$\lforest \neq \epsilon$ and $\rforest = \epsilon$:] In this case, we obtain
$\nodes = \big\{ (1, \lnodeidx, 1, 1) \big| \lnodeidx \in \{1, \ldots, \siz{\lforest} + 1 \} \big\}$
and
\begin{equation*}
\edges = \big\{ \big((1, \lnodeidx, 1, 1), (1, \lnodeidx+1, 1, 1)\big) \big| \lnodeidx \in \{1, \ldots, \siz{\lforest} \} \big\}.
\end{equation*}
Accordingly, the only possible path through $\graph_{\cost, \lforest, \rforest}$ is
$\pth = (1, 1, 1, 1), (1, 2, 1, 1), \ldots, (1, \siz{\lforest}+1, 1, 1)$.
And indeed it holds $\map_\pth = \emptyset$.
\end{description}

It remains to show both claims for the case of non-empty forests. For both claims, we apply an
induction over the added size of both input forests, for which the base case is already provided
by our considerations above. For the induction, we assume that both claims hold for inputs forests
$\lforest$ and $\rforest$ with $|\lforest| + |\rforest| \leq k$.

Now, we consider two input forests $\lforest$ and $\rforest$ with $\lnodelim + \rnodelim = k + 1$,
where $\lnodelim = |\lforest|$ and $\rnodelim = |\rforest|$.

Regarding the first claim, let $\pth = \rgnode_0, \ldots, \rgnode_\pathlim$ be a path through
$\graph_{\cost, \lforest, \rforest}$, let $\lforest' := \lforest[2, \siz{\lforest}]$,
let $\rforest' := \rforest[2, \siz{\rforest}]$, and consider the following cases regarding $\rgnode_1$.
\begin{description}
\item[$\rgnode_1 = (1, 2, 1, 1)$:] In this case, it must hold
$\fdist_\cost(\lforest, \rforest) = \cost(\lnode_1, \gap) + \fdist_\cost(\lforest', \rforest)$,
otherwise $(\rgnode_0, \rgnode_1) \notin \edges$. Now, if $\lforest'$ is empty, then
$\pth$ must have the form $\pth = (1, 1, 1, 1), (1, 2, 1, 1),\allowbreak (1, 2, 1, 2), \ldots, \allowbreak (1, 2, 1, \siz{\rforest}+1)$,
and $\emptyset$ must be a co-optimal mapping between $\lforest'$ and $\rforest$.
Accordingly, $\emptyset = \map_\pth$ must also be a co-optimal mapping between $\lforest$ and $\rforest$,
because $\cost(\emptyset, \lforest, \rforest) = \cost(\lnode_1, \gap) + \cost(\emptyset, \lforest', \rforest)
= \cost(\lnode_1, \gap) + \fdist_\cost(\lforest', \rforest) = \fdist_\cost(\lforest, \rforest)$.

If $\lforest'$ is \emph{not} empty, then the path $\rgnode_1, \ldots, \rgnode_\pathlim$
is isomorphic to a path from $(1, 1, 1, 1)$ to $(1, |\lforest'|+1, 1, |\rforest|+1)$ in $\graph_{\cost, \lforest', \rforest}$.
Accordingly, per induction hypothesis, $\map_{\pth'}$ is a co-optimal mapping between $\lforest'$ and $\rforest$.
Further, we obtain per construction $\map_\pth = \{ (\lnodeidx+1, \rnodeidx) | (\lnodeidx, \rnodeidx) \in \map_{\pth'} \}$.
Accordingly, it holds: $\Cost(\map_\pth, \lforest, \rforest) = \cost(\lnode_1, \gap) + \Cost(\map_{\pth'}, \lforest', \rforest)
= \cost(\lnode_1, \gap) + \fdist_\cost(\lforest', \rforest) = \fdist_\cost(\lforest, \rforest)$, which means that
$\map_\pth$ is co-optimal, as claimed.
\item[$\rgnode_1 = (1, 1, 1, 2)$:] In this case, it must hold
$\fdist_\cost(\lforest, \rforest) = \cost(\gap, \rnode_1) + \fdist_\cost(\lforest, \rforest')$,
otherwise $(\rgnode_0, \rgnode_1) \notin \edges$. Now, if $\rforest'$ is empty, then
$\pth$ must have the form $\pth = (1, 1, 1, 1), (1, 1, 1, 2),\allowbreak (1, 2, 1, 2), \ldots, (1, \siz{\lforest}+1, 1, 2)$,
and $\emptyset$ must be a co-optimal mapping between $\lforest$ and $\rforest'$.
Accordingly, $\emptyset = \map_\pth$ must also be a co-optimal mapping between $\lforest$ and $\rforest$,
because $\cost(\emptyset, \lforest, \rforest) = \cost(\gap, \rnode_1) + \cost(\emptyset, \lforest, \rforest')
= \cost(\gap, \rnode_1) + \fdist_\cost(\lforest, \rforest') = \fdist_\cost(\lforest, \rforest)$.

If $\rforest'$ is \emph{not} empty, then the path $\rgnode_1, \ldots, \rgnode_\pathlim$
is isomorphic to a path from $(1, 1, 1, 1)$ to $(1, |\lforest|+1, 1, |\rforest'|+1)$ in $\graph_{\cost, \lforest, \rforest'}$.
Accordingly, by virtue of our induction hypothesis, $\map_{\pth'}$ is a co-optimal mapping between $\lforest$ and $\rforest'$.
Further, we obtain per construction $\map_\pth = \{ (\lnodeidx, \rnodeidx+1) | (\lnodeidx, \rnodeidx) \in \map_{\pth'} \}$.
Accordingly, it holds: $\Cost(\map_\pth, \lforest, \rforest) = \cost(\gap, \rnode_1) + \Cost(\map_{\pth'}, \lforest, \rforest')
= \cost(\gap, \rnode_1) + \fdist_\cost(\lforest, \rforest') = \fdist_\cost(\lforest, \rforest)$, which means that
$\map_\pth$ is co-optimal, as claimed.
\item[$\rgnode_1 = (1, 2, 1, 2)$:] In this case, it must hold
$\fdist_\cost(\lforest, \rforest) = \cost(\lnode_1, \rnode_1) + \fdist_\cost(\lforest', \rforest')$.
Now, if $\lforest'$ is empty, then
$\pth$ must have the form $\pth = (1, 1, 1, 1), (1, 2, 1, 2), \ldots, (1, 2, 1, \siz{\rforest}+1)$,
and $\emptyset$ must be a co-optimal mapping between $\lforest'$ and $\rforest'$.
Accordingly, $\{(1, 1) \} = \map_\pth$ must also be a co-optimal mapping between $\lforest$ and $\rforest$,
because $\cost(\{(1, 1) \}, \lforest, \rforest) = \cost(\lnode_1, \rnode_1) + \cost(\emptyset, \lforest', \rforest')
= \cost(\lnode_1, \rnode_1) + \fdist_\cost(\lforest', \rforest') = \fdist_\cost(\lforest, \rforest)$.

If $\rforest'$ is empty, then
$\pth$ must have the form $\pth = (1, 1, 1, 1), (1, 2, 1, 2), \ldots, (1, \siz{\lforest}+1, 1, 2)$,
and $\emptyset$ must be a co-optimal mapping between $\lforest$ and $\rforest'$.
Accordingly, $\{(1, 1) \} = \map_\pth$ must also be a co-optimal mapping between $\lforest$ and $\rforest$,
because $\cost(\{(1, 1) \}, \lforest, \rforest)\allowbreak = \cost(\lnode_1, \rnode_1) + \cost(\emptyset, \lforest', \rforest')
= \cost(\lnode_1, \rnode_1) + \fdist_\cost(\lforest', \rforest') = \fdist_\cost(\lforest, \rforest)$.

If neither $\lforest'$ nor $\rforest'$ are empty, then the path $\rgnode_1, \ldots, \rgnode_\pathlim$
is isomorphic to a path from $(1, 1, 1, 1)$ to $(1, |\lforest'|+1, 1, |\rforest'|+1)$ in
$\graph_{\cost, \lforest', \rforest'}$. Accordingly, by virtue of
our induction hypothesis, $\map_{\pth'}$ is a co-optimal mapping between $\lforest'$ and $\rforest'$.
Further, we obtain per construction $\map_\pth = \{(1, 1)\} \cup \{ (\lnodeidx+1, \rnodeidx+1) | (\lnodeidx, \rnodeidx) \in \map_{\pth'} \}$.
Accordingly, it holds: $\Cost(\map_\pth, \lforest, \rforest) = \cost(\lnode_1, \rnode_1) + \Cost(\map_{\pth'}, \lforest', \rforest')
= \cost(\lnode_1, \rnode_1) + \fdist_\cost(\lforest', \rforest') = \fdist_\cost(\lforest, \rforest)$, which means that
$\map_\pth$ is co-optimal, as claimed.
\end{description}
Other cases can not occur such that our induction is concluded.

Regarding the second claim, let $\map$ be a co-optimal mapping between $\lforest$ and $\rforest$,
i.e.\ $\Cost(\map, \lforest, \rforest) = \fdist_\cost(\lforest, \rforest)$,
and distinguish the following cases.
\begin{description}
\item[$1 \in \unmappedLeft(\map, \lforest, \rforest)$:] In this case it holds
$\Cost(\map, \lforest, \rforest) = \cost(\lnode_1, \gap) + \Cost(\map', \lforest', \rforest)$
with $\map' = \{ (\lnodeidx - 1, \rnodeidx) | (\lnodeidx, \rnodeidx) \in \map \}$.
It must hold that $\map'$ is a co-optimal mapping between $\lforest'$ and $\rforest$. Otherwise,
we would obtain $\fdist_\cost(\lforest, \rforest) \leq 
\fdist_\cost(\lforest', \rforest) + \cost(\lnode_1, \gap) < \Cost(\map', \lforest', \rforest) + \cost(\lnode_1, \gap)
= \Cost(\map, \lforest, \rforest) = \fdist_\cost(\lforest, \rforest)$,
which is a contradiction. This also implies that
$\fdist_\cost(\lforest, \rforest) = \cost(\lnode_1, \gap) + \fdist_\cost(\lforest', \rforest)$,
which in turn implies that $\big((1, 1, 1, 1), (1, 2, 1, 1)\big) \in \edges$.

Now, if $\lforest' = \epsilon$, $\map$ must be $\emptyset$, and we can construct the path
$\pth = (1, 1, 1, 1), (1, 2, 1, 1),\allowbreak \ldots, (1, 2, 1, \siz{\rforest}+1)$, which is a
path through $\graph_{\cost, \lforest, \rforest}$ such that $\map_\pth = \emptyset$.

If $\lforest'$ is not empty, our induction hypothesis implies that there exists a path
$\pth'$ from $(1, 1, 1, 1)$ to $(1, |\lforest'|+1, 1, |\rforest|+1)$ in
$\graph_{\cost, \lforest', \rforest}$ such that $\map_{\pth'} = \map'$. Therefore, we can construct
an isomorphic path $\tilde \pth$ between $(1, 2, 1, 1)$ and $(1, \siz{\lforest}+1, \siz{\rforest}+1)$
in $\graph_{\cost, \lforest, \rforest}$. Accordingly, $\pth := (1, 1, 1, 1), \tilde \pth$
must be a path through $\graph_{\cost, \lforest, \rforest}$, and per construction it must
hold that $\map_\pth = \map$.
\item[$1 \in \unmappedRight(\map, \lforest, \rforest)$:] In this case it holds
$\Cost(\map, \lforest, \rforest) = \cost(\gap, \rnode_1) + \Cost(\map', \lforest, \rforest')$
with $\map' = \{ (\lnodeidx, \rnodeidx - 1) | (\lnodeidx, \rnodeidx) \in \map \}$.
It must hold that $\map'$ is a co-optimal mapping between $\lforest$ and $\rforest'$. Otherwise,
we would obtain $\fdist_\cost(\lforest, \rforest) \leq 
\fdist_\cost(\lforest, \rforest') + \cost(\gap, \rnode_1) < \Cost(\map', \lforest, \rforest') + \cost(\gap, \rnode_1)
= \Cost(\map, \lforest, \rforest) = \fdist_\cost(\lforest, \rforest)$,
which is a contradiction. This also implies that
$\fdist_\cost(\lforest, \rforest) = \cost(\gap, \rnode_1) + \fdist_\cost(\lforest, \rforest')$,
which in turn implies that $\big((1, 1, 1, 1), (1, 1, 1, 2)\big) \in \edges$.

Now, if $\rforest' = \epsilon$, $\map$ must be $\emptyset$, and we can construct the path
$\pth = (1, 1, 1, 1), (1, 1, 1, 2),\allowbreak \ldots, (1, \siz{\lforest}+1, 1, 2)$, which is a
path through $\graph_{\cost, \lforest, \rforest}$ such that $\map_\pth = \emptyset$.

If $\lforest'$ is not empty, our induction hypothesis implies that there exists a path
$\pth'$ from $(1, 1, 1, 1)$ to $(1, |\lforest|+1, 1, |\rforest'|+1)$ in
$\graph_{\cost, \lforest, \rforest'}$ such that $\map_{\pth'} = \map'$.
Therefore, we can construct an isomorphic path $\tilde \pth$ between $(1, 2, 1, 1)$ and
$(1, \siz{\lforest}+1, \siz{\rforest}+1)$ in $\graph_{\cost, \lforest, \rforest}$.
Accordingly, $\pth := (1, 1, 1, 1), \tilde \pth$
must be a path through $\graph_{\cost, \lforest, \rforest}$, and per construction it must
hold that $\map_\pth = \map$.
\item[$\exists (1, \rnodeidx), (\lnodeidx, 1) \in \map$:]
In this case, $\lnodeidx = \rnodeidx = 1$, which we can show as follows. 
Consider the case $\rnodeidx > 1$. In that case,
$\lnodeidx < 1$, which is impossible. Similarly, if $\lnodeidx > 1$, it must hold $\rnodeidx < 1$,
which is impossible. Therefore $\lnodeidx = 1$ and $\rnodeidx = 1$.

In this case it holds
$\Cost(\map, \lforest, \rforest) = \cost(\lnode_1, \rnode_1) + \Cost(\map', \lforest', \rforest')$
with $\map' = \{ (\lnodeidx-1, \rnodeidx - 1) | (\lnodeidx, \rnodeidx) \in \map \setminus \{ (1, 1) \} \}$.
It must hold that $\map'$ is a co-optimal mapping between $\lforest'$ and $\rforest'$. Otherwise,
we would obtain $\fdist_\cost(\lforest, \rforest) \leq 
\fdist_\cost(\lforest', \rforest') + \cost(\lnode_1, \rnode_1) < \Cost(\map', \lforest', \rforest') + \cost(\lnode_1, \rnode_1)
= \Cost(\map, \lforest, \rforest) = \fdist_\cost(\lforest, \rforest)$,
which is a contradiction. This also implies that
$\fdist_\cost(\lforest, \rforest) = \cost(\lnode_1, \rnode_1) + \fdist_\cost(\lforest', \rforest')$,
which in turn implies that $\big((1, 1, 1, 1), (1, 2, 1, 2)\big) \in \edges$.

Now, if $\lforest' = \epsilon$, $\map$ must be $\{(1, 1)\}$, and we can construct the path
$\pth = (1, 1, 1, 1),\allowbreak (1, 2, 1, 2), \ldots, \allowbreak(1, 2, 1, \siz{\rforest}+1)$, which is a
path through $\graph_{\cost, \lforest, \rforest}$ such that $\map_\pth = \{(1, 1)\}$.

If $\rforest' = \epsilon$, $\map$ must be $\{(1, 1)\}$, and we can construct the path
$\pth = (1, 1, 1, 1), (1, 2, 1, 2),\allowbreak \ldots, (1, \siz{\lforest}+1, 1, 2)$, which is a
path through $\graph_{\cost, \lforest, \rforest}$ such that $\map_\pth = \{(1, 1)\}$.

If neither $\lforest'$ nor $\rforest'$ is empty, our induction hypothesis implies that there exists a path
$\pth'$ through $\graph_{\cost, \lforest', \rforest'}$ such that $\map_{\pth'} = \map'$.
Therefore, we can construct
an isomorphic path $\tilde \pth$ between $(1, 2, 1, 2)$ and $(1, \siz{\lforest}+1, \siz{\rforest}+1)$
in $\graph_{\cost, \lforest, \rforest}$. Accordingly, $\pth := (1, 1, 1, 1), \tilde \pth$
must be a path through $\graph_{\cost, \lforest, \rforest}$, and per construction it must
hold that $\map_\pth = \map$.
\end{description}
As no other cases can occur, this concludes the proof.

\end{proof}
\end{thm}

\subsection{Finding a single co-optimal mapping}

\begin{algorithm}
\caption{A recursive backtracing algorithm for the TED, which infers a co-optimal mapping
between the input trees $\ltree$ and $\rtree$.
Refer to our \href{https://gitlab.ub.uni-bielefeld.de/bpaassen/python-edit-distances/-/blob/master/edist/ted.pyx}{project web site}
for a reference implementation.}
\label{alg:backtrace}
\begin{algorithmic}[1]
\Function{backtrace}{Two trees $\ltree$ and $\rtree$, the matrices $\mat \dist$ and $\mat \fdist$
after executing Algorithm~\ref{alg:ted}, and a cost function $\cost$}
\State $\map \gets \emptyset$.
\State \Call{backtr}{$\ltree$, $\rtree$, $\mat \dist$, $\mat \fdist$, $\cost$, $\map$, 1, 1}.
\State \Return $\map$.
\EndFunction
\Function{backtr}{Trees $\ltree$ and $\rtree$, $\mat \dist$, $\mat \fdist$, $\cost$, partial mapping $\map$, node indices $\lkeyrootidx$ and $\rkeyrootidx$}
\If{$\lkeyrootidx > 1 \vee \rkeyrootidx > 1$} \Comment{Update $\mat \fdist$}
	\For{$\lnodeidx \gets \rleaf_{\ltree}(\lkeyrootidx), \ldots, \lkeyrootidx$}
\State $\mat \fdist_{\lnodeidx, \rleaf_\rtree(\rkeyrootidx) + 1} \gets \mat \fdist_{\lnodeidx + 1, \rleaf_\rtree(\rkeyrootidx) + 1} + \cost(\lnode_{\lnodeidx}, \gap)$.
\Comment Equation~\ref{eq:del}
\EndFor
\For{$\rnodeidx \gets \rleaf_\rtree(\rkeyrootidx), \ldots, \rkeyrootidx$}
\State $\mat \fdist_{\rleaf_\ltree(\lkeyrootidx) + 1, \rnodeidx} \gets \mat \fdist_{\rleaf_\ltree(\lkeyrootidx) + 1, \rnodeidx + 1} + \cost(\gap, \rnode_{\rnodeidx})$.
\Comment Equation~\ref{eq:ins}
\EndFor
\For{$\lnodeidx \gets \rleaf_\ltree(\lkeyrootidx), \ldots, \lkeyrootidx$}
\For{$\rnodeidx \gets \rleaf_\rtree(\rkeyrootidx), \ldots, \rkeyrootidx$}
\If{$\rleaf_{\ltree}(\lnodeidx) = \rleaf_{\ltree}(\lkeyrootidx) \wedge \rleaf_{\rtree}(\rnodeidx) = \rleaf_{\rtree}(\rkeyrootidx)$}
\State $\mat \fdist_{\lnodeidx, \rnodeidx} \gets \mat \dist_{\lnodeidx, \rnodeidx} + \mat \fdist_{\rleaf_{\ltree}(\lkeyrootidx) + 1, \rleaf_{\rtree}(\rkeyrootidx) + 1}$.
\Else
\State $\mat \fdist_{\lnodeidx, \rnodeidx} \gets \min \{
\mat \fdist_{\lnodeidx + 1, \rnodeidx    } + \cost(\lnode_{\lnodeidx}, \gap),$
\State $\quad \mat \fdist_{\lnodeidx    , \rnodeidx + 1} + \cost(\gap, \rnode_{\rnodeidx}),$
\State $\quad \mat \fdist_{\rleaf_{\ltree}(\lnodeidx) + 1, \rleaf_{\rtree}(\rnodeidx) + 1} + \mat \dist_{\lnodeidx, \rnodeidx} \}$. \Comment Equation~\ref{eq:fdist_decompose}
\EndIf
\EndFor
\EndFor
\EndIf
\State $\lnodeidx \gets \lkeyrootidx$, $\rnodeidx \gets \rkeyrootidx$.
\Comment Start finding a path through the edit graph between $\ltree_\lkeyrootidx$ and $\rtree_\rkeyrootidx$.
\While{$\lnodeidx \leq \rleaf_{\ltree}(\lkeyrootidx) \wedge \rnodeidx \leq \rleaf_{\rtree}(\rkeyrootidx)$}
	\If{$\big(\rleaf_{\ltree}(\lnodeidx) = \rleaf_{\ltree}(\lkeyrootidx) \wedge \rleaf_{\rtree}(\rnodeidx) = \rleaf_{\rtree}(\rkeyrootidx)\big)
	\vee \big( \cost(\lnode_\lnodeidx, \rnode_\rnodeidx) = \cost(\lnode_\lnodeidx, \gap) + \cost(\gap, \rnode_\rnodeidx) \big)$}
		\If{$\mat \fdist_{\lnodeidx, \rnodeidx} = \mat \fdist_{\lnodeidx + 1, \rnodeidx + 1} + \cost(\lnode_\lnodeidx, \rnode_\rnodeidx)$}
		\State $\map \gets \map \cup \{ (\lnodeidx, \rnodeidx) \}$. \Comment{replacement is optimal}
		\State $\lnodeidx \gets \lnodeidx + 1$, $\rnodeidx \gets \rnodeidx + 1$.
		\State \textbf{continue}.
		\EndIf
	\Else
		\If{$\mat \fdist_{\lnodeidx, \rnodeidx} = \mat \fdist_{\rleaf_{\ltree}(\lnodeidx) + 1, \rleaf_{\rtree}(\rnodeidx) + 1} + \mat \dist_{\lnodeidx, \rnodeidx}$}
		\State \Call{backtr}{$\ltree$, $\rtree$, $\mat \dist$, $\mat \fdist$, $\cost$, $\map$, $\lnodeidx$, $\rnodeidx$}. \Comment{Recursively edit subtree $\ltree_\lnodeidx$ into $\rtree_\rnodeidx$.}
		\State $\lnodeidx \gets \rleaf_{\ltree}(\lnodeidx)+1$,
		$\rnodeidx \gets \rleaf_{\rtree}(\rnodeidx)+1$.
		\State \textbf{continue}.
		\EndIf
	\EndIf
	\If{$\mat \fdist_{\lnodeidx, \rnodeidx} = \mat \fdist_{\lnodeidx + 1, \rnodeidx} + \cost(\lnode_\lnodeidx, \gap)$}
		\State $\lnodeidx \gets \lnodeidx + 1$. \Comment{deletion is optimal}
	\ElsIf{$\mat \fdist_{\lnodeidx, \rnodeidx} = \mat \fdist_{\lnodeidx, \rnodeidx + 1} + \cost(\gap, \rnode_\rnodeidx)$}
		\State $\rnodeidx \gets \rnodeidx + 1$. \Comment{insertion is optimal}
	\EndIf
\EndWhile
\EndFunction
\end{algorithmic}
\end{algorithm}

Now that we have proven that finding a co-optimal mapping is equivalent to finding a path through
the co-optimal edit graph, it is relatively simple to construct an algorithm which identifies
one such mapping.

\begin{thm}\label{thm:backtrace}
Given two input trees $\ltree$ and $\rtree$ as well as a cost function $\cost$ that is non-negative,
self-equal, and conforms to the triangular inequality,
Algorithm~\ref{alg:backtrace} computes a co-optimal mapping $\map$ between $\ltree$ and $\rtree$.
Further, Algorithm~\ref{alg:backtrace} runs in
$\effic((\lnodelim + \rnodelim) \cdot \lnodelim \cdot \rnodelim)$ time complexity and
$\effic(\lnodelim \cdot \rnodelim)$ space complexity.

\begin{proof}
Note: This is only a sketch of a proof. For a rigorous proof, we would 
have to properly match the actions of Algorithm~\ref{alg:backtrace} with 
the co-optimal edit graph.

Algorithm~\ref{alg:backtrace} starts at $(1, 1, 1, 1)$ and then travels along the
co-optimal edit graph, implicitly constructing it as needed. In particular, lines 28-33
cover the (replacement) edges defined in Equations~\ref{eq:coopt_graph_rep1} and~\ref{eq:coopt_graph_rep2},
and lines 35-39 cover the (subtree replacement) edges defined via Equation~\ref{eq:coopt_graph_rep3}.
Lines 41-42 cover the (deletion) egdes defined via Equation~\ref{eq:coopt_graph_del}, and lines 43-45 cover
the (insertion) edges defined via Equation~\ref{eq:coopt_graph_ins}. The backwards connections
defined via Equations~\ref{eq:coopt_graph_up1}, \ref{eq:coopt_graph_up2}, \ref{eq:coopt_graph_up3}
are taken when returning from a recursive call in line 36.
Importantly, note that the update in lines 7-25 always ensures that
we consider to correct subforest edit distance matrix $\mat \fdist$ when
stepping into a new level of recursion.

Further note that Algorithm~\ref{alg:backtrace} directly constructs a co-optimal mapping from
the path via line 30 which adds a tuple $(\lnodeidx, \rnodeidx)$ to the mapping whenever we use
a replacement edge, as suggested by Theorem~\ref{thm:coopt_path}.

Regarding space complexity, the only data structures we require are
the matrices $\mat \fdist$ and $\mat \dist$ from
before, as well as the trees $\ltree$ and $\rtree$, the cost function
$\cost$, and the (partial) mapping $\map$ which results in $\effic(\lnodelim \cdot \rnodelim)$ space complexity.
Regarding runtime, we note that each iteration of the while loop 
in lines 27-46 of Algorithm~\ref{alg:backtrace} advances
either $\lnodeidx$ or $\rnodeidx$. Accordingly, the while loop can run
at most $\effic(\lnodelim + \rnodelim)$ times.
At worst, we need to perform a recursion in each step, in which case
a section of the matrix $\mat \fdist$ is updated. Each of these updates takes at worst $\effic(\lnodelim \cdot \rnodelim)$
operations, such that we obtain $\effic((\lnodelim + \rnodelim) \cdot \lnodelim \cdot \rnodelim)$
overall.
\end{proof}
\end{thm}

As an example, consider the co-optimal edit graph in Figure~\ref{fig:coopt_graph}, corresponding
to the subforest edit distances in Figure~\ref{fig:ted}.
We begin by calling the function \texttt{BACKTR} with the arguments
$\ltree = \sym{a}(\sym{b}(\sym{c}, \sym{d}), \sym{e})$, $\rtree = \sym{f}(\sym{g})$,
$\mat \dist$ and $\mat \fdist$ from the forward computation
of Algorithm~\ref{alg:ted}, the cost function $\cost$,
an empty partial mapping $\map = \emptyset$, and the roots of both
trees $\lkeyrootidx = 1$ and $\rkeyrootidx = 1$.
Because $\lkeyrootidx = \rkeyrootidx = 1$, the if condition in line 7
does not apply. Line 26, then, sets $\lnodeidx = \rnodeidx = 1$.
Now, we begin to construct the mapping. First, we note that the
condition in line 28 applies, meaning that $\lnodeidx$ and $\rnodeidx$
are both in the scope of our currently considered keyroots.
Further, we observe that
$5 = \mat \fdist_{1, 1} = \mat \fdist_{2, 2} + \cost(\lnode_1, \rnode_1) = 4 + 1$,
that is: replacing $\lnode_1 = \sym{a}$ with $\rnode_1 = \sym{f}$ is part of a co-optimal mapping.
Therefore, the condition in line 29 applies, we add $(1, 1)$ to the mapping $\map$ in line 30, and increment both $\lnodeidx$ and $\rnodeidx$.
We now return to line 28. Now, we find that $\rleaf_{\ltree}(2) \neq
\rleaf_{\ltree}(1)$, such that the condition does not apply anymore.
However, we do find that
$4 = \mat \fdist_{2, 2} = \mat \fdist_{5, 3} + \mat \dist_{2, 2} = 1 + 3$,
that is:
replacing the subtree $\sym{b}(\sym{c}, \sym{d})$ with the subtree $\sym{g}$ is part of a co-optimal mapping.
Accordingly, the condition in line 35 applies and we call
\texttt{BACKTR} recursively with $\ltree$, $\rtree$, $\mat \dist$,
$\mat \fdist$, $\cost$ as before, as well as
$\map = \{(1, 1)\}$, $\lkeyrootidx = 2$ and $\rkeyrootidx = 2$.

Now, the condition in line 7 applies, such that the update in lines
8-24 is executed. More specifically, the update ensures that
$\mat \fdist_{\lnodeidx, \rnodeidx} =
\fdist_\cost(\ltree[\lnodeidx, \rleaf_{\ltree}(2)], \rtree[\rnodeidx, \rleaf_{\rtree}(2)]) + \fdist_\cost([\rleaf_{\ltree}(2) + 1, 5], [\rleaf_{\rtree}(2) + 1, 2] =
\fdist_\cost(\ltree[\lnodeidx, 4], \rtree[\rnodeidx, 2]) + \fdist_\cost(\sym{e}, \epsilon)$ for all
$\lnodeidx \in \{ 2, 3, 4, 5\}$ and $\rnodeidx \in \{2, 3\}$.
In this example, this only changes the
entry $\mat \fdist_{5, 2}$ which would be set to $\fdist_\cost(\ltree[\lnodeidx, 4], \rtree[\rnodeidx, 2) + \fdist_\cost(\sym{e}, \epsilon)
= 1 + 1 = 2$.

Line 26 now sets $\lnodeidx = \lkeyrootidx = 2$ and $\rnodeidx = \rkeyrootidx = 2$ and we enter the while loop in line 27.
Now, the condition in line 28 applies again and we observe that
$4 = \mat \fdist_{2, 2} = \mat \fdist_{3, 3} + \cost(\lnode_2, \rnode_2) = 3 + 1$,
that is: replacing $\lnode_2 = \sym{b}$ with $\rnode_2 = \sym{g}$ is part of a co-optimal mapping. Therefore, we add
$(2, 2)$ to the mapping $\map$, and increment both $\lnodeidx$ as well as $\rnodeidx$. Now, we find that $3 = \rnodeidx > \rleaf_{\rtree}(\rkeyrootidx) = \rleaf_{\rtree}(2)= 2$.
Therefore, we stop the loop in line 27 and return to the higher
recursion level.

In the higher recursion, we now execute line 37, meaning that
$\lnodeidx$ becomes $\rleaf_{\ltree}(2) + 1 = 4 + 1 = 5$
and $\rnodeidx$ becomes $\rleaf_{\rtree}(2) + 1 = 2 + 1 = 3$.
Next, we return to line 27 and notice that $3 = \rnodeidx > \rleaf_{\rtree}(\rkeyrootidx) = \rleaf_{\rtree}(1) = 2$.
Therefore, our algorithm stops and we return to the
$\texttt{BACKTRACE}$ in line 4. We return the mapping
$\map = \{ (1, 1), (2, 2) \}$, which is indeed
a co-optimal mapping between $\ltree$ and $\rtree$.

Note that Algorithm~\ref{alg:backtrace} always prefers replacements if multiple edits are
co-optimal. As such, Algorithm~\ref{alg:backtrace} will prefer to map the nodes close to
the root of both trees to each other, and delete/insert nodes closer to the leaves.
The other possible co-optimal mappings for this example are
$\{ (1, 1), (3, 2) \}$,  $\{ (1, 1), (4, 2)\}$, $\{(1, 1), (5, 2) \}$, $\{(2, 1), (3, 2)\}$,
and $\{ (2, 1), (4, 2) \}$ (also refer to Figure~\ref{fig:coopt_mappings}),
all of which correspond to exactly one path in Figure~\ref{fig:coopt_graph}.

\subsection{Counting co-optimal mappings}

As we have already seen, it is infeasible to list all co-optimal mappings in general (see Theorem~\ref{thm:num_coopts}).
Interestingly, though, we can still \emph{count} the number of such mappings efficiently.
We will first consider the problem of counting the number of paths in a general DAG, and then return
to the co-optimal edit graph, specifically.

\begin{thm} \label{thm:count_paths_forward}
Let $\graph = (\nodes, \edges)$ be a DAG with ordering relation $<$ and let $\rgnode_1, \ldots,
\rgnode_\rnodelim$ be the nodes in $\nodes$ as ordered according to $<$. Then, Algorithm~\ref{alg:count_paths_forward}
returns a $\rnodelim \times 1$ vector $\vec \forw$, such that $\forw_\lnodeidx$ is exactly the
number of paths leading from $\rgnode_1$ to $\rgnode_\lnodeidx$.
Further, Algorithm~\ref{alg:count_paths_forward} runs in $\effic(\rnodelim)$ time and space complexity.
\begin{proof}

To prove this result, we first show two lemmata:
\begin{enumerate}
\item Algorithm~\ref{alg:count_paths_forward} visits all reachable nodes from $\rgnode_1$ in ascending order,
and no other nodes.
\item When Algorithm~\ref{alg:count_paths_forward} visits node $\rgnode_\lnodeidx$, $\forw_\lnodeidx$
contains exactly the number of paths from $\rgnode_1$ to $\rgnode_\lnodeidx$.
\end{enumerate}
We call a note \emph{visited}, if it is pulled from $Q$. We proof both lemmata by induction over $\lnodeidx$.

\begin{enumerate}
\item Our base case is $\rgnode_1$, which is indeed visited first.

Now, assume that the claim holds for all reachable nodes $\leq \rgnode$. Consider the smallest node
$\lgnode > \rgnode$ which is reachable from $1$. Then, there is a path $\lgnode_0, \ldots, \lgnode_\pathlim$
with $\lgnode_0 = \rgnode_1$ and $\lgnode_\pathlim = \lgnode$. Because $\graph$ is a DAG, $\lgnode_{\pathlim - 1} < \lgnode$.
Further, because $\lgnode_0, \ldots, \lgnode_{\pathlim - 1}$ is a path from $\rgnode_1$ to $\lgnode_{\pathlim - 1}$,
$\lgnode_{\pathlim - 1}$ is reachable from $\rgnode_1$. Because $\lgnode$ is per definition the smallest node
larger than $\rgnode$ which is reachable from $\rgnode_1$, it must hold $\lgnode_{\pathlim - 1} \leq \rgnode$.
Therefore, per induction, $\lgnode_{\pathlim - 1}$ has been visited before. This implies that $\lgnode \in Q$.
Because we select the minimum from $Q$ in each iteration, and because all elements smaller than $\lgnode$
have been visited before (and are not visited again due to the DAG property), $\lgnode$ will be
visited next. Therefore, still all reachable nodes from $\rgnode_1$ are visited in ascending order, and
all nodes that are visited are reachable nodes.
\item Again, our base case is $\rgnode_1$, which is visited first. As it is visited, $\forw_1 = 1$.
Indeed, there is only one path from $\rgnode_1$ to $\rgnode_1$, which is the trivial path.

Now, assume that the claim holds for all reachable nodes $\leq \rgnode$. Then, consider the smallest node
$\rgnode_\lnodeidx > \rgnode$ which is reachable from $\rgnode_1$. Further, let $\rgnode_{\lnodeidx_1}, \ldots, \rgnode_{\lnodeidx_\lnodelim}$
be all nodes which are reachable from $\rgnode_1$, such that $(\rgnode_{\lnodeidx_\rnodeidx}, \rgnode_\lnodeidx) \in \edges$.
Because $\graph$ is a DAG, $\rgnode_{\lnodeidx_\rnodeidx} < \rgnode_\lnodeidx$. Further, because
$\rgnode_{\lnodeidx_\rnodeidx}$ is reachable from $1$ and $\rgnode_\lnodeidx$ is per definition
the smallest node larger than $\rgnode$ which is reachable from $\rgnode_1$, it must hold
$\rgnode_{\lnodeidx_\rnodeidx} \leq \rgnode$. Therefore, per induction, $\forw_{\lnodeidx_\rnodeidx}$
is equal to the number of paths from $\rgnode_1$ to $\rgnode_{\lnodeidx_\rnodeidx}$. For any such path
$\pth$, the concatenation $\pth \concat \rgnode_\lnodeidx$ is a path from $\rgnode_1$ to $\rgnode_\lnodeidx$.
Conversely, we can decompose any path $\pth'$ from $\rgnode_1$ to $\rgnode_\lnodeidx$ as
$\pth' = \pth \concat \rgnode_\lnodeidx$ where $\pth$ is a path from $\rgnode_1$ to some node
$\rgnode_{\lnodeidx_\rnodeidx}$.
Accordingly, the number of paths from $\rgnode_1$ to $\rgnode_\lnodeidx$ is exactly
$\sum_{\rnodeidx = 1}^\lnodelim \forw_{\lnodeidx_\rnodeidx}$.

Finally, because of the first lemma,
we know that all $\rgnode_{\lnodeidx_\rnodeidx}$ have been visited already (without duplicates), and that on each of these visits,
$\forw_{\lnodeidx_\rnodeidx}$ has been added to $\forw_\lnodeidx$. Therefore, we obtain
$\forw_\lnodeidx = \sum_{\rnodeidx = 1}^\lnodelim \forw_{\lnodeidx_\rnodeidx}$.
\end{enumerate}

Because Lemma 1 implies that we do not visit any node smaller than $\rgnode_\lnodeidx$ after $\rgnode_\lnodeidx$ has
been visited, the value $\forw_\lnodeidx$ does not change after $\rgnode_\lnodeidx$ is visited. Therefore,
$\forw_{\lnodeidx}$ still contains the number of paths from $\rgnode_1$ to $\rgnode_\lnodeidx$ at the end of the
algorithm.

Regarding runtime, it follows from the first lemma that, per iteration,
exactly one reachable node is processed and will not be visited again. In the worst case,
all nodes in the graph are reachable, which yields $\effic(\rnodelim)$ iterations.
In each iteration we need to retrieve the minimum of $Q$ and insert all $\rgnode$ into $Q$,
for which $(\lgnode, \rgnode) \in \edges$.
Both is possible in constant time if a suitable data structure for $Q$ is used. If one uses
a tree structure for $Q$, the runtime rises to $\effic(\rnodelim \cdot \log(\rnodelim))$.
The space complexity is $\effic(\rnodelim)$ because $\vec \forw$ has $\rnodelim$ entries
and $Q$ can not exceed $\rnodelim$ entries.
\end{proof}
\end{thm}

\begin{algorithm}
\caption{An algorithm to count the number of paths between $\rgnode_1$ and $\rgnode_\lnodeidx$
in a DAG $\graph = (\{\rgnode_1, \ldots, \rgnode_\rnodelim\}, \edges)$ with ordering relation $<$.}
\label{alg:count_paths_forward}
\begin{algorithmic}
\Function{count-paths-forward}{A DAG $\graph = (\{\rgnode_1, \ldots, \rgnode_\rnodelim\}, \edges)$ with ordering relation $<$.}
\State $\vec \forw \gets \vec 0^\rnodelim$.
\State $\forw_1 \gets 1$.
\State $Q \gets \{ \rgnode_1 \}$.
\While{$Q \neq \emptyset$}
	\State $\rgnode_\lnodeidx \gets \min_< Q$.
	\State $Q \gets Q \setminus \{ \rgnode_\lnodeidx \}$.
	\For{$(\rgnode_\lnodeidx, \rgnode_\rnodeidx) \in \edges$}
		\State $\forw_\rnodeidx \gets \forw_\rnodeidx + \forw_\lnodeidx$.
		\State $Q \gets Q \cup \{ \rgnode_\rnodeidx \}$.
	\EndFor
\EndWhile
\State \Return $\vec \forw$.
\EndFunction
\end{algorithmic}
\end{algorithm}

As an example, consider the DAG in Figure~\ref{fig:coopt_graph} with the
ordering indices shown in blue.
Assuming a sorted set for $Q$, Algorithm~\ref{alg:count_paths_forward} would initialize
$\forw_1 \gets 1$ and $Q \gets \{(1, 1, 1, 1)\}$ and would then behave as follows.
\begin{enumerate}
\item $\rgnode_\lnodeidx = \rgnode_1 = (1, 1, 1, 1)$, $\forw_2 \gets 1$, $\forw_3 \gets 1$, $Q \gets \{(1, 2, 1, 1), (1, 2, 1, 2)\}$.
\item $\rgnode_\lnodeidx = \rgnode_2 = (1, 2, 1, 1)$, $\forw_4 \gets 1$, $Q \gets \{ (1, 2, 1, 2), (2, 3, 1, 2) \}$.
\item $\rgnode_\lnodeidx = \rgnode_3 = (1, 2, 1, 2)$, $\forw_5 \gets 1$, $\forw_6 \gets 1$, $Q \gets \{ (2, 3, 1, 2), (1, 3, 1, 2), (2, 3, 1, 3) \}$.
\item $\rgnode_\lnodeidx = \rgnode_4 = (2, 3, 1, 2)$, $\forw_7 \gets 1$, $\forw_9 \gets 1$, $Q \gets \{ (1, 3, 1, 2), (2, 3, 1, 3), (2, 4, 1, 2), (3, 4, 1, 3) \}$.
\item $\rgnode_\lnodeidx = \rgnode_5 = (1, 3, 1, 2)$, $\forw_8 \gets 1$, $\forw_9 \gets 1 + 1$, $Q \gets \{ (2, 3, 1, 3), (2, 4, 1, 2), (1, 4, 1, 2), (3, 4, 1, 3) \}$.
\item $\rgnode_\lnodeidx = \rgnode_6 = (2, 3, 1, 3)$, $\forw_{10} \gets 1$, $Q \gets \{ (2, 4, 1, 2), (1, 4, 1, 2), (3, 4, 1, 3), (2, 4, 1, 3) \}$.
\item $\rgnode_\lnodeidx = \rgnode_7 = (2, 4, 1, 2)$, $\forw_{12} \gets 1$, $Q \gets \{ (1, 4, 1, 2), (3, 4, 1, 3), (2, 4, 1, 3), (2, 5, 1, 3) \}$.
\item $\rgnode_\lnodeidx = \rgnode_8 = (1, 4, 1, 2)$, $\forw_{11} \gets 1$, $\forw_{13} \gets 1$, $Q \gets \{ (3, 4, 1, 3), (2, 4, 1, 3), (1, 5, 1, 2), (2, 5, 1, 3), (1, 5, 1, 3) \}$.
\item $\rgnode_\lnodeidx = \rgnode_9 = (3, 4, 1, 3)$, $\forw_{10} \gets 1 + 2$, $Q \gets \{ (2, 4, 1, 3), (1, 5, 1, 2), (2, 5, 1, 3), (1, 5, 1, 3) \}$.
\item $\rgnode_\lnodeidx = \rgnode_{10} = (2, 4, 1, 3)$, $\forw_{12} \gets 1 + 3$, $Q \gets \{ (1, 5, 1, 2), (2, 5, 1, 3), (1, 5, 1, 3) \}$.
\item $\rgnode_\lnodeidx = \rgnode_{11} = (1, 5, 1, 2)$, $\forw_{14} \gets 1$, $Q \gets \{ (2, 5, 1, 3), (1, 5, 1, 3), (1, 6, 1, 3) \}$.
\item $\rgnode_\lnodeidx = \rgnode_{12} = (2, 5, 1, 3)$, $\forw_{13} \gets 1 + 4$, $Q \gets \{ (1, 5, 1, 3), (1, 6, 1, 3) \}$.
\item $\rgnode_\lnodeidx = \rgnode_{13} = (1, 5, 1, 3)$, $\forw_{14} \gets 1 + 5$, $Q \gets \{ (1, 6, 1, 3) \}$.
\item $\rgnode_\lnodeidx = \rgnode_{14} = (1, 6, 1, 3)$, $Q \gets \emptyset$.
\end{enumerate}
The resulting $\forw$-values for all nodes are indicated in Figure~\ref{fig:forward_backward}.

\begin{figure}
\begin{center}
\begin{tikzpicture}[xscale=2.8,yscale=1.5]

\node[align=left] (abcde_vs_fg) at (0, 0) {$\forw_1 = 1$ \\ $\back_1 = 6$};
\node[align=left] (bcde_vs_fg)  at (0,-1) {$\forw_2 = 1$ \\ $\back_2 = 2$};
\node[align=left] (bcde_vs_g)   at (2,-1) {$\forw_3 = 1$ \\ $\back_3 = 4$};
\node[align=left] (cde_vs_g)    at (2,-2) {$\forw_5 = 1$ \\ $\back_5 = 3$};
\node[align=left] (de_vs_g)     at (2,-3) {$\forw_8 = 1$ \\ $\back_8 = 2$};
\node[align=left] (e_vs_g)      at (2,-4) {$\forw_{11} = 1$ \\ $\back_{11} = 1$};
\node[align=left] (e_vs_eps)    at (3,-4) {$\forw_{13} = 5$ \\ $\back_{13} = 1$};
\node[align=left] (eps_vs_eps)  at (3,-5) {$\forw_{14} = 6$ \\ $\back_{14} = 1$};

\begin{scope}[shift={(-1,-1)}]

\node[align=left] (cd_vs_g)     at (1,-1) {$\forw_4 = 1$ \\ $\back_4 = 2$};
\node[align=left] (cd_vs_eps)   at (2,-1) {$\forw_6 = 1$ \\ $\back_6 = 1$};
\node[align=left] (d_vs_g)      at (1,-2) {$\forw_7 = 1$ \\ $\back_7 = 1$};
\node[align=left] (d_vs_eps)    at (2,-2) {$\forw_{10} = 3$ \\ $\back_{10} = 1$};
\node[align=left] (eps_vs_eps2) at (2,-3) {$\forw_{12} = 4$ \\ $\back_{12} = 1$};

\end{scope}

\node[align=center] (eps_vs_eps3) at (3,-3) {$\forw_9 = 2$ \\ $\back_9 = 1$};

\path[edge, semithick, draw=skyblue3, fill=skyblue1]%
(abcde_vs_fg) edge[transform canvas={xshift=-1mm}] (bcde_vs_fg)
(bcde_vs_fg)  edge[transform canvas={xshift=-1mm}] (cd_vs_g)
(cd_vs_g)     edge (d_vs_g)
(d_vs_g)      edge (eps_vs_eps2)
(eps_vs_eps2) edge[bend right=40,transform canvas={yshift=-3mm}] (e_vs_eps)
(e_vs_eps)    edge[transform canvas={xshift=-4mm}] (eps_vs_eps);

\path[edge, semithick, draw=orange3, fill=orange1]%
(abcde_vs_fg) edge[transform canvas={xshift=+1mm}] (bcde_vs_fg)
(bcde_vs_fg)  edge[transform canvas={xshift=+1mm}] (cd_vs_g)
(cd_vs_g)     edge[out=315,in=135] (eps_vs_eps3)
(eps_vs_eps3) edge[bend right,transform canvas={yshift=-1mm}] (d_vs_eps)
(d_vs_eps)    edge[transform canvas={xshift=-1mm}] (eps_vs_eps2)
(eps_vs_eps2) edge[bend right=40, transform canvas={yshift=-1mm}] (e_vs_eps)
(e_vs_eps)    edge[transform canvas={xshift=-2mm}] (eps_vs_eps);

\path[edge, semithick, draw=plum3, fill=plum1]%
(abcde_vs_fg) edge[transform canvas={yshift=-3mm}] (bcde_vs_g)
(bcde_vs_g)   edge (cd_vs_eps)
(cd_vs_eps)   edge (d_vs_eps)
(d_vs_eps)    edge[transform canvas={xshift=+1mm}] (eps_vs_eps2)
(eps_vs_eps2) edge[bend right=40, transform canvas={yshift=+1mm}] (e_vs_eps)
(e_vs_eps)    edge (eps_vs_eps);

\path[edge, semithick, draw=scarletred3, fill=scarletred1]%
(abcde_vs_fg) edge[transform canvas={yshift=-1mm}] (bcde_vs_g)
(bcde_vs_g)   edge[transform canvas={xshift=-2mm}] (cde_vs_g)
(cde_vs_g)    edge[transform canvas={xshift=-1mm}] (de_vs_g)
(de_vs_g)     edge (e_vs_g)
(e_vs_g)      edge (eps_vs_eps);

\path[edge, semithick, draw=chameleon3, fill=chameleon1]%
(abcde_vs_fg) edge[transform canvas={yshift=+1mm}] (bcde_vs_g)
(bcde_vs_g)   edge (cde_vs_g)
(cde_vs_g)    edge[transform canvas={xshift=+1mm}] (de_vs_g)
(de_vs_g)     edge (e_vs_eps)
(e_vs_eps)    edge[transform canvas={xshift=+2mm}] (eps_vs_eps);

\path[edge, semithick, draw=butter3, fill=butter1]%
(abcde_vs_fg) edge[transform canvas={yshift=+3mm}] (bcde_vs_g)
(bcde_vs_g)   edge[transform canvas={xshift=+2mm}] (cde_vs_g)
(cde_vs_g)    edge (eps_vs_eps3)
(eps_vs_eps3) edge[bend right,transform canvas={yshift=+1mm}] (d_vs_eps)
(d_vs_eps)    edge[transform canvas={xshift=+3mm}] (eps_vs_eps2)
(eps_vs_eps2) edge[bend right=40, transform canvas={yshift=+3mm}] (e_vs_eps)
(e_vs_eps)    edge[transform canvas={xshift=+4mm}] (eps_vs_eps);

\end{tikzpicture}
\end{center}
\caption{An illustration of all possible paths from $(1, 1, 1, 1)$ to $(1, 6, 1, 3)$ in the DAG from Figure~\ref{fig:coopt_graph}.
The paths are diffenteriated by color.
The number of (unique) paths from $(1, 1, 1, 1)$ to each node $\rgnode_\lnodeidx$ is denoted as $\forw_\lnodeidx$, and the
number of (unique) paths from each node $\rgnode_\lnodeidx$ to $(1, 6, 1, 3)$ is denoted as $\back_\lnodeidx$.}
\label{fig:forward_backward}
\end{figure}

Interestingly, we can also invert this computation to compute the number of paths which lead from
any node $\rgnode$ to the last node in a DAG.

\begin{thm} \label{thm:count_paths_backward}
Let $\graph = (\nodes, \edges)$ be a DAG with ordering relation $<$ and let $\rgnode_1, \ldots,
\rgnode_\rnodelim$ be the nodes in $\nodes$ as ordered according to $<$. Then, Algorithm~\ref{alg:count_paths_backward}
returns a $\rnodelim \times 1$ vector $\vec \back$, such that $\back_\lnodeidx$ is exactly the
number of paths leading from $\rgnode_\lnodeidx$ to $\rgnode_\rnodelim$.
Further, Algorithm~\ref{alg:count_paths_backward} runs in $\effic(\rnodelim)$ time and space complexity.
\begin{proof}
Note that the structure of this proof is exactly symmetric to Theorem~\ref{thm:count_paths_forward}.

To prove this result, we first show two lemmata:
\begin{enumerate}
\item Algorithm~\ref{alg:count_paths_backward} visits all nodes from which $\rgnode_\rnodelim$ is reachable
in descending order, and no other nodes.
\item When Algorithm~\ref{alg:count_paths_backward} visits node $\rgnode_\lnodeidx$, $\back_\lnodeidx$ contains
exactly the number of paths from $\rgnode_\lnodeidx$ to $\rgnode_\rnodelim$.
\end{enumerate}
We call a note \emph{visited}, if it is pulled from $Q$. We proof both lemmata by induction over $\lnodeidx$
in descending order.

\begin{enumerate}
\item Our base case is $\rgnode_\rnodelim$, which is indeed visited first.

Now, assume that the claim holds for nodes $\geq \rgnode$ such that $\rgnode_\rnodelim$ is reachable from $\rgnode$.
Consider now the largest $\rgnode_\lnodeidx$, such that $\rgnode > \rgnode_\lnodeidx$ and $\rgnode_\rnodelim$ is reachable
from $\rgnode_\lnodeidx$. Then, there is a path $\lgnode_0, \ldots, \lgnode_\pathlim$
with $\lgnode_0 = \rgnode_\lnodeidx$ and $\lgnode_\pathlim = \rgnode_\rnodelim$.
Because $\graph$ is a DAG, $\lgnode_1 > \rgnode_\rnodeidx$.
Further, because $\lgnode_1, \ldots, \lgnode_\pathlim$ is a path from $\lgnode_1$ to $\rgnode_\rnodelim$,
$\rgnode_\rnodelim$ is reachable from $\lgnode_1$. Because $\rgnode_\lnodeidx$ is per definition the
largest node smaller than $\rgnode$ from which $\rgnode_\rnodelim$ is reachable, $\lgnode_1 \geq \rgnode$.
Therefore, per induction, $\lgnode_1$ has been visited before. This implies that $\rgnode_\lnodeidx \in Q$.
Because we select the maximum from $Q$ in each iteration, and because all elements larger than $\rgnode_\lnodeidx$
have been visited before (and are not visited again due to the DAG property), $\rgnode_\lnodeidx$ will be
visited next. Therefore, still all nodes from which $\rgnode_\rnodelim$ is reachable are visited in descending order,
and all nodes which are visited are nodes from which $\rgnode_\rnodelim$ is reachable.
\item Again, our base case is $\rgnode_\rnodelim$, which is visited first. As it is visited,
we have $\back_\rnodelim = 1$. And indeed there is only one path from $\rgnode_\rnodelim$ to $\rgnode_\rnodelim$,
namely the trivial path.

Now, assume that the claim holds for all nodes $\geq \rgnode$ from which $\rgnode_\rnodelim$ is reachable.
Then, consider the largest node $\rgnode_\lnodeidx < \rgnode$ from which $\rgnode_\rnodelim$ is reachable.
Further, let $\rgnode_{\lnodeidx_1}, \ldots, \rgnode_{\lnodeidx_\lnodelim}$ be all nodes from which
$\rgnode_\rnodelim$ is reachable, such that $(\rgnode_\lnodeidx, \rgnode_{\lnodeidx_\rnodeidx}) \in \edges$.
Because $\graph$ is a DAG, $\rgnode_{\lnodeidx_\rnodeidx} > \rgnode_\lnodeidx$ for all $\rnodeidx$.
Further, because $\rgnode_\rnodelim$ is reachable from $\rgnode_{\lnodeidx_\rnodeidx}$
and $\rgnode_\lnodeidx$ is the largest node smaller than $\rgnode$ from which $\rgnode_\rnodelim$
is reachable, it must hold $\rgnode_{\lnodeidx_\rnodeidx} \geq \rgnode$.
Therefore, per induction, $\back_{\lnodeidx_\rnodeidx}$ is the number of paths from $\rgnode_{\lnodeidx_\rnodeidx}$
to $\rgnode_\rnodelim$. For any such path $\pth$, the concatenation $\rgnode_\lnodeidx \concat \pth$
is a path from $\rgnode_\lnodeidx$ to $\rgnode_\rnodelim$. Conversely, we can decompose any path
$\pth'$ from $\rgnode_\lnodeidx$ to $\rgnode_\rnodelim$ as $\pth' = \rgnode_\lnodeidx \concat \pth$
where $\pth$ is a path from $\rgnode_{\lnodeidx_\rnodeidx}$ to $\rgnode_\rnodelim$ for some $\rnodeidx$.
Accordingly, the number of paths from $\rgnode_\lnodeidx$ to $\rgnode_\rnodelim$ is exactly
$\sum_{\rnodeidx = 1}^\lnodelim \back_{\lnodeidx_\rnodeidx}$.

Finally, because of the first lemma, we know that all $\rgnode_{\lnodeidx_\rnodeidx}$ have been visited already (without duplicates),
and that on each of these visits, $\back_{\lnodeidx_\rnodeidx}$ has been added to $\back_\lnodeidx$.
Therefore, we obtain $\back_\lnodeidx = \sum_{\rnodeidx=1}^\lnodelim \back_{\lnodeidx_\rnodeidx}$.
\end{enumerate}

Because Lemma 1 implies that we do not visit any node larger than $\rgnode_\lnodeidx$ after $\rgnode_\lnodeidx$ has
been visited, the value $\back_\lnodeidx$ does not change after $\lnodeidx$ is visited. Therefore,
$\back_{\lnodeidx}$ still contains the number of paths from $\rgnode_\lnodeidx$ to $\rgnode_\pathlim$ at the end of the
algorithm.

Regarding runtime, it follows from the first lemma that, per iteration,
exactly one reachable node is processed and will not be visited again. In the worst case,
all nodes in the graph are reachable, which yields $\effic(\rnodelim)$ iterations.
In each iteration we need to retrieve the maximum of $Q$ and insert all $\lgnode$ into $Q$, for which
$(\lgnode, \rgnode) \in \edges$.
Both is possible in constant time if a suitable data structure for $Q$ is used. If one uses
a tree structure for $Q$, the runtime rises to $\effic(\rnodelim \cdot \log(\rnodelim))$.
The space complexity is $\effic(\rnodelim)$ because $\vec \back$ has $\rnodelim$ entries
and $Q$ can not exceed $\rnodelim$ entries.
\end{proof}
\end{thm}

\begin{algorithm}
\caption{An algorithm to count the number of paths between each node $\rgnode_\lnodeidx$ and node $\rgnode_\rnodelim$
in a DAG $\graph = (\{\rgnode_1, \ldots, \rgnode_\rnodelim\}, \edges)$, where $\rgnode_1, \ldots, \rgnode_\rnodelim$ is the ordered
node list according to the DAGs ordering relation $<$.}
\label{alg:count_paths_backward}
\begin{algorithmic}
\Function{count-paths-backward}{A DAG $\graph = (\{\rgnode_1, \ldots, \rgnode_\rnodelim\}, \edges)$, an ordering relation $<$.}
\State $\vec \back \gets \rnodelim \times 1$ vector of zeros.
\State $\back_\rnodelim \gets 1$.
\State $Q \gets \{ \rgnode_\rnodelim \}$.
\While{$Q \neq \emptyset$}
	\State $\rgnode_\rnodeidx \gets \max_< Q$.
	\State $Q \gets Q \setminus \{ \rgnode_\rnodeidx \}$.
	\For{$(\rgnode_\lnodeidx, \rgnode_\rnodeidx) \in \edges$}
		\State $\back_\lnodeidx \gets \back_\lnodeidx + \back_\rnodeidx$.
		\State $Q \gets Q \cup \{ \rgnode_\lnodeidx \}$.
	\EndFor
\EndWhile
\State \Return $\vec \back$.
\EndFunction
\end{algorithmic}
\end{algorithm}

As an example, consider the DAG on the right in Figure~\ref{fig:coopt_graph}.
The resulting $\back$-values for all nodes are indicated in Figure~\ref{fig:forward_backward}.

Beyond the utility of counting the number of paths in linear time, the combination of both algorithms
also permits us to compute how often a certain edge of the graph occurs in paths from $\rgnode_1$ to
$\rgnode_\rnodelim$.

\begin{thm}\label{thm:forward_backward}
Let $\graph = (\nodes = \{\rgnode_1, \ldots, \rgnode_\rnodelim\}, \edges)$ be a DAG with ordering
relation $<$ where $\rgnode_1, \ldots, \rgnode_\rnodelim$ are ordered according to $<$. Further,
let $\vec \forw$ be the result of Algorithm~\ref{alg:count_paths_forward} for $\graph$,
and let $\vec \back$ be the result of Algorithm~\ref{alg:count_paths_backward} for $\graph$.
Then, for any edge $(\rgnode_\lnodeidx, \rgnode_\rnodeidx) \in \edges$ it holds:
$\forw_\lnodeidx \cdot \back_\rnodeidx$ is precisely the number of paths from $\rgnode_1$ to
$\rgnode_\rnodelim$ which contain $\rgnode_\lnodeidx, \rgnode_\rnodeidx$, that is, paths
$\pth = \lgnode_1, \ldots, \lgnode_\pathlim$ such that an $\pathidx \in \{1, \ldots, \pathlim-1\}$
exists for which $\lgnode_\pathidx = \rgnode_\lnodeidx$ and $\lgnode_{\pathidx+1} = \rgnode_\rnodeidx$.

\begin{proof}
Let $(\rgnode_\lnodeidx, \rgnode_\rnodeidx) \in \edges$ and let $\lnodelim$ be the number of paths
from $\rgnode_1$ to $\rgnode_\rnodelim$ which traverse $(\lgnode, \rgnode)$. Further, let
$\lgnode_1, \ldots, \lgnode_{\pathlim'}$ be a path from $\rgnode_1$ to $\rgnode_\lnodeidx$ and let
$\lgnode_{\pathlim'+1}, \ldots, \lgnode_\pathlim$ be a path from $\rgnode_\rnodeidx$ to
$\rgnode_\rnodelim$. Then, because $(\rgnode_\lnodeidx, \rgnode_\rnodeidx) \in \edges$,
$\lgnode_1, \ldots, \lgnode_\pathlim$ is a path from $\rgnode_1$ to $\rgnode_\rnodelim$ in
$\graph$ which traverses $(\rgnode_\lnodeidx, \rgnode_\rnodeidx)$. We know by virtue of
Theorem~\ref{thm:count_paths_forward} that the number of paths from
$\rgnode_1$ to $\rgnode_\lnodeidx$ is $\forw_\lnodeidx$, and we know by virtue of
Theorem~\ref{thm:count_paths_backward} that the number of paths from $\rgnode_\rnodeidx$ to
$\rgnode_\rnodelim$ is $\back_\rnodeidx$. Now, as we noted before, any combination of a path
counted in $\forw_\lnodeidx$ and a path counted in $\back_\rnodeidx$ is a path from $\rgnode_1$
to $\rgnode_\rnodelim$, and any of these combinations is unique. Therefore, we obtain
$\lnodelim \geq \forw_\lgnode \cdot \back_\rgnode$.

Further, we note that we can decompose \emph{any} path from $\rgnode_1$ to $\rgnode_\rnodelim$ as
illustrated above, such that $\lnodelim \leq \forw_\lgnode \cdot \back_\rgnode$.
\end{proof}
\end{thm}

For the example DAG from Figure~\ref{fig:coopt_graph} we show all possible paths from $(1, 1, 1, 1)$ to
$(1, 6, 1, 3)$ in Figure~\ref{fig:forward_backward}. For every edge in this DAG you can verify that,
indeed, the number of traversing paths is equivalent to the $\forw$ value of the source node
times the $\back$ value of the target node.

Note that the number of paths which traverses a certain edge reveals crucial information about the
co-optimal mappings. In particular, if we consider an edge of the form
$\big((\lkeyrootidx, \lnodeidx, \rkeyrootidx, \rnodeidx), (\lkeyrootidx', \lnodeidx+1, \rkeyrootidx', \rnodeidx+1)\big)$,
the number of paths from $(1, 1, 1, 1)$ to $(1, |\lforest|+1, 1, |\rforest|+1)$ which traverse
this edge is an estimate of the number of co-optimal mappings which contain the tuple $(\lnodeidx, \rnodeidx)$.
Unfortunately, this estimate is not necessarily exact, because there may be multiple paths through
the co-optimal edit graph which correspond to the same co-optimal mapping.

In particular, excessive paths occure whenever $\cost(\lnode_\lnodeidx, \gap)
+ \cost(\gap, \rnode_\rnodeidx) = \cost(\lnode_\lnodeidx, \rnode_\rnodeidx)$. In these cases,
deletion, replacement, and insertion are all co-optimal, and thus there exist three paths from
$(\lkeyrootidx, \lnodeidx, \rkeyrootidx, \rnodeidx)$ to $(\lkeyrootidx, \lnodeidx+1, \rkeyrootidx, \rnodeidx+1)$,
one which uses a deletion first and then an insertion, one which uses only a replacement, and one
which uses an insertion first and then a deletion. The first and last of these paths correspond
to the same co-optimal mapping, leading to overcounting. We can avoid this overcounting by
covering this special case explicitly. This results in a new forward-counting-Algorithm~\ref{alg:edit_paths_forward},
a new backward-counting-Algorithm~\ref{alg:edit_paths_backward}, and a forward-backward
Algorithm~\ref{alg:forward_backward} which characterize the number of co-optimal mappings rather
than the number of co-optimal paths.

\begin{thm}
Let $\ltree$ and $\rtree$ be trees over some alphabet $\alphabet$ and let $\cost$ be a 
cost function that is non-negative, self-equal, and conforms to the triangular inequality.
Further, let $\graph_{\ltree, \rtree, \cost} = (\nodes, \edges)$ be the co-optimal edit graph corresponding
to $\ltree$, $\rtree$, and $\cost$. Then, the first output argument of
Algorithm~\ref{alg:forward_backward} is a $|\ltree| \times |\rtree|$ matrix $\FB$ such that
$\FB_{\lnodeidx, \rnodeidx}$ is exactly the number of co-optimal mappings which contain $(\lnodeidx, \rnodeidx)$.
Further, the second output argument of Algorithm~\ref{alg:forward_backward} is the number of
co-optimal mappings.
Finally, Algorithm~\ref{alg:forward_backward} has $\effic(|\ltree|^6 \cdot |\rtree|^6)$
time and $\effic(|\ltree|^2 \cdot |\rtree|^2)$ space complexity.

\begin{proof}
For the technical details of this proof, refer to my dissertation \parencite{Paassen2019thesis}.
Here, I provide a sketch of the proof.

First, we observe that Algorithm~\ref{alg:edit_paths_forward} is analogous to Algorithm~\ref{alg:count_paths_forward},
and that Algorithm~\ref{alg:edit_paths_backward} is analogous to Algorithm~\ref{alg:count_paths_backward}.
The latter analogy holds because we just postpone adding the contributions to $\back_\lnodeidx$ to
the visit of $\back_\lnodeidx$ itself, but all contributions are still collected. We further speed
up the process by considering only cells of the dynamic programming matrix which are actually reachable
from $(1, 1, 1, 1)$. Another non-obvious part of the analogy is that we go into recursion to compute the
number of co-optimal paths for a subtree replacement. In this regard, we note that we can extend
each path from $(1, 1, 1, 1)$ to $(\lkeyrootidx, \lnodeidx, \rkeyrootidx, \rnodeidx)$ to a path to
$(\lkeyrootidx, \rleaf_{\ltree}(\lnodeidx) + 1, \rkeyrootidx, \rleaf_{\rtree}(\rnodeidx) + 1)$
by using one of the possible paths in the co-optimal edit graph corresponding to $\ltree_\lnodeidx$
and $\rtree_\rnodeidx$. However, this would over-count the paths which delete the node $\lnode_\lnodeidx$
or insert the node $\rnode_\rnodeidx$, which we prevent by setting $\mat \fdist_{1, 2}' = \mat \fdist_{2, 1}' = \infty$.
The same argument holds for the backwards case: We can extend any path from $(\lkeyrootidx, \rleaf_{\ltree}(\lnodeidx) + 1, \rkeyrootidx, \rleaf_{\rtree}(\rnodeidx) + 1)$
to $(1, |\ltree| + 1, 1, |\rtree| + 1)$ to a path from $(\lnodeidx, \rnodeidx)$ to $(|\ltree| + 1, |\rtree| + 1)$
by using one of the possible paths in the co-optimal edit graph corresponding to $\ltree_\lnodeidx$
and $\rtree_\rnodeidx$.

Finally, algorithm~\ref{alg:forward_backward} computes the products of $\forw$ and $\back$-values
according to Theorem~\ref{thm:forward_backward}. The only special case is, once again, the case
of subtree replacements. In that case, we can again argue that, for any combination of a path which
leads from $(1, 1, 1, 1)$ to $(\lkeyrootidx, \lnodeidx, \rkeyrootidx, \rnodeidx)$, and a path which
leads from $(\lkeyrootidx, \rleaf_{\ltree}(\lnodeidx) + 1, \rkeyrootidx, \rleaf_{\rtree}(\rnodeidx) + 1)$
to $(1, |\ltree| + 1, 1, |\rtree| + 1)$, we can construct a path from $(1,1, 1, 1)$ to $(1, |\ltree| + 1, 1, |\rtree| + 1)$
by inserting a \enquote{middle piece} which corresponds to a path in the co-optimal edit graph
for $\ltree_\lnodeidx$ and $\rtree_\rnodeidx$. Therefore, for
$\fb = \Forw_{\lnodeidx, \rnodeidx} \cdot \Back_{\rleaf_{\ltree}(\lnodeidx) + 1, \rleaf_{\rtree}(\rnodeidx) + 1}$,
$\fb \cdot \FB'_{\lnodeidx', \rnodeidx'}$
is an additional contribution to the count of $(\lnodeidx' + \lnodeidx - 1, \rnodeidx' + \rnodeidx - 1)$.

Now, consider the efficiency claims. First, we analyze Algorithm~\ref{alg:count_paths_forward}.
In the worst case, lines 27-31 need to be executed in each possible iteration.
In that case, $\mat \fdist'$ and $\mat \dist'$ need to be computed via Algorithm~\ref{alg:ted},
which requires $\effic(\siz{\ltree}^2\cdot\siz{\rtree}^2)$ steps and
$\effic(\siz{\ltree} \cdot \siz{\rtree})$ space.
Including the recursive calls, this can occur $\effic(\siz{\ltree}^2\cdot\siz{\rtree}^2)$
times at worst such that Algorithm~\ref{alg:count_paths_forward} has an overall
runtime complexity of $\effic(\siz{\ltree}^4 \cdot \siz{\rtree}^4)$.

Regarding space complexity, each level of recursion needs to maintain a constant
number of matrices of size $\effic(\siz{\ltree}\cdot \siz{\rtree})$. A worst,
there can be $\effic(\siz{\ltree}\cdot \siz{\rtree})$ levels of recursion active
at the same time, implying a space complexity of $\effic(\siz{\ltree}^2\cdot \siz{\rtree}^2)$.

Now, note that Algorithm~\ref{alg:count_paths_backward}, by construction, iterates over the
same elements as Algorithm~\ref{alg:count_paths_forward} and has the same structure,
such that the complexity results carry over.

Finally, regarding Algorithm~\ref{alg:edit_paths_forward} itself, we find that, in the
worst case, lines 15-23 get executed in every possible iteration. These lines
include a recursive call to Algorithm~\ref{alg:edit_paths_forward}, and in each
such recursive call, Algorithm~\ref{alg:count_paths_forward} and Algorithm~\ref{alg:count_paths_backward}
get executed. With the same argument as before, we perform at most
$\effic(\siz{\ltree}^2\cdot\siz{\rtree}^2)$ of such recursive calls, yielding an overall
runtime complexity of $\effic(\siz{\ltree}^6 \cdot \siz{\rtree}^6)$ in the worst case.

Regarding space complexity, each level of recursion needs to maintain a constant number
of matrices of size $\effic(\siz{\ltree}\cdot \siz{\rtree})$. A worst,
there can be $\effic(\siz{\ltree}\cdot \siz{\rtree})$ levels of recursion active
at the same time, implying a space complexity of $\effic(\siz{\ltree}^2\cdot \siz{\rtree}^2)$.
\end{proof}
\end{thm}

Note that the version of the algorithm presented here is dedicated to minimize space
complexity. By additionally tabulating $\FB$ for all subtrees, space complexity
rises to $\effic(\siz{\ltree}^4 \cdot \siz{\rtree}^4)$ in the worst case, but
runtime complexity is reduced to $\effic(\siz{\ltree}^3 \cdot \siz{\rtree}^3)$\footnote{This is the
version we implemented in our \href{https://gitlab.ub.uni-bielefeld.de/bpaassen/python-edit-distances/-/blob/master/edist/ted.pyx}{reference implementation}}.
Another point to note is that the worst case for this algorithm is quite unlikely.
First, both input trees would have to be left- or right-heavy. Second, in every step of
the computation, multiple options have to be co-optimal, which only occurs in
degenerate cases where, for example, the deletion or insertion cost for all symbols
is zero.

\begin{algorithm}
\caption{A variation of the forward path-counting Algorithm~\ref{alg:count_paths_forward} for
the TED.}
\label{alg:edit_paths_forward}
\begin{algorithmic}[1]
\Function{forward}{Two trees $\ltree$ and $\rtree$, the matrices $\mat \dist$ and $\mat \fdist$
after executing Algorithm~\ref{alg:ted}, and a cost function $\cost$}
\State Initialize $\Forw$ as a $(\siz{\ltree} + 1) \times (\siz{\rtree} +1 )$ matrix of zeros.
\State $\Forw_{1, 1} \gets 1$, $Q \gets \{ (1, 1) \}$
\State $C \gets \emptyset$.
\While{$Q \neq \emptyset$}
	\State $(\lnodeidx, \rnodeidx) \gets \min Q$. \Comment Lexicographic ordering
	\State $Q \gets Q \setminus \{ (\lnodeidx, \rnodeidx) \}$.
	\State $C \gets C \cup \{ (\lnodeidx, \rnodeidx) \}$.
	\If{$\lnodeidx \leq \siz{\ltree} \wedge \mat \fdist_{\lnodeidx, \rnodeidx} = \cost(\lnode_\lnodeidx, \gap) + \mat \fdist_{\lnodeidx + 1, \rnodeidx}$}
		\State $\Forw_{\lnodeidx + 1, \rnodeidx} \gets \Forw_{\lnodeidx + 1, \rnodeidx} +  \Forw_{\lnodeidx, \rnodeidx}$.
		\State $Q \gets Q \cup \{ (\lnodeidx + 1, \rnodeidx) \}$.
	\EndIf
	\If{$\rnodeidx \leq \siz{\rtree} \wedge \mat \fdist_{\lnodeidx, \rnodeidx} = \cost(\gap, \rnode_\rnodeidx) + \mat \fdist_{\lnodeidx, \rnodeidx + 1}$}
		\State $\Forw_{\lnodeidx, \rnodeidx + 1} \gets \Forw_{\lnodeidx, \rnodeidx + 1} + \Forw_{\lnodeidx, \rnodeidx}$.
		\State $Q \gets Q \cup \{ (\lnodeidx, \rnodeidx + 1) \}$.
	\EndIf
	\If{$\lnodeidx = \siz{\ltree} + 1 \vee \rnodeidx = \siz{\rtree} + 1 \vee \cost(\lnode_\lnodeidx, \rnode_\rnodeidx) = \cost(\lnode_\lnodeidx, \gap) + \cost(\gap, \rnode_\rnodeidx)$}
		\State \textbf{continue}
	\EndIf
	\If{$\rleaf_{\ltree}(\lnodeidx) = \rleaf_{\ltree}(1) \wedge \rleaf_{\rtree}(\rnodeidx) = \rleaf_{\rtree}(1)$}
		\If{$\mat \fdist_{\lnodeidx, \rnodeidx} = \mat \fdist_{\lnodeidx + 1, \rnodeidx + 1} + \cost(\lnode_\lnodeidx, \rnode_\rnodeidx)$}
			\State $\Forw_{\lnodeidx + 1, \rnodeidx + 1} \gets \Forw_{\lnodeidx + 1, \rnodeidx + 1} + \Forw_{\lnodeidx, \rnodeidx}$
			\State $Q \gets Q \cup \{ (\lnodeidx + 1, \rnodeidx + 1) \}$.
		\EndIf
	\Else
		\If{$\mat \fdist_{\lnodeidx, \rnodeidx} = \mat \fdist_{\rleaf_{\ltree}(\lnodeidx) + 1, \rleaf_{\rtree}(\rnodeidx) + 1} + \mat \dist_{\lnodeidx, \rnodeidx}$}
			\State Compute $\mat \fdist'$ and $\mat \dist'$ via Algorithm~\ref{alg:ted} for the subtrees $\ltree^\lnodeidx$ and $\rtree^\rnodeidx$.
			\State $\mat \fdist'_{1, 2} \gets \infty$. $\mat \fdist'_{2, 1} \gets \infty$.
			\State $(Q', \Forw') \gets $ \Call{forward}{$\ltree^\lnodeidx$, $\rtree^\rnodeidx$, $\mat \dist'$, $\mat \fdist'$, $\cost$}.
			\State $\Forw_{\rleaf_{\ltree}(\lnodeidx) + 1, \rleaf_{\rtree}(\rnodeidx) + 1} \gets \Forw_{\rleaf_{\ltree}(\lnodeidx) + 1, \rleaf_{\rtree}(\rnodeidx) + 1} 
			+ \Forw'_{\siz{\ltree^\lnodeidx} + 1, \siz{\rtree^\rnodeidx} + 1} \cdot \Forw_{\lnodeidx, \rnodeidx}$.
			\State $Q \gets Q \cup \{ (\rleaf_{\ltree}(\lnodeidx) + 1, \rleaf_{\rtree}(\rnodeidx) + 1) \}$.
		\EndIf
	\EndIf
\EndWhile
\State \Return $(C, \Forw)$.
\EndFunction
\end{algorithmic}
\end{algorithm}

\begin{algorithm}
\caption{A variation of the backward path-counting Algorithm~\ref{alg:count_paths_backward} for
the TED.}
\label{alg:edit_paths_backward}
\begin{algorithmic}[1]
\Function{backward}{Two trees $\ltree$ and $\rtree$, the matrices $\mat \dist$ and $\mat \fdist$
after executing Algorithm~\ref{alg:ted}, a cost function $\cost$, and a set of tuples $C$ as returned by Algorithm~\ref{alg:edit_paths_forward}.}
\State Initialize $\Back$ as a $(\siz{\ltree} + 1) \times (\siz{\rtree} +1 )$ matrix of zeros.
\State $\Back_{\siz{\ltree} + 1, \siz{\rtree} + 1} \gets 1$.
\While{$C \neq \emptyset$}
	\State $(\lnodeidx, \rnodeidx) \gets \max C$. \Comment Lexicographic ordering
	\State $C \gets C \setminus \{ (\lnodeidx, \rnodeidx) \}$.
	\If{$\lnodeidx \leq \siz{\ltree} \wedge \mat \fdist_{\lnodeidx, \rnodeidx} = \cost(\lnode_\lnodeidx, \gap) + \mat \fdist_{\lnodeidx + 1, \rnodeidx}$}
		\State $\Back_{\lnodeidx, \rnodeidx} \gets \Back_{\lnodeidx, \rnodeidx} + \Back_{\lnodeidx + 1, \rnodeidx}$
	\EndIf
	\If{$\rnodeidx \leq \siz{\rtree} \wedge \mat \fdist_{\lnodeidx, \rnodeidx} = \cost(\gap, \rnode_\rnodeidx) + \mat \fdist_{\lnodeidx, \rnodeidx + 1}$}
		\State $\Back_{\lnodeidx, \rnodeidx} \gets \Back_{\lnodeidx, \rnodeidx} + \Back_{\lnodeidx, \rnodeidx + 1}$
	\EndIf
	\If{$\lnodeidx = \siz{\ltree} + 1 \vee \rnodeidx = \siz{\rtree} + 1 \vee \cost(\lnode_\lnodeidx, \rnode_\rnodeidx) = \cost(\lnode_\lnodeidx, \gap) + \cost(\gap, \rnode_\rnodeidx)$}
		\State \textbf{continue}
	\EndIf
	\If{$\rleaf_{\ltree}(\lnodeidx) = \rleaf_{\ltree}(1) \wedge \rleaf_{\rtree}(\rnodeidx) = \rleaf_{\rtree}(1)$}
		\If{$\mat \fdist_{\lnodeidx, \rnodeidx} = \mat \fdist_{\lnodeidx + 1, \rnodeidx + 1} + \cost(\lnode_\lnodeidx, \rnode_\rnodeidx)$}
			\State $\Back_{\lnodeidx, \rnodeidx} \gets \Back_{\lnodeidx, \rnodeidx} + \Back_{\lnodeidx + 1, \rnodeidx + 1}$
		\EndIf
	\Else
		\If{$\mat \fdist_{\lnodeidx, \rnodeidx} = \mat \fdist_{\rleaf_{\ltree}(\lnodeidx) + 1, \rleaf_{\rtree}(\rnodeidx) + 1} + \mat \dist_{\lnodeidx, \rnodeidx}$}
			\State Compute $\mat \fdist'$ and $\mat \dist'$ via Algorithm~\ref{alg:ted} for the subtrees $\ltree^\lnodeidx$ and $\rtree^\rnodeidx$.
			\State $\mat \fdist'_{1, 2} \gets \infty$. $\mat \fdist'_{2, 1} \gets \infty$.
			\State $(Q', \Forw') \gets $ \Call{forward}{$\ltree^\lnodeidx$, $\rtree^\rnodeidx$, $\mat \dist'$, $\mat \fdist'$, $\cost$}.
			\State $\Back_{\lnodeidx, \rnodeidx} \gets \Back_{\lnodeidx, \rnodeidx} + \Back_{\rleaf_{\ltree}(\lnodeidx) + 1, \rleaf_{\rtree}(\rnodeidx) + 1} 
			\cdot \Forw'_{\siz{\ltree^\lnodeidx} + 1, \siz{\rtree^\rnodeidx} + 1}$.
		\EndIf
	\EndIf
\EndWhile
\State \Return $\Back$.
\EndFunction
\end{algorithmic}
\end{algorithm}

\begin{algorithm}
\caption{A forward-backward algorithm to compute the number of times the tuple $(\lnodeidx, \rnodeidx)$
occurs in co-optimal mappings for paths in the co-optimal edit graphs between two input trees
$\ltree$ and $\rtree$. The second output is the overall number of co-optimal mappings.
Refer to our \href{https://gitlab.ub.uni-bielefeld.de/bpaassen/python-edit-distances/-/blob/master/edist/ted.pyx}{project web site}
for a reference implementation.}
\label{alg:forward_backward}
\begin{algorithmic}[1]
\Function{cooptimals}{Two trees $\ltree$ and $\rtree$, the matrices
$\mat \dist$ and $\mat \fdist$ after executing algorithm~\ref{alg:ted},
and a cost function $\cost$}
\State $(C, \Forw) \gets $ \Call{forward}{$\ltree$, $\rtree$, $\mat \dist$, $\mat \fdist$, $\cost$}. \Comment Refer to Algorithm~\ref{alg:edit_paths_forward}.
\State $\Back \gets $ \Call{backward}{$\ltree$, $\rtree$, $\mat \dist$, $\mat \fdist$, $\cost$, $C$}. \Comment Refer to Algorithm~\ref{alg:edit_paths_backward}.
\State Initialize $\FB$ as a $\siz{\ltree} \times \siz{\rtree}$ matrix of zeros.
\For{$(\lnodeidx, \rnodeidx) \in C$}
	\If{$\lnodeidx = \siz{\ltree}+1 \vee \rnodeidx = \siz{\rtree}+1$}
		\State \textbf{continue}
	\EndIf
	\If{$\big(\rleaf_{\ltree}(\lnodeidx) = \siz{\ltree} \wedge \rleaf_{\rtree}(\rnodeidx) = \siz{\rtree}\big)
	\vee \cost(\lnode_\lnodeidx, \rnode_\rnodeidx) = \cost(\lnode_\lnodeidx, \gap) + \cost(\gap, \rnode_\rnodeidx)$}
		\If{$\mat \fdist_{\lnodeidx, \rnodeidx} = \mat \fdist_{\lnodeidx + 1, \rnodeidx + 1} + \cost(\lnode_\lnodeidx, \rnode_\rnodeidx)$}
			\State $\FB_{\lnodeidx, \rnodeidx} \gets \FB_{\lnodeidx, \rnodeidx} + \Forw_{\lnodeidx, \rnodeidx} \cdot \Back_{\lnodeidx + 1, \rnodeidx + 1}$.
		\EndIf
	\Else
		\If{$\mat \fdist_{\lnodeidx, \rnodeidx} = \mat \fdist_{\rleaf_{\ltree}(\lnodeidx) + 1, \rleaf_{\rtree}(\rnodeidx) + 1} + \mat \dist_{\lnodeidx, \rnodeidx}$}
			\State $\fb \gets \Forw_{\lnodeidx, \rnodeidx} \cdot \Back_{\rleaf_{\ltree}(\lnodeidx) + 1, \rleaf_{\rtree}(\rnodeidx) + 1}$.
			\State Compute $\mat \fdist'$ and $\mat \dist'$ via Algorithm~\ref{alg:ted} for the subtrees $\ltree^\lnodeidx$ and $\rtree^\rnodeidx$.
			\State $\mat \fdist'_{1, 2} \gets \infty$. $\mat \fdist'_{2, 1} \gets \infty$.
			\State $(\FB', k) \gets$ \Call{cooptimals}{$\ltree^\lnodeidx$, $\rtree^\rnodeidx$, $\mat \fdist'$, $\mat \dist'$, $\cost$}.
			\For{$\lnodeidx' \gets 1, \ldots, \siz{\ltree^\lnodeidx}$}
				\For{$\rnodeidx' \gets 1, \ldots, \siz{\rtree^\rnodeidx}$}
					\State $\FB_{\lnodeidx + \lnodeidx' - 1, \rnodeidx + \rnodeidx' - 1} \gets \FB_{\lnodeidx + \lnodeidx' - 1, \rnodeidx + \rnodeidx' - 1} + \FB'_{\lnodeidx', \rnodeidx'} \cdot \fb$.
				\EndFor
			\EndFor
		\EndIf
	\EndIf
\EndFor
\State \Return $(\FB, \Forw_{\siz{\ltree}+1, \siz{\rtree}+1})$.
\EndFunction
\end{algorithmic}
\end{algorithm}

\begin{table}
\label{tab:freqmat}
\caption{The forward matrix $\Forw$, the backward matrix $\Back$, and the matrix $\FB$
for the trees $\ltree = \sym{a}(\sym{b}(\sym{c}, \sym{d}), \sym{e})$
and $\rtree = \sym{f}(\sym{g})$ from Figure~\ref{fig:ted}, as returned by Algorithms~\ref{alg:edit_paths_forward},
\ref{alg:edit_paths_backward}, and~\ref{alg:forward_backward} respectively.
The color coding follows Figure~\ref{fig:ted}.}
\begin{center}
\begin{tabular}{ccccc}
$\Forw_{\lnodeidx, \rnodeidx}$ & $\rnodeidx$    & $1$ & $2$ & $3$\\
$\lnodeidx$ & $\lnode_\lnodeidx$ \textbackslash $\rnode_\rnodeidx$ & $\sym{f}$ & $\sym{g}$ & $\gap$ \\
\cmidrule(lr){1-1} \cmidrule(lr){2-2} \cmidrule(lr){3-5}
$1$         & {\color{aluminium6}$\sym{a}$} & \tikzmark{a11}$1$ &     &   \\
$2$         & {\color{skyblue3}$\sym{b}$}   & \tikzmark{a21}$1$ & \tikzmark{a22}$1$ &  \\
$3$         & {\color{orange3}$\sym{c}$}    & & \tikzmark{a32}$1$ &  \\
$4$         & {\color{skyblue3}$\sym{d}$}   & & \tikzmark{a42}$1$ & \tikzmark{a43}$1$ \\
$5$         & {\color{aluminium6}$\sym{e}$} & & \tikzmark{a52}$1$ & \tikzmark{a53}$5$ \\
$6$         & $\gap$    & & & \tikzmark{a63}$6$
\end{tabular}
\begin{tikzpicture}[overlay, remember picture]
\draw[edge, ->, class0, semithick, fill=none]%
(pic cs:a11)+(0,+5pt) to[bend right] (pic cs:a21);
\draw[edge, ->, class0, semithick, fill=none]%
(pic cs:a11)+(+5pt,+5pt) to (pic cs:a22);
\draw[edge, ->, class1, semithick, fill=none, transform canvas={yshift=-3pt}]%
(pic cs:a21)+(+5pt,+5pt) to (pic cs:a53);
\draw[edge, ->, class1, semithick, fill=none, transform canvas={yshift=+3pt}]%
(pic cs:a21)+(+5pt,+5pt) to (pic cs:a53);
\draw[edge, ->, class1, semithick, fill=none, transform canvas={yshift=+6pt}]%
(pic cs:a22)+(+5pt,0) to (pic cs:a53);
\draw[edge, ->, class1, semithick, fill=none, transform canvas={xshift=+5pt}]%
(pic cs:a22)+(0,+5pt) to[bend left] (pic cs:a32);
\draw[edge, ->, class2, semithick, fill=none]%
(pic cs:a32)+(+5pt,+5pt) to (pic cs:a43);
\draw[edge, ->, class2, semithick, fill=none, transform canvas={xshift=+5pt}]%
(pic cs:a32)+(0,+5pt) to[bend left] (pic cs:a42);
\draw[edge, ->, class1, semithick, fill=none, transform canvas={xshift=+5pt}]%
(pic cs:a42)+(0,+5pt) to[bend left] (pic cs:a52);
\draw[edge, ->, class1, semithick, fill=none]%
(pic cs:a42)+(+5pt,+5pt) to (pic cs:a53);
\draw[edge, ->, class1, semithick, fill=none, transform canvas={xshift=+5pt}]%
(pic cs:a43)+(0,+5pt) to[bend left] (pic cs:a53);
\draw[edge, ->, class0, semithick, fill=none]%
(pic cs:a52)+(+5pt,+5pt) to (pic cs:a63);
\draw[edge, ->, class0, semithick, fill=none, transform canvas={xshift=+5pt}]%
(pic cs:a53)+(0,+5pt) to[bend left] (pic cs:a63);
\end{tikzpicture}
\hspace{0.5cm}%
%
\begin{tabular}{ccccc}
$\Back_{\lnodeidx, \rnodeidx}$ & $\rnodeidx$    & $1$ & $2$ & $3$\\
$\lnodeidx$ & $\lnode_\lnodeidx$ \textbackslash $\rnode_\rnodeidx$ & $\sym{f}$ & $\sym{g}$ & $\gap$ \\
\cmidrule(lr){1-1} \cmidrule(lr){2-2} \cmidrule(lr){3-5}
$1$         & {\color{aluminium6}$\sym{a}$} & \tikzmark{b11}$6$ &     &   \\
$2$         & {\color{skyblue3}$\sym{b}$}   & \tikzmark{b21}$2$ & \tikzmark{b22}$4$ &  \\
$3$         & {\color{orange3}$\sym{c}$}    & & \tikzmark{b32}$3$ &  \\
$4$         & {\color{skyblue3}$\sym{d}$}   & & \tikzmark{b42}$2$ & \tikzmark{b43}$1$ \\
$5$         & {\color{aluminium6}$\sym{e}$} & & \tikzmark{b52}$1$ & \tikzmark{b53}$1$ \\
$6$         & $\gap$    & & & \tikzmark{b63}$1$
\end{tabular}
\begin{tikzpicture}[overlay, remember picture]
\draw[edge, <-, class0, semithick, fill=none]%
(pic cs:b11)+(0,+5pt) to[bend right] (pic cs:b21);
\draw[edge, <-, class0, semithick, fill=none]%
(pic cs:b11)+(+5pt,+5pt) to (pic cs:b22);
\draw[edge, <-, class1, semithick, fill=none, transform canvas={yshift=-3pt}]%
(pic cs:b21)+(+5pt,+5pt) to (pic cs:b53);
\draw[edge, <-, class1, semithick, fill=none, transform canvas={yshift=+3pt}]%
(pic cs:b21)+(+5pt,+5pt) to (pic cs:b53);
\draw[edge, <-, class1, semithick, fill=none, transform canvas={yshift=+6pt}]%
(pic cs:b22)+(+5pt,0) to (pic cs:b53);
\draw[edge, <-, class1, semithick, fill=none, transform canvas={xshift=+5pt}]%
(pic cs:b22)+(0,+5pt) to[bend left] (pic cs:b32);
\draw[edge, <-, class2, semithick, fill=none]%
(pic cs:b32)+(+5pt,+5pt) to (pic cs:b43);
\draw[edge, <-, class2, semithick, fill=none, transform canvas={xshift=+5pt}]%
(pic cs:b32)+(0,+5pt) to[bend left] (pic cs:b42);
\draw[edge, <-, class1, semithick, fill=none, transform canvas={xshift=+5pt}]%
(pic cs:b42)+(0,+5pt) to[bend left] (pic cs:b52);
\draw[edge, <-, class1, semithick, fill=none]%
(pic cs:b42)+(+5pt,+5pt) to (pic cs:b53);
\draw[edge, <-, class1, semithick, fill=none, transform canvas={xshift=+5pt}]%
(pic cs:b43)+(0,+5pt) to[bend left] (pic cs:b53);
\draw[edge, <-, class0, semithick, fill=none]%
(pic cs:b52)+(+5pt,+5pt) to (pic cs:b63);
\draw[edge, <-, class0, semithick, fill=none, transform canvas={xshift=+5pt}]%
(pic cs:b53)+(0,+5pt) to[bend left] (pic cs:b63);
\end{tikzpicture}

\vspace{0.5cm}
\begin{tabular}{cccc}
$\FB_{\lnodeidx, \rnodeidx}$ & $\rnodeidx$    & $1$ & $2$ \\
$\lnodeidx$ & $\lnode_\lnodeidx$ \textbackslash $\rnode_\rnodeidx$ & $\sym{f}$ & $\sym{g}$  \\
\cmidrule(lr){1-1} \cmidrule(lr){2-2} \cmidrule(lr){3-4}
$1$ & {\color{aluminium6}$\sym{a}$} & {\color{aluminium6} $4$} & {\color{aluminium6} $0$} \\
$2$ & {\color{skyblue3}$\sym{b}$}   & {\color{skyblue3} $2$} & {\color{skyblue3} $1$} \\
$3$ & {\color{orange3}$\sym{c}$}    & $0$ & {\color{skyblue3} $1$} $+$ {\color{orange3} $1$} \\
$4$ & {\color{skyblue3}$\sym{d}$}   & $0$ & {\color{skyblue3} $1$} $+$ {\color{skyblue3} $1$} \\
$5$ & {\color{aluminium6}$\sym{e}$} & $0$ & {\color{aluminium6} $1$} \\
\end{tabular}
\end{center}
\end{table}

For the example TED calculation in Figure~\ref{fig:ted}, the results for $\Forw$, $\Back$, and 
$\FB$ according to Algorithm~\ref{alg:forward_backward} are shown in
Table~\ref{tab:freqmat}. By comparing with the co-optimal mappings in Figure~\ref{fig:coopt_mappings},
you can verify that this matrix does indeed sum up all co-optimal mappings.

Interestingly, the matrix $\FB$ has further helpful properties. By considering the sum over all columns and
subtracting it from the total number of co-optimal mappings we obtain the number of co-optimal mappings
in which a certain node in $\ltree$ is deleted. In our example, $a$ is deleted in $2$ co-optimal mappings,
$b$ in $3$ co-optimal mappings, $c$ in $4$ co-optimal mappings, $d$ in $4$ co-optimal mappings, and
$e$ in $5$ co-optimal mappings. Conversely, by summing up over all rows and subtracting the result
from the total number of co-optimal mappings we obtain the number of co-optimal mappings in which
a certain node of $\rtree$ is inserted. In this example, neither $f$ nor $g$ are inserted in any
co-optimal mapping.

Another interesting property is that the matrix $\FB$ represents the
frequency of certain pairings of nodes in co-optimal mappings, if we divide all entries by the
total number of co-optimal mappings. This version of the matrix also offers an alternative view on the
tree edit distance itself. 

\begin{thm}
Let $\ltree$ and $\rtree$ be trees over some alphabet $\alphabet$, and let $\cost$ be a 
cost function over $\alphabet$. Further, let $\FB$ and $k$ be the two outputs of Algorithm~\ref{alg:forward_backward}
for $\ltree$, $\rtree$, and $\cost$, and let $\freqmat_\cost(\ltree, \rtree) := \frac{1}{k} \cdot \FB$.

Then, the following equation holds:
\begin{align}
\dist_\cost(\ltree, \rtree) = &\sum_{\lnodeidx = 1}^{|\ltree|} \sum_{\rnodeidx = 1}^{|\rtree|}
	\freqmat_{\cost}(\ltree, \rtree)_{\lnodeidx, \rnodeidx} \cdot \cost(\lnode_\lnodeidx, \rnode_\rnodeidx) \\
&+ \sum_{\lnodeidx = 1}^{|\ltree|} p^{\mathrm{del}}_\lnodeidx \cdot \cost(\lnode_\lnodeidx, \gap) 
+ \sum_{\rnodeidx = 1}^{|\rtree|} p^{\mathrm{ins}}_\rnodeidx \cdot \cost(\gap, \rnode_\rnodeidx) & \text{where} \notag \\
p^{\mathrm{del}}_\lnodeidx := &1 - \sum_{\rnodeidx = 1}^{|\rtree|} \freqmat_{\cost}(\ltree, \rtree)_{\lnodeidx, \rnodeidx} \notag \\
p^{\mathrm{ins}}_\rnodeidx := &1 - \sum_{\lnodeidx = 1}^{|\ltree|} \freqmat_{\cost}(\ltree, \rtree)_{\lnodeidx, \rnodeidx} \notag 
\end{align}

\begin{proof}
Per construction, $\FB$ is equivalent to the number
of co-optimal mappings $\map$, such that $(\lnodeidx, \rnodeidx) \in \map$, and $k$ is equivalent
to the number of co-optimal mappings overall. The cost of each co-optimal mapping
is per definition $\dist_\cost(\ltree, \rtree)$. Therefore, summing over the cost of all
these mappings and dividing by the number of mappings is also equal to $\dist_\cost(\ltree, \rtree)$.
\end{proof}
\end{thm}

This alternative representation of the TED is particularly useful if one wishes to learn the parameters of the
tree edit distance, as all the computational complexity of the tree edit distance is encapsulated in the
matrix $\freqmat_{\cost}(\ltree, \rtree)_{\lnodeidx, \rnodeidx}$ and all learned parameters are linearly
multiplied with this matrix. This trick has been originally suggested by \textcite{Bellet2012} to learn
optimal parameters for the string edit distance.

Ths concludes our tutorial. For further reading, I recommend the robust tree edit distance by
\textcite{Pawlik2011}, as well as the metric learning approaches by \textcite{Bellet2012} and \textcite{Paassen2018ICML}.

\printbibliography

\end{document}

%% file: symbols.tex
\newtheorem{thm}{Theorem}

\theoremstyle{definition}
\newtheorem{dfn}{Definition}

\newcommand{\R}{\mathbb{R}}
\newcommand{\N}{\mathbb{N}}
\newcommand{\mat}[1]{\bm{#1}}

\newcommand{\effic}{\mathcal{O}}
\newcommand{\concat}{\oplus}

\newcommand{\node}{x}
\newcommand{\lnode}{x}
\newcommand{\rnode}{y}

\newcommand{\tree}{{\bar x}}
\newcommand{\ltree}{{\bar x}}
\newcommand{\rtree}{{\bar y}}
\newcommand{\ztree}{{\bar z}}
\newcommand{\trees}{\mathcal{T}}
\newcommand{\lforest}{X}
\newcommand{\rforest}{Y}

\newcommand{\siz}[1]{|#1|}
\newcommand{\childidx}{r}
\newcommand{\childlim}{R}
\newcommand{\lchildidx}{l}

\newcommand{\rchildidx}{r}

\newcommand{\alphabet}{\mathcal{X}}
\newcommand{\sym}[1]{\text{\texttt{#1}}}
\newcommand{\gap}{-}
\newcommand{\pre}{\pi}
\newcommand{\paridx}{p}

\newcommand{\edit}{\delta}
\newcommand{\edits}{\Delta}
\newcommand{\script}{\bar \delta}
\newcommand{\editidx}{t}
\newcommand{\editlim}{T}
\newcommand{\del}{\mathrm{del}}
\newcommand{\ins}{\mathrm{ins}}
\newcommand{\rep}{\mathrm{rep}}
\newcommand{\map}{M}
\newcommand{\unmappedLeft}{I}
\newcommand{\unmappedRight}{J}

\newcommand{\dist}{d}
\newcommand{\cost}{c}
\newcommand{\Cost}{C}
\newcommand{\lnodeidx}{i}
\newcommand{\lnodelim}{m}
\newcommand{\rnodeidx}{j}
\newcommand{\rnodelim}{n}
\newcommand{\fdist}{D}
\newcommand{\rleaf}{rl}
\newcommand{\keyroot}{\mathrm{k}}
\newcommand{\lkeyrootidx}{k}
\newcommand{\rkeyrootidx}{l}
\newcommand{\keyroots}{\mathcal{K}}

\newcommand{\freqmat}{\mat P}
\newcommand{\forw}{\alpha}
\newcommand{\Forw}{\mat A}
\newcommand{\back}{\beta}
\newcommand{\Back}{\mat B}
\newcommand{\fb}{\gamma}
\newcommand{\FB}{\mat \Gamma}

\newcommand{\graph}{\mathcal{G}}
\newcommand{\nodes}{V}
\newcommand{\edges}{E}
\newcommand{\lgnode}{u}
\newcommand{\rgnode}{v}
\newcommand{\pth}{p}
\newcommand{\pathidx}{t}
\newcommand{\pathlim}{T}

%% file: tikz_styles.tex
\tikzstyle{point}=[circle, inner sep=0pt, minimum size=3mm, line width=0.5mm, anchor=center]
\tikzstyle{textnode}=[draw=none, fill=none]
\tikzstyle{proto}=[diamond, inner sep=0pt, minimum size=5mm, line width=0.5mm, anchor=center]
\tikzstyle{edge}=[->, >=stealth', shorten <=2pt, shorten >=2pt, auto, line width=0.5mm]
\tikzstyle{class0color}=[aluminium6]
\tikzstyle{class0}=[draw=aluminium6, fill=aluminium4, text=aluminium6]
\tikzstyle{class1color}=[skyblue3]
\tikzstyle{class1}=[draw=skyblue3, fill=skyblue1, text=skyblue3]
\tikzstyle{class2color}=[orange3]
\tikzstyle{class2}=[draw=orange3, fill=orange1, text=orange3]